\documentclass[12pt]{article}

\usepackage[T1]{fontenc}			
\usepackage[english]{babel}
\usepackage[dvips]{epsfig}		
\usepackage{amsmath}						
\usepackage{amssymb}					
\usepackage{array}						
\usepackage{multicol}					
\usepackage{xspace}						
\usepackage{graphics}					
\usepackage{indentfirst} 			
\usepackage{latexsym}					
\usepackage{setspace}					
\usepackage{makeidx}					
\usepackage{vmargin}					
\usepackage{amsmath,amsthm,amssymb,stmaryrd,wasysym}
\usepackage{color}
\usepackage{wrapfig}
\usepackage{ulem}
\usepackage{multirow}
\usepackage{stackrel}
\usepackage{mathrsfs}
\usepackage{fancyref}
\usepackage{graphicx}
\usepackage{enumerate}
\usepackage{mdwlist}
\usepackage{caption}
\usepackage{fancybox}
\usepackage{pifont}
\usepackage[all]{xy}
\usepackage{makecell}
\usepackage{epigraph}
\usepackage{pifont}
\usepackage{dsfont}
\usepackage{rotating}
\input amssym.def

\definecolor{linkcolor}{rgb}{0,0,0.6}
\usepackage[	pdftex,colorlinks=true,	
			pdfstartview=FitV,
			linkcolor=linkcolor,
			citecolor=linkcolor,
			urlcolor=linkcolor,
			hyperindex=true,
			hyperfigures=false]
			{hyperref}

\usepackage{fancyhdr}
\makeindex

\def\be{\begin{equation}}
\def\ee{\end{equation}}
\def\bqn{\begin{eqnarray}}
\def\eqn{\end{eqnarray}}
\newcommand*\xbar[1]{
  \hbox{
    \vbox{
      \hrule height 0.5pt 
      \kern0.5ex         
      \hbox{
        \kern-0.25em      
        \ensuremath{#1}
        \kern-0.25em
      }
    }
  }
} 

\newcommand{\bea}{\begin{eqnarray}}
\newcommand{\eea}{\end{eqnarray}}
\newcommand{\p}{\partial}

\newcommand{\mR}{\mathbb R}

\newcommand{\bi}{\begin{itemize}}
\newcommand{\ei}{\end{itemize}}
\newcommand{\nn}{\nonumber}
\newcommand{\tvert}{\, |\, }
\newcommand{\pl}{\left(}
\newcommand{\pr}{\right)}
\newcommand{\acl}{\left \{ }
\newcommand{\acr}{\right \} }
\newcommand{\crl}{\left[}
\newcommand{\crr}{\right]}

\newcommand{\bR}{\rm{\bf{R}}}
\newcommand{\bb}{\rm{\bf{b}}}
\newcommand{\bbf}{\rm{\bf{f}}}

\newcommand{\w}{\wedge}

\newcommand{\half}{\frac{1}{2}}

\newcommand{\Lag}{\mathcal{L}}

\newcommand{\alg}{\mathfrak g}
\newcommand{\alh}{\mathfrak h}
\newcommand{\alp}{\mathfrak i}

\newcommand{\C}{\mathscr{C}}
\newcommand\fonc[1]{C^\infty\pl#1\pr}

\newcommand{\nr}{nonrelativistic }
\newcommand{\rel}{relativistic }

\newcommand{\Kcp}{Koszul connection}

\newcommand{\BEw}{Bargmann-Eisenhart wave }

\newcommand{\gws}{gravitational waves }

\newcommand{\vf}{\field{T\M}}

\newcommand{\Mx}{$\pl\M,\xi\pr$ }
\newcommand{\M}{\mathscr{M}}

\newcommand{\tA}{\tilde{A}}

\newcommand{\Om}{\Omega}

\newcommand{\un}{^{-1}}

\newcommand{\ie}{\textit{i.e.}\, }
\newcommand{\cf}{\textit{cf.}\, }
\newcommand{\eg}{\textit{e.g.}\, }

\newcommand{\etcp}{\textit{etc.}}

\newcommand{\br}[2]{\crl #1,#2 \crr}

\newcommand\pset[1]{\left \lbrace #1\right \rbrace}
\newcommand\bprop[1]{\begin{prop}#1\end{prop}}
\newcommand\bpropp[2]{\begin{prop}[#1]#2\end{prop}}
\newcommand\bdefi[2]{\begin{defi}[#1]#2\end{defi}}

\newcommand\blem[2]{\begin{lem}[#1]#2\end{lem}}
\newcommand\bccor[1]{\begin{cor}#1\end{cor}}
\newcommand\bcor[2]{\begin{cor}[#1]#2\end{cor}}

\newcommand\bexa[2]{\begin{exa}[#1]#2\end{exa}}

\newcommand{\Ker}{\text{\rm Ker }}
\newcommand{\Rad}{\text{\rm Rad }}
\newcommand{\Ann}{\text{\rm Ann }}
\newcommand{\End}{\text{\rm End}}

\newcommand{\barg}{\mathfrak{bar}}

\newcommand{\CARR}{{\text{\rm Carr}}}
\newcommand{\CARRZ}{{\text{\rm Carr}}_0}

\newcommand\N[1]{\overset{N}{#1}}

\newcommand\ovA[1]{\overset{A}{#1}}
\newcommand\ovbA[1]{\overset{A}{\bar #1}}
\newcommand\ovAp[1]{\overset{A'}{#1}}

\newcommand\bcase[1]{\bea\begin{cases}#1\end{cases}\eea}

\newcommand\bN[1]{\overset{\bar N}{#1}}

\newcommand\form[1]{\Omega^1\pl #1\pr}
\newcommand\forma[2]{\Omega^{#1}\pl #2\pr}
\newcommand\bforma[1]{\field{\vee^2\, #1}}

\newcommand\bform{\field{\vee^2\, T\M}}
\newcommand\bforms{\field{\vee^2\, T^*\M}}
\newcommand\field[1]{\Gamma\pl #1\pr}
\newcommand{\ff}{\form{\M}}

\newcommand{\fo}{field of observers }

\newcommand\Prop[1]{Proposition \ref{#1}}
\newcommand\Defi[1]{Definition \ref{#1}}

\newcommand\Span[1]{\text{\rm Span}\hspace{1.2mm} #1}
\newcommand\Spann[2]{\text{\rm Span}\left\lbrace #1,#2\right\rbrace}
\newcommand\Spannn[1]{\text{\rm Span}\left\lbrace #1\right\rbrace}

\newcommand\bptt[2]{\bigotimes #1\otimes\bigotimes #2	}

\newcommand{\vectasymdu}{\field{\w^2T^*\M\otimes T\M}}

\newcommand{\Milneg}{\field{\Ker\psi}}
\newcommand{\GammaV}{\mathscr V\pl\M, \xi, \psi, \gamma\pr}
\newcommand{\GammaVr}{\mathscr V\pl\M, g\pr}

\newcommand{\lmn}{{}^\lambda_{\mu\nu}}
\newcommand{\lmna}{{}^\lambda_{[\mu\nu]}}

\newcommand{\Lagxi}{\Lag_\xi}

\newcommand\fieldinv[1]{\Gamma_\text{\rm inv}\pl#1\pr}
\newcommand{\fAnnxi}{\field{\Ann \xi}}
\newcommand{\FO}{FO\pl\M,\psi\pr}
\newcommand{\LF}{LP\pl\M,\xi,\psi\pr}
\newcommand{\EC}{EC\pl\M,\xi\pr}
\newcommand{\PC}{PC\pl\M,\xi\pr}

\newcommand\NA[1]{\overset{N,A}{#1}}
\newcommand\NAp[1]{\overset{N',A'}{#1}}
\newcommand{\fV}{\field{\Ker\psi/\Span \xi}}

\newcommand\inv{\text{\rm inv}}

\newcommand{\LFinv}{LP_\inv\pl\M,\xi,\psi\pr}

\newcommand\foncinv[1]{C^\infty_\inv\pl#1\pr}

\newcommand{\leib}{{\mathfrak{leib}}}
\newcommand{\carr}{{\mathfrak{carr}}}

\newcolumntype{M}[1]{>{\centering}m{#1}}
\newcommand*{\longhookrightarrow}{\ensuremath{\lhook\joinrel\relbar\joinrel\rightarrow}}

\theoremstyle{definition}

\definecolor{rougef}{rgb}{0.56,0,0}		
\definecolor{vertf}{rgb}{0,0.5,0}		
\definecolor{bleuf}{rgb}{0,0,0.8}

\newcommand\btitle[1]{\newblock ``\textit{{#1}}''}

\newcommand\barxiv[1]{\newblock \href{http://arXiv.org/abs/#1}{\texttt{arXiv:#1}}}

\newlength{\blength}
\renewcommand{\proof}[1]{\vspace{-.05cm}
\begin{list}{\bf Proof:}
{\listparindent=\parindent\parsep=0pt \labelwidth=-0.5cm
\labelsep=\parindent \addtolength{\labelsep}{-\blength}
\addtolength{\labelsep}{0.75cm}
\itemindent=-\blength
\addtolength{\itemindent}{\parindent} \leftmargin=1.0cm}
\item
#1~$\qedsymbol$\end{list}
\vspace{.0cm}}

\theoremstyle{plain}
\newtheorem{thm}{Theorem}[section]
\newtheorem{lem}[thm]{Lemma}

\newtheorem{cor}[thm]{Corollary}
\newtheorem{prop}[thm]{Proposition}

\newtheorem{defi}[thm]{Definition}
\theoremstyle{definition}
\newtheorem{notation}[thm]{Notation}

\newtheorem{exa}[thm]{Example}

\begin{document}

\thispagestyle{empty}

 \begin{centering}

{\large {\bfseries 
Connections and dynamical trajectories \\\vspace{1mm} 
in generalised Newton-Cartan gravity II.\\\vspace{1mm}
An ambient perspective
}
\\\vspace{2mm}
}

Xavier Bekaert${}^*$ \& Kevin Morand${}^*{}^\flat{}^\sharp$\\
\vspace{2.5mm}
${}^*${Institut Denis Poisson,\\
Universit\'e de Tours, Universit\'e d'Orl\'eans, CNRS\\
Parc de Grandmont, 37200 Tours, France}\\
\vspace{2.5mm}
${}^\flat$Departamento de Ciencias F\'isicas, Universidad Andres Bello
\\ Republica 220, Santiago de Chile\\
\vspace{2.5mm}
${}^\sharp$Departamento de F\'isica, Universidad T\'ecnica Federico Santa Mar\'ia\\
Centro Cient\'ifico-Tecnol\'ogico de Valpara\'iso, Casilla 110-V, Valpara\'iso, Chile\\
\vspace{1mm}

\vspace{1mm}{\tt \footnotesize Xavier.Bekaert@lmpt.univ-tours.fr}\\
{\tt \footnotesize Kevin.Morand@lmpt.univ-tours.fr}

\end{centering}
\begin{abstract}
\noindent
Connections compatible with degenerate metric structures are known to possess peculiar features: on the one hand, the compatibility conditions involve restrictions on the torsion; on the other hand, torsionfree compatible connections are not unique, the arbitrariness being encoded in a tensor field whose type depends on the metric structure. Nonrelativistic structures typically fall under this scheme, the paradigmatic example being a contravariant degenerate metric whose kernel is spanned by a one-form. Torsionfree compatible  (\ie Galilean) connections are characterised by the gift of a two-form (the force field). Whenever the two-form is closed, the connection is said Newtonian. Such a \nr spacetime is known to admit an ambient description as the orbit space of a gravitational wave with parallel rays.
The leaves of the null foliation are endowed with a \nr structure dual to the Newtonian one, dubbed Carrollian spacetime. 
We propose a generalisation of this unifying framework by introducing a new non-Lorentzian ambient metric structure of which we study the geometry. We characterise the space of (torsional) connections preserving such a metric structure which is shown to project to (resp. embed) the most general class of (torsional) Galilean (resp. Carrollian) connections.

\end{abstract}

\vspace{1.5cm}

\vspace{.5cm}

\vspace{4.5cm}

\pagebreak
\pagenumbering{gobble}
\tableofcontents

\pagebreak

\pagenumbering{arabic}

\setcounter{page}{1}

\section{Introduction}

\noindent In contradistinction with the familiar \rel case, the geometry of \nr spacetimes is characterised by {\it degenerate} metric structures. The princeps example of \nr spacetimes originates from Newtonian mechanics, in which time and space are assumed to be separate entities, each one being imparted with an absolute status. Both of these features can be nicely encoded in a geometric fashion \cite{Cartan1923,Friedrichs1928} by identifying the \nr spacetime as a manifold $\bar\M$ endowed with a metric structure ($\bar h^{\mu\nu}$, $\bar\psi_\mu$), where $\bar h^{\mu\nu}$ stands for a twice-contravariant degenerate metric whose radical is spanned by a nowhere-vanishing 1-form $\bar\psi_\mu$ (hence $\bar h^{\mu\nu}\bar\psi_\mu=0$). {Nowadays,} this construction is {often} referred to as {\it Newton-Cartan geometry} \cite{footnote1}. The 1-form $\bar\psi$ plays here the role of an {\it absolute clock} while the {co}metric $\bar h$ provides $\bar\M$ with a notion of spatial distance and will hence be referred to as a collection of {\it absolute rulers}. An equivalent definition of absolute rulers can be given in the form of a positive-definite metric $\bar\gamma$ acting on the kernel of $\bar\psi$. When causality is assumed (\ie when the absolute clock satisfies the Frobenius integrability condition $\bar\psi_{[\mu} \p_\lambda\bar\psi_{\nu]}=0$), the kernel of $\bar\psi$ defines an integrable distribution whose integral submanifolds foliate the spacetime $\bar\M$. Each of these integral hypersurfaces can then be thought of as the {\it absolute space} at a given time on which a notion of distance is provided by the collection of absolute rulers $\bar\gamma$. The ``metric'' structure of Newtonian gravity is thus embodied in a triplet
{$\pl\bar\M,\bar\psi,\bar\gamma\pr$, referred to as a {\it Leibnizian structure} in \cite{Bernal2003,Bekaert2014b}. }

Strictly speaking, metric structures (degenerate or not) do not provide a notion of parallelism. The latter can be implemented in the guise of a compatible connection. However, connections compatible with degenerate metric structures are known \cite{Crampin1968} to differ from the usual, nondegenerate case by at least two related aspects:
on the one hand, the compatibility conditions involve restrictions on the torsion so that not all torsion tensors are admissible; on the other hand, only a subset of metric structures admits a torsionfree compatible connection. Moreover, {even} when the existence of such a connection is ensured, its uniqueness is not and two different torsionfree compatible connections differ by a tensor field whose type depends on the metric structure. As for Leibnizian structures, the compatibility condition with the absolute clock ($\bar\nabla_\mu\bar\psi_\nu=0$) enforces the following constraint on the torsion tensor: $\bar\psi_\lambda \bar\Gamma\lmna=\p_{[\mu}\bar\psi_{\nu]}$. Consequently, only Leibnizian structures with {\it closed} absolute clock admit torsionfree compatible connections (this subset was dubbed {\it Augustinian} structures in  \cite{Bekaert2014b}). Consequently, when dealing with \nr metric structures with non-closed absolute clocks ($d\bar\psi\neq0$), torsionfreeness and compatibility become mutually exclusive features. Furthermore, the previous constraint fixes the ``timelike'' part of the torsion tensor so that the gift of the latter is not sufficient to uniquely define a compatible connection and must be supplemented by a 2-form $\bar F_{\mu\nu}$.

When the absolute clock is closed, torsionfree compatible connections are allowed and are then completely characterised by the gift of a 2-form $\bar F$. The appearance of such a 2-form in the \nr context is in fact quite natural if one acknowledges the fact that in Newtonian mechanics, spacetime is a mere ``container'' whose structure is not rich enough to prescribe the motion of particles. In particular,
when the external forces are of the Lorentz type (\ie $\ddot{x}^\mu=\bar h^{\mu\nu} \bar F_{\nu\rho}\,\dot{x}^\rho$), the dynamics of a single particle is specified by a force-field $\bar F$ (interpreted as the Faraday tensor in the electromagnetic case).
{\it Newtonian} connections are torsionfree Galilean connections for which $\bar F$ is closed, so that locally there exists a 1-form $\bar A$ such that $\bar F=d\bar A$.
Remarkably, this closedness property is equivalent to an algebraic condition on the curvature of the connection: the so-called Duval-K\"unzle condition \cite{Kunzle1972,Duval1977}.
A distinctive feature of Newtonian connections is that the corresponding geodesic equation can be derived via a variational principle from a Lagrangian built in terms of the Leibnizian structure and the potential 1-form.
As shown in \cite{Trumper1983,Bekaert2014b}, such a Lagrangian defines a (possibly nondegenerate) {\it Lagrangian metric} on $\bar\M$ which reduces on the absolute spaces to the absolute collections of rulers $\bar\gamma$, and furthermore constitutes the necessary {and sufficient} structure needed to supplement a Leibnizian structure in order to {define a unique} Newtonian connection.
\pagebreak

\noindent Figure \ref{diagcinematique3} sums up the interrelations between the different  {\it dramatis person{}\ae} constituting the kinematical content of Newton-Cartan geometry, in complete analogy with the relativistic case:
\begin{figure}[ht]
\centering
   \includegraphics[width=0.4\textwidth]{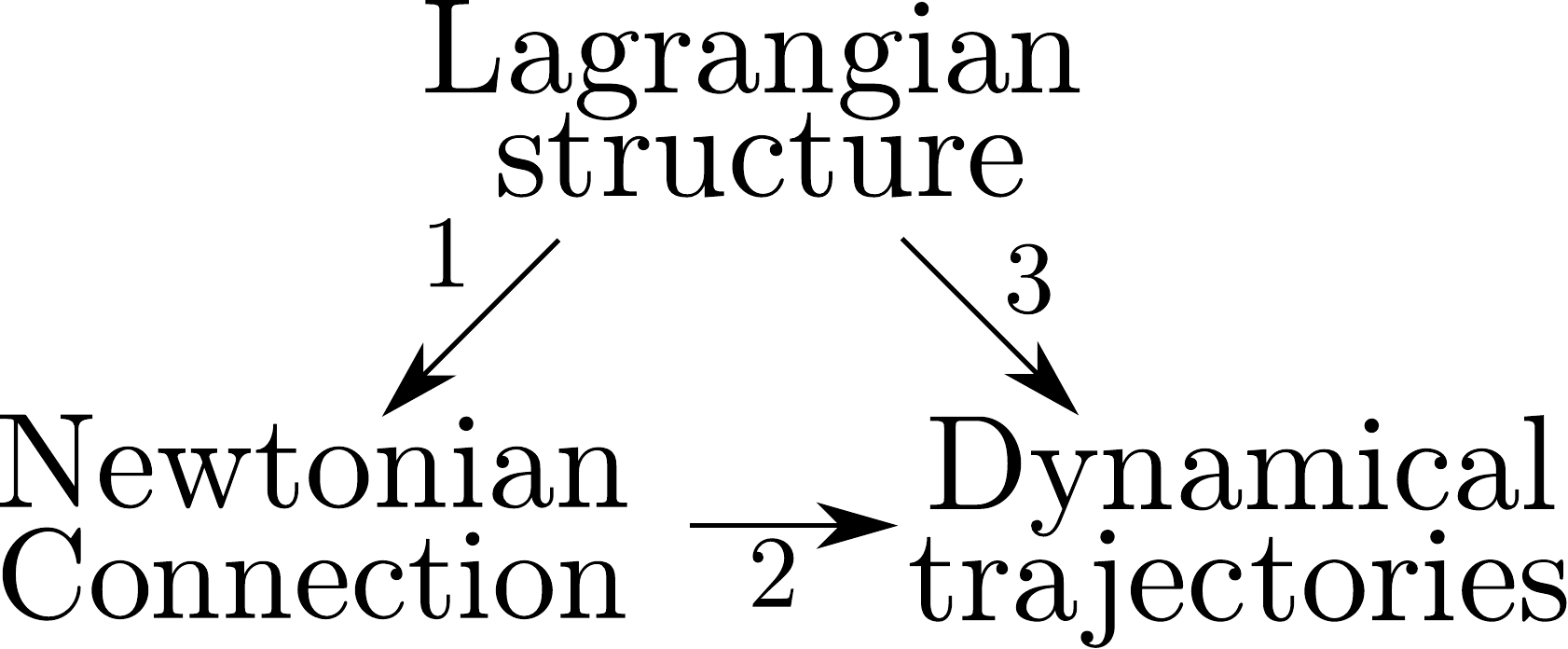}
  \caption{Kinematical content of Newton-Cartan geometry\label{diagcinematique3}}
\end{figure}

\noindent Besides its purely mathematical interest, Newton-Cartan geometry has recently known a {revival} of interest triggered first by condensed-matter applications \cite{condmatt,Jensen2014} following the seminal work by Son \cite{Son2013} (\cf also \cite{Carter1994} for an early inspiring work on superfluids).
These structures have also made novel appearances in the active fields of Lifshitz and Schr\"odinger holography \cite{holography,Christensen2013}
 and in the context of Ho{\v{r}}ava-Lifshitz gravity \cite{horava}. Most of these recent works focus (with the exception of the work \cite{Geracie2015a}
which considers the general case) on a special class of torsional Galilean connection dubbed Torsional Newton-Cartan (TNC) geometries.

\paragraph{Newtonian structures embedded inside Bargmannian structures}
~\\~\\
Although Newtonian structures have been advocated to live on their own, new light can be shed on such structures by embedding them inside relativistic ones, thus providing naturality to seemingly peculiar \nr structures and properties by importing them from usual relativistic ones. The origin of this perspective on \nr physics can be traced back to an early work of Eisenhart \cite{Eisenhart1928} establishing that the dynamical trajectories of a holonomic mechanical system with $d$ degrees of freedom can be put in correspondence with the affine geodesics of a specific $d+2$ dimensional relativistic spacetime. Thus, \nr dynamical trajectories can always be ``lifted'' to geodesics (hence the denomination of ``Eisenhart lift'') and conversely, any \rel geodesic can be projected onto a \nr dynamical trajectory. The class of \rel spacetimes allowing the Eisenhart lift can be characterised by the existence of a light-like vector field which is parallel with respect to the Levi-Civita connection. This class of metrics had previously been considered in \cite{Brinkmann} in a different context and later received the interpretation of \gws with parallel rays \cite{Duval:1990hj}, where the rays are the integral curves of the light-like and parallel vector field (dubbed {\it wave vector field} in the following). Such spacetimes will be referred to as {\it Bargmann-Eisenhart waves}, \cf \cite{Bekaert2013c}.

This bridge between \nr physics and \rel spacetimes has been independently rediscovered afterwards by Duval and collaborators \cite{Duval:1984cj,Duval:1990hj}
who generalised the ``ambient'' approach of Eisenhart in order to provide an account of the different levels of the kinematical and dynamical contents of Newtonian gravity as embedded inside Bargmann-Eisenhart waves.

\begin{figure}[ht]
\centering
   \includegraphics[width=0.7\textwidth]{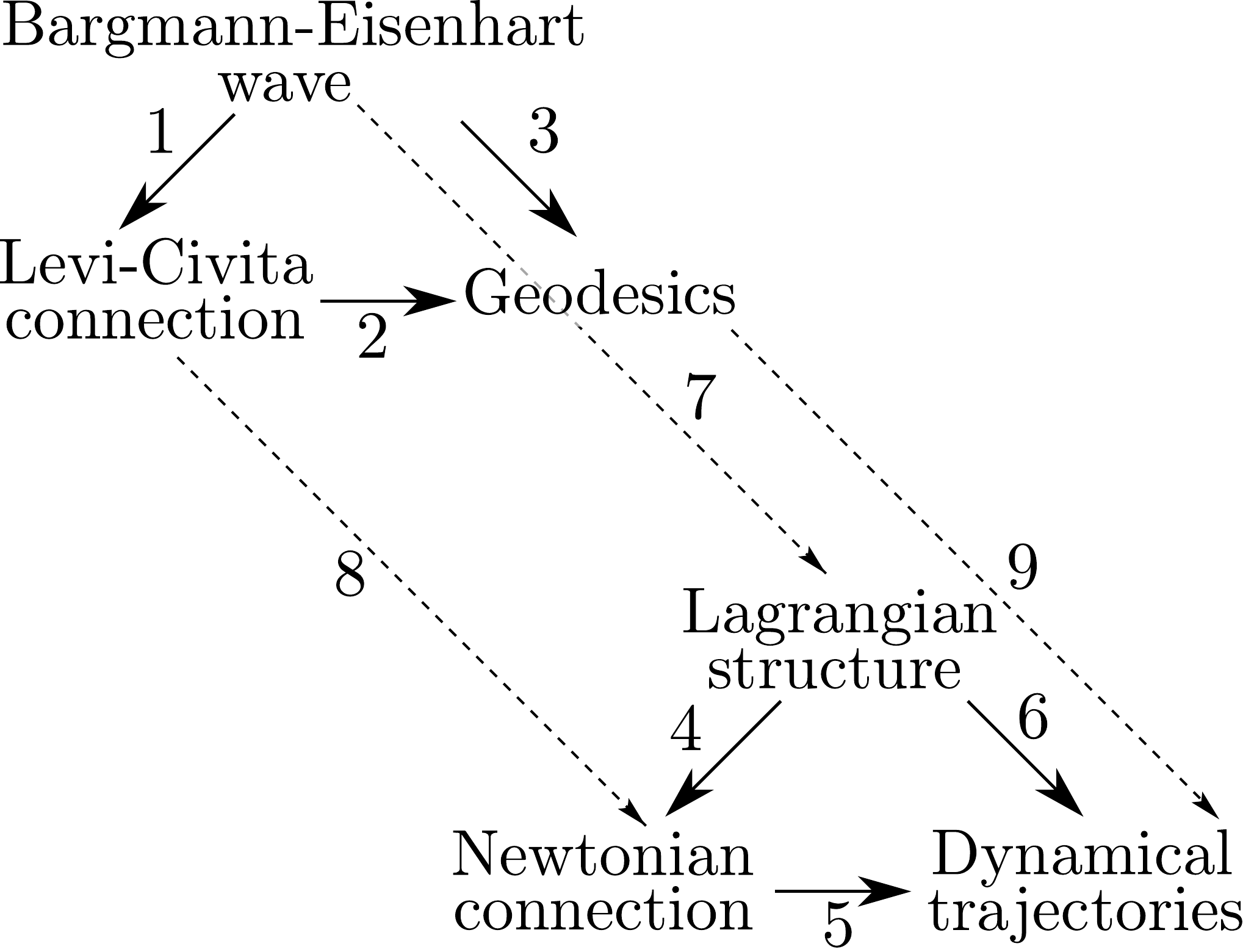}
  \caption{Newtonian structures embedded inside Bargmannian structures\label{diagnewtbarg}}
\end{figure}

\noindent The main idea underlying the ambient approach to \nr physics consists in performing a dimensional reduction of a \BEw along the null rays, thus differing from the usual Kaluza-Klein framework in which the reduction typically occurs along a spacelike direction (or even from the timelike dimensional reduction for stationary spacetimes). The quotient manifold $\bar\M$ obtained by dimensional reduction of a Bargmann-Eisenhart wave along the null rays thus inherits a structure of Newtonian spacetime, a fact that can be appreciated at the different levels of structure, as depicted in Figure \ref{diagnewtbarg}.
Nonrelativistic structures thus appear as ``shadows'' of relativistic ones. In this aspect, the ambient formalism is reminiscent of Plato's allegory of the Cave \cite{Plato} {(\cf \cite{Minguzzi2006,Bekaert2013c} for more on this analogy)}
which depicts material phenomena as mere shadows of pure Forms. Explicitly, the {metric} of a \BEw projects onto the quotient manifold (dubbed {\it Platonic screen} in the following) as the Lagrangian (Arrow 7) while the Levi-Civita connection associated with the \rel metric structure defines a \nr Newtonian connection (Arrow 8) related to the {corresponding} Lagrangian structure. The Eisenhart lift, understood as a correspondence between \nr dynamical trajectories and \rel geodesics, is symbolised by Arrow 9.

Since its introduction in \cite{Duval:1984cj,Duval:1990hj}, the {ambient} formalism has been successfully used to approach a wide range of nonrelativistic problems, such as Chern-Simons electrodynamics \cite{Duval1994}, fluid dynamics \cite{Hassa'ine2000}, Newton-Hooke cosmology \cite{Gibbons2003}, Schr\"odinger symmetry \cite{Duval2009}, Kohn's theorem \cite{Gibbons2011}, hidden symmetries \cite{Cariglia2012a}, \textit{etc}.

\paragraph{Nonrelativistic theories of gravitation and kinematical algebras}
~\\~\\
\noindent The key to the ambient approach followed in \cite{Duval:1984cj,Duval:1990hj} lies in the interplay between algebraic and geometric structures, the origin of which, in a \nr context, can be traced back to the seminal work of K\"unzle \cite{Kunzle1972} in which Leibnizian structures were obtained as $G$-structures for the homogeneous Galilei group. Similarly, Bargmann-Eisenhart waves are defined as $G$-structures for the homogeneous Bargmann group \cite{footnote2} (\cf \cite{Duval:1984cj,Duval:1990hj}). It is indeed hard to overstate the relevance of the Bargmann group (the central extension of the Galilei group \cite{Bargmann1952}) when the geometrisation of \nr physics is concerned, both in the intrinsic or ambient fashion. From an intrinsic viewpoint, the Bargmann group has {proved to be very useful in order to reformulate} the Duval-K\"unzle condition and thus to deal with Newtonian connections (\cf \cite{Duval1977} for an approach in terms of affine connections, \cite{Andringa2011,Geracie2015a} in the context of gauging procedures and \cite{Morand:2016rrt} in the formalism of Cartan geometries). From an ambient standpoint, {these approaches rely} crucially on the group-theoretical avatar of the ambient formalism, namely the embedding of \nr symmetry groups (\eg the Bargmann \cite{Bargmann1952} and Schr\"odinger \cite{Niederer1972} groups) inside their relativistic homologues (Poincar\'e \cite{Gomis1978a} and conformal \cite{Burdet1973} groups, respectively). These \nr groups can thus be obtained from their \rel counterparts by the group-theoretical analogue of a light-like dimensional reduction, namely as subgroups preserving a light-like direction.

\noindent Besides its importance for Newton-Cartan geometry, group-theoretical input has been used in order to describe other \nr structures. In a seminal paper \cite{Bacry1968}, Bacry and L\'evy-Leblond displayed a classification of the so-called ``kinematical'' algebras, namely algebras which encode the infinitesimal symmetries of a free particle. Their classification distinguishes between ``relativistic'' (Poincar\'e, (anti) de Sitter), ``nonrelativistic'' algebras (Galilei, Newton-Hooke) and ``ultrarelativistic'' algebras (Carroll).
As the Galilei group can be thought of as a \nr avatar of the Poincar\'e group (\ie in the limit $c\to \infty$), Newton-Hooke algebras can be seen as a \nr equivalent of the (anti) de Sitter algebras. From an ambient standpoint, the Newton-Hooke spacetime can be obtained as a light-like dimensional reduction from an Hpp-wave, whose isometry group coincides precisely with the central extension of the Newton-Hooke group  \cite{Gibbons2003}. Dual to the Galilean case, the Carroll group, introduced in \cite{Levy-Leblond1965}, can be seen as an ``ultrarelativistic'' limit of the Poincar\'e group (\ie in the limit $c\to 0$). Recently, the Carroll group has known a new topicality following the works \cite{Duval2014e,Duval2014d,Duval2014b} (\cf also \cite{Bergshoeff2014b,Carroll}) unravelling the relation with the Bondi-Metzner-Sachs (BMS) group which can be characterised as conformal extension of the Carroll group. Furthermore, the work \cite{Duval2014e} provided an interesting interpretation of the duality between the Galilei and Carroll groups in the ambient framework: both groups can be obtained from the Bargmann group, by projection and embedding respectively. This duality has a natural geometric counterpart: on the one hand, Newtonian manifolds can be obtained from a Bargmann-Eisenhart wave by projection on the Platonic screen; on the other hand, the leaves {orthogonal to} the wave vector field of a Bargmann-Eisenhart wave are endowed with a Carrollian structure (\ie a nowhere-vanishing vector field $\tilde \xi$ spanning the radical of a twice-covariant degenerate metric $\tilde \gamma$) inherited from the Bargmann-Eisenhart structure $(g,\xi)$ while the Levi-Civita connection associated with {the ambient metric $g$ induces on each leaf a connection compatible} with the Carrollian structure (and is hence referred to as a {\it Carrollian} connection).

We note that the nonrelativistic (resp. ultrarelativistic) structures induced by a Bargmann-Eisenhart wave are {\it not} the most general one can consider. On the one hand, the metric structure obtained by projection on the Platonic screen is Augustinian {(\ie $d\bar\psi=0$)
and the projected connection is Newtonian.}
On the other hand, the extrinsic curvature of the leaves foliating a Bargmann-Eisenhart wave
{vanishes}, so that the induced Carrollian structure is necessarily invariant along $\tilde\xi$. Furthermore, the twice-covariant tensor encoding the arbitrariness in the choice of a Carrollian connection must be equal to the ``transverse extrinsic curvature'' (in the terminology of \cite{Barrabes1991}).

In the present work we propose a generalisation of the ambient setup of \cite{Duval:1984cj,Duval:1990hj} which will allow {to embed} the most {general \nr and ultrarelativistic structures in} an ambient manifold. This is achieved via the introduction of a new ambient metric structure (dubbed {\it ambient Leibnizian structure} in the following) of which we study the geometry. Explicitly, an ambient Leibnizian structure is defined as a quadruplet {$\pl\M,\xi,\psi,\gamma\pr$} where $\xi$ is a nowhere-vanishing vector field on the manifold $\M$, {the 1-form $\psi$ is an absolute clock annihilating $\xi$ and the twice-covariant ``metric'' $\gamma$ is defined on the kernel of $\psi$ and its radical is spanned by $\xi$.} This new metric structure, despite being neither Leibnizian nor Carrollian, provides a unifying ambient framework allowing to embed the most general classes of both Leibnizian and Carrollian structures.

\subsection{Outline of the paper}

\noindent  We now give an outline of the paper and summarise our main results:
~\\

\noindent Section \ref{Ambient geometry} consists in a review of standard material regarding principal $\mR$-bundles in order to fix some terminology. The term {\it Platonic screen} is introduced to designate the orbit space {$\bar\M$ (\ie the base space) of such a principal $\mR$-bundle $\M$}. We then review various isomorphisms between spaces of ambient and base structures.

\noindent In Section \ref{Ambient metric structures}, we introduce the notion of ambient Leibnizian structure. The algebra \big(coined the Leibniz algebra $\leib(d+2)$\big) of {infinitesimal} automorphisms preserving a flat ambient Leibnizian structure is
shown to be {a semidirect sum of the Abelian ideal ${\mathbb R}^{d+2}$ of infinitesimal translations with the (inhomogeneous) Carroll algebra $\carr(d+1)$ as ``homogeneous'' subalgebra (\ie fixing a point)}, so that ambient Leibnizian structures can be thought of as $G$-structures for the \textit{inhomogeneous} Carroll group. Interestingly, the Leibniz algebra unifies the Bargmann and Carroll algebras in that it contains both of them as specific subalgebras. 
The necessary and sufficient conditions for an ambient Leibnizian structure to be projectable on the Platonic screen are given.

\noindent Section \ref{Ambient Galilean connections} is dedicated to the study of connections compatible with an ambient Leibnizian structure, dubbed {\it ambient Galilean} connections.
Section \ref{Torsionfree connections} deals with the torsionfree case. The equivalence problem (\ie the search for the necessary data supplementing the metric structure in order to unambiguously fix the compatible connection) is considered. The arbitrariness in the choice of an ambient Galilean connection is then shown to be encoded in a couple constituted by a 2-form and a symmetric covariant rank-2 tensor, corresponding to the {respective arbitrariness in the choice of a torsionfree connection respectively Galilean or Carrollian.} Furthermore, {this solution to} the equivalence problem allows us to construct a surjective map from the space of \nr torsionfree ambient Galilean connections to the space of torsionfree Galilean connections as well as an injective map from the space of torsionfree Carrollian connections to the one of torsionfree ambient Galilean connections. In other words, any torsionfree Galilean (resp. Carrollian) manifold can be obtained by projection of (resp. {embedding} inside) an ambient Galilean manifold \cite{footnote3}.
\noindent These results are {generalised to} the torsional case in Section \ref{Torsional connections}. We characterise the affine space of torsional ambient Galilean connections, {of which the expression in components} is given by eq.\eqref{eqtorsionalambientGalilean}. We introduce a collection of privileged origins {(dubbed {\it torsional special ambient  connections} in the following) in this infinite-dimensional affine space and construct a surjective map from the space of torsional ambient Galilean manifolds to the space of \nr torsional Galilean manifolds.
~\\

\noindent In a forthcoming publication \cite{Morand2017}, the subclass of ambient Galilean connections compatible with a Lorentzian metric extending the ambient Leibnizian structure will be presented.
In the torsionfree case, this reproduces the results of \cite{Duval:1984cj,Duval:1990hj} stating that the compatibility condition forces the (previously arbitrary) 2-form to be closed, so that the projected torsionfree connection is necessarily Newtonian. Moreover, the corresponding restrictions on the Carrollian {structure will be identified:} the symmetric twice-covariant tensor must coincide with the transverse extrinsic curvature of the considered wavefront worldvolume.

\noindent This class of Lorentzian manifolds will be shown to possess the sufficient arbitrariness to embed the whole class of \nr Galilean manifolds. In particular, 
{\it torsionfree} Galilean connections (with generic force field $\bar F$) will be shown to arise as projection of {\it torsional} (Lorentzian metric compatible) connections where the force field is inherited from the ``light-like part''  of the ambient torsion. 
~\\

\noindent Appendix \ref{sectionappendix} is devoted to Carrollian manifolds. We first review the construction of \cite{Henneaux1979,Duval2014e} related to Carrollian metric structures and then investigate Carrollian connections. We characterise the affine spaces of such connections in the torsionfree and torsional case and display component expressions for the most general Carrollian connection.

\subsection{Notations}

{
\noindent We will mostly follow the notations of \cite{Bekaert2014b}.
~\\

\noindent Let $V$ be a vector space and $v,w\in V$ two vectors. We will denote by $v\vee w=\half\pl v\otimes w+w\otimes v\pr$ \big(respectively $v \wedge w=\half\pl v\otimes w-w\otimes v\pr$\big) the (anti)symmetric product, and similarly for higher products.  
The (anti)symmetrisation of indices is performed with weight one and is denoted by round (respectively, square) brackets,
\eg $\Phi_{(\mu\nu)}:=\half\pl\Phi_{\mu\nu}+\Phi_{\nu\mu}\pr$ and
$\Phi_{[\mu\nu]}:=\half\pl\Phi_{\mu\nu}-\Phi_{\nu\mu}\pr$.

\noindent Let $\cal V$ be a vector bundle over $\M$ with typical fibre the vector space $V$.
By $\Gamma({\cal V})$, we will denote the space of its sections, \ie globally defined $V$-valued fields on $\M$.
\noindent In contrast with \cite{Bekaert2014b}, the manifold $\M$ will stand for the ambient relativistic spacetime manifold of dimension $d+2$. Barred quantities will most often denote nonrelativistic objects defined on the $\pl d+1\pr$-dimensional nonrelativistic spacetime $\bar\M$ while ultrarelativistic objects on the $(d+1)$-ultrarelativistic spacetime $\tilde \M$ will be topped with a tilde. 

\noindent Bijections will be denoted as $\stackrel{\sim}{\to}$ while our notation for injections (resp. surjections) will be $\hookrightarrow$ (resp. $\twoheadrightarrow$).
}
\pagebreak
\section{Ambient spacetimes as principal bundles}\label{Ambient spacetimes}

\noindent This section introduces the basic ambient setup, \ie the ambient manifold and its various metric structures prior to the introduction of a compatible connection. In Section \ref{Ambient geometry}, we recall the geometric definition of an ambient manifold as a principal $\mathbb R$-bundle $\M$ with fundamental vector field $\xi$ and introduce some terminology (proposed in \cite{Bekaert2013c}) for the corresponding base space $\bar\M$ and sections. More importantly, we discuss the necessary and sufficient conditions for (covariant and contravariant) tensor fields on $\M$ to be projectable on $\bar\M$. These criteria will be repeatedly used in the paper.
In Section \ref{Ambient metric structures} we introduce the notion of ``ambient Leibnizian structure'' which unifies both Leibnizian and Carrollian metric structures. We define the  ``Leibniz algebra'' as the isometry algebra of the flat ambient Leibnizian structure and show that the former admits both the Bargmann and the Carroll algebras as subalgebras. 
We conclude the section by discussing related concepts such as ``Leibnizian bases and pairs'', projectors and ambient transverse (co)-metrics. 

\subsection{Ambient geometry}\label{Ambient geometry}
 
\noindent Let $\M$ be a manifold endowed with a complete and nowhere-vanishing vector field $\xi\in\vf$. The integral curve $\gamma_x:\mR\to \M:\lambda\mapsto\gamma_x\pl\lambda\pr$ of the vector field $\xi$, passing through the point $x\in \M$, is the unique solution to the differential equation
\bea\frac{d\gamma_x\pl\lambda\pr}{d\lambda}=\xi|_{\gamma_x\pl\lambda\pr}\label{diff}\eea 
 with $\lambda\in\mR$ and initial condition $\gamma_x\pl0\pr=x$ while the left-hand side stands for the directional derivative along $\gamma_x$ $\big($\ie $\frac{d\gamma_x\pl\lambda\pr}{d\lambda}\pl f\pr=\frac{d\pl f\circ\gamma_x\pr}{d\lambda}$ for all $f\in\fonc{\M}$$\big)$. 
Since $\xi$ is assumed to be complete, its integral curves exist for all values of the parameter $\lambda$ and, consequently, the flow \bea\varphi_\xi:\M\times \mR\to \M:\pl x,\lambda\pr\mapsto\gamma_x\pl\lambda\pr\eea is global thus inducing a well-defined right-action of the additive Lie group $\mR$ on $\M$. Since $\xi$ is also assumed to be nowhere-vanishing, this action is \textit{free}. 
The integral curve $I_x:=\acl \gamma_x\pl\lambda\pr\in \M\tvert \lambda\in \mR \acr$ through $x\in \M$ is thus the orbit of $x$ under the $\mR$-action $\varphi_\xi$. 

\bdefi{Platonic screen \cite{Bekaert2013c}}{Let $\M$ be a manifold endowed with a complete and nowhere-vanishing vector field $\xi\in\vf$. The Platonic screen of $\M$ is the orbit space $\bar \M$ of the action $\varphi_\xi$, \ie the set $\bar \M:=\M/\mR=\acl I_x\tvert x\in \M\acr$ of all integral curves on $\M$.
}
\noindent The quotient manifold Theorem (\cf\eg Theorem 21.10 in \cite{Lee2003}) ensures that if the  $\mR$-action  $\varphi_\xi$ on $\M$ is also proper (which we will always assume), then the Platonic screen is a 
manifold. More precisely, the projection map onto orbits of $\varphi_\xi$, denoted $\pi:\M\to \bar \M:x\mapsto I_x$ is a submersion and therefore defines the principal fiber bundle:
\bea
\mR\longhookrightarrow\M\stackrel{\pi}{\longrightarrow}\bar\M
\label{diagrammprincipalbundle}
\eea
whose fibers are the integral curves $I_x$ and whose fundamental vector field is $\xi$. The vertical subspace $V_x\subset T_x\M$ at a point $x\in\M$ is therefore spanned by $\xi_x$, \ie $V_x\cong\Span{\xi_x}$, so that $\pi_*\xi_x=0$ for all $x\in\M$, where $\pi_*:T_x\M\to T_{\pi\pl x\pr}\bar\M$ designates the pushforward of the map $\pi$ at $x$. 

\noindent To emphasise the geometrical meaning of the ambient manifold which is the main leitmotiv of this paper, let us summarise the following facts:
\bpropp{Ambient structure}{Let $\M$ be a manifold. The following structures on $\M$ are in bijective correspondence:
\begin{enumerate}
 \item a congruence $I$ of parameterised open curves from $\mR$ to $\M$, 
 \item a complete nowhere-vanishing vector field $\xi$ on $\M$,
 \item a principal $\mR$-bundle $(\M,\xi)$ with fundamental vector field $\xi$.
\end{enumerate}
The relation between these geometrical structures is as follows: the congruence $I$ is the family of integral curves of the vector field $\xi$ as well as the vertical foliation of the manifold $\M$.
}
\noindent Without loss of generality, such a principal $\mR$-bundle $\M$ can be assumed to be trivial, \ie it is isomorphic to $\bar\M\times \mR$ (\cf\eg Proposition 16.14.5 of \cite{Dieudonne1970}). This ensures the existence of global cross sections which can be seen as embeddings of the Platonic screen $\bar\M$
inside the ambient manifold $\M$. We adopt the following terminology:
\bdefi{Screen worldvolume \cite{Bekaert2013c}}{A global section $\sigma:\bar\M\hookrightarrow\M$ of a principal $\mR$-bundle $\pi:\M\to\bar\M$ is called a screen worldvolume. 
}
\noindent Such an ambient structure can be endowed with a notion of horizontality through the gift of a {\it principal connection} for the principal $\mR$-bundle $(\M,\xi)$ \ie a 1-form $A\in\ff$ on $\M$ satisfying the following two conditions 
\begin{enumerate}
\item $A\pl\xi\pr=1$
\item $\Lagxi A=0$. 
\end{enumerate}
The space of principal connections will be denoted $\PC$. In the following it will prove useful to relax the second condition and consider {\it Ehresmann connections} defined as follows: 
\bdefi{Ehresmann connection}{\label{defiEhresmannconnection}Let $\pl \M,\xi\pr$ be an ambient structure. An Ehresmann connection is a 1-form $A\in\ff$ on $\M$ satisfying $A\pl \xi\pr=1$. }
\noindent Such an Ehresmann connection can be seen as dual to the notion of fields of observers (\cf Definition 2.14 in \cite{Bekaert2014b}). 
The space of Ehresmann connections on $\pl \M,\xi\pr$ will be denoted $\EC$. 

\noindent Let us briefly characterise the (co)tangent bundles of the Platonic screen $\bar\M$. 
More precisely, let us point out the two canonical isomorphisms of vector bundles: $T\bar\M\cong T\M/\mbox{Span}\,\xi$
and $T^*\bar\M\cong\Ann \xi$, where $\Ann \xi_x:=\{\alpha_x\in T^*_x\M\,|\,\alpha_x(\xi_x)=0\}$.
The first isomorphism relies on the equivalence relation of tangent vectors to $\M$ at $x\in\M$:
\bea X_x\sim Y_x\quad\Leftrightarrow \quad X_x\,=\,Y_x\,+\,\lambda\, \xi_x\,,\qquad \text{ for some }\lambda\in\mathbb R\,.\eea
Such equivalence classes are in bijective correspondence with tangent vectors to $\bar\M$ at $\pi(x)$. 
Indeed, the kernel of the surjective pushforward $\pi_*:T\M\twoheadrightarrow T\bar\M$ of the projection $\pi:\M\twoheadrightarrow\bar\M$ is the vertical bundle Span$\,\xi$.
The second isomorphism is provided by the pullback $\pi^*:T^*\bar\M\hookrightarrow T^*\M$ whose image is Ann$\,\xi$\,.

In the remainder of the present Section, we intend to make use of the previous characterisation of (co)tangent bundles of $\bar\M$ in order to discriminate among the fields living on the ambient manifold $\M$ those admitting a well-defined projection on the Platonic screen $\bar\M$. These are standard results holding for principal bundles that we review for completeness. 

\paragraph{Projection and lift of a function}\noindent Given an ambient structure $\pl\M,\xi\pr$, we will call {\it invariant} a section $f\in\field{\bptt{T\M}{T^*\M}}$ satisfying $\Lag_\xi f=0$ and will denote $\fieldinv{\bptt{T\M}{T^*\M}}$ the space of invariant sections of $\bptt{T\M}{T^*\M}$. In particular, an invariant function on the principal $\mR$-bundle $\M$ is a function $f\in C^\infty(\M)$ such that $\Lag_\xi f=0$. 
\bdefi{Lift of a function}{\label{defiliftfunction}Let $\bar f\in\fonc{\bar\M}$ be a function on the Platonic screen $\bar\M$. Via pullback by the projection map, the function $\bar f$ defines a unique invariant function $f:=\pi^*\bar f= \bar f\circ\pi\in\foncinv{\M}$ on the ambient manifold $\M$, called the lift of $\bar f$.}
\noindent There is a bijective correspondence between invariant functions on the principal $\mR$-bundle $\M$ and functions on the Platonic screen $\bar\M$, that is: $C_\text{\rm inv}^\infty(\M)\cong C^\infty(\bar\M)$\,. 

\paragraph{Projection and lifts of a vector field}\label{parvectorfield}
\bdefi{Projectable and invariant vector fields}{\label{defiprojeq}The vector field $X\in\vf$ on $\M$ is said projectable (respectively invariant) if it satisfies $\Lag_\xi X=f\,\xi$, for some function $f\in\fonc{\M}$ (respectively $\Lag_\xi X=0$). 
}
\noindent As suggested by its name, the pushforward $\pi_* X\in\Gamma(T\bar\M)$ by the projection map of a projectable vector field $X\in\Gamma(T\M)$ is a well defined vector field on the Platonic screen $\bar\M$.
 \bprop{\label{propLieproj}If $X$ and $Y\in\vf$ are two projectable (resp. invariant) vector fields on $\M$, then their Lie bracket $\br{X}{Y}\in\vf$ is projectable (resp. invariant). }
\bdefi{Lift of a vector field}{\label{liftvectofield}Let $\bar X\in\field{T\bar\M}$ be a vector field on the Platonic screen $\bar\M$. A vector field $X\in\vf$ satisfying the conditions:
\begin{itemize}
\item $X$ is projectable (resp. invariant)
\item $\pi_*X=\bar X$
\end{itemize}
is called a lift (resp. invariant lift) of $\bar X$ in $\M$. 
}
\noindent Clearly, the arbitrariness in the possible lifts of a given vector field is a vertical vector field:
\bprop{\label{propdifferlift}Let $\bar X\in\field{T\bar\M}$ be a vector field on the Platonic screen $\bar\M$. Let $X$ and $X'\in\vf$ be two (invariant) lifts of $\bar X$ in $\M$. Then, there exists a (invariant) function $f\in\fonc{\M}$ such that $X'=X+f\,\xi$. }
\noindent This construction of invariant lifts provides the following isomorphism: $$\Gamma_\text{\rm inv}\pl T\M/\Span{\xi}\pr\cong \Gamma(T\bar\M)\,.$$ 

\paragraph{Projection of a 1-form}
\noindent A necessary condition in order for a 1-form $\alpha\in\form{\M}$ to project onto a well-defined 1-form $\bar \alpha\in\form{\bar \M}$ is to be a section of the vector subbundle Ann $\xi\subset T^*\M$, \ie to satisfy $\alpha\pl\xi\pr=0$. 
\bpropp{Projectable 1-form}{\label{propform}
A 1-form $\alpha\in\form{\M}$ is projectable on the Platonic screen $\bar \M$ if and only if the two following conditions are satisfied: 
\begin{enumerate}
 \item the 1-form annihilates the fundamental vector field, \ie $\alpha\pl \xi\pr=0$, 
 \item the 1-form is invariant, $\mathcal L_\xi \alpha=0$. 
\end{enumerate}
\noindent The projection $\bar\alpha\in\form{\bar\M}$  of the 1-form is then defined as the 1-form satisfying the relation $\pi^*\bar\alpha=\alpha$.}
\noindent This discussion explains the following isomorphism: $\Gamma_\text{\rm inv}\pl \Ann{\xi}\pr\cong \Gamma(T^*\bar\M)\,.$ 

\noindent The previous conditions generalise straightforwardly to a covariant metric, so that the following Proposition holds: 
\begin{prop}\label{propbilinearform}
A covariant metric $g\in\bforma{T^*\M}$ defined on the ambient structure $\pl\M,\xi\pr$ is projectable on the Platonic screen $\bar \M$ if and only if the two following conditions are satisfied: 
\begin{enumerate}
 \item $\xi\in\Rad g$, \ie $g\pl \xi\pr=0$, 
 \item $\mathcal L_\xi g=0$. 
\end{enumerate}
\end{prop}
\noindent If such conditions are met, the projection $\bar g\in \bforma{T^*\bar \M}$ is defined as $\pi^*\bar g=g$. 

\noindent It should be noted that \Prop{propbilinearform} prevents any possibility to define a covariant metric on the Platonic screen of an ambient structure by projecting a Lorentzian metric, since only {\it degenerate} covariant metrics are projectable. In order to circumvent this drawback, we will be led to endow ambient structures with degenerate metric structures.  

\paragraph{Projection of a Koszul connection}
\noindent We now address the issue of projectable connections by providing a set of sufficient conditions ensuring that a given Koszul connection on the ambient manifold $\M$ admits a well-defined canonical projection on the Platonic screen. 

\noindent Let $\nabla:\Gamma\pl T\M\pr\to \End\big(\field{T\M}\big)$ be a Koszul connection on $\M$ (\cf footnote 7 of \cite{Bekaert2014b} for a reminder of the defining properties of a Koszul connection). A Koszul connection $\bar\nabla:\Gamma\big(T\bar\M\big)\to \End\big(\field{T\bar\M}\big)$ can be canonically defined on the Platonic screen $\bar\M$ by making the following diagram commute:
\bea
\xymatrix{
\pl X,Y\pr\ar@{->>}[d]^{\pi_*}\ar[r]^\nabla&\nabla_XY\ar@{->>}[d]^{\pi_*}\\
\pl\bar X,\bar Y\pr\ar[r]^{\bar\nabla}&\bar\nabla_{\bar X}\bar Y \label{diagdefibarnabla}
}
\eea
provided the following conditions are satisfied:
\begin{enumerate}[a)]
\item The ambient vector field $\nabla_XY\in\field{T\M}$ must be projectable. 
\item The vector field $\bar\nabla_{\bar X}\bar Y:=\pi_*\pl\nabla_XY\pr$ must be independent of the choice of representatives $X$ and $Y$. 
\item The defined derivative operator $\bar\nabla$ must satisfy the axioms of a \Kcp. 
\end{enumerate}
\noindent A Koszul connection $\nabla$ on $\M$ such that conditions a)-c) are satisfied will be referred to as \textit{projectable}. We now provide a set of sufficient conditions ensuring projectability:
\bprop{\label{propinvKoszul}Let $\pl \M,\xi\pr$ be an ambient structure endowed with a Koszul connection $\nabla:\Gamma\pl T\M\pr\to \End\big(\field{T\M}\big)$ and denote $T\in\field{\w^2T^*\M\otimes T\M}$ the torsion tensor associated with $\nabla$. If $\nabla$ satisfies the following conditions:
\begin{enumerate}
\item $\nabla \xi=0$
\item $T\pl\xi,\cdot\pr=0$
\item $\Lag_\xi \nabla=0$
\end{enumerate}
then $\nabla$ is projectable.  A Koszul connection $\nabla$ satisfying conditions \textit{1-3} will be said invariant. 
}
\noindent The conditions for invariance can be stated in a more explicit way as follows:
{\it
\begin{enumerate}
\item $\nabla_X \xi=0$ for all $X\in\vf$
\item $T\pl\xi,X\pr=0$ for all $X\in\vf$
\item $\pl\Lag_\xi\nabla\pr_XY:=\br{\xi}{\nabla_XY}-\nabla_{\br{\xi}{X}}Y-\nabla_X\br{\xi}{Y}=0$ for all $X,Y\in\vf$. 
\end{enumerate}
}

\subsection{Ambient Leibnizian structures}\label{Ambient metric structures}

\noindent The present Section deals with ambient structures endowed with non-Riemannian metric structures. 
\noindent From now on, we will focus on ambient structures \Mx endowed with an absolute clock (\cf Definition 2.10 in \cite{Bekaert2014b}), denoted $\psi$, which is assumed to annihilate the fundamental vector field $\xi$, \ie $\psi\in\field{\Ann \xi}$. We start by defining an ambient analogue of the notion of absolute rulers (\cf Definition 2.11 in \cite{Bekaert2014b}). 

\bdefi{Ambient absolute rulers}{\label{defirelspatialmetric}
A collection of ambient absolute rulers on $\M$ is a field of positive semi-definite contravariant symmetric bilinear forms $h\in\field{\vee^2(\Ann\,\xi)^*}$ acting on 1-forms annihilating $\xi$ and such that its radical is spanned by the absolute clock $\psi$ \ie
\be\Rad h\cong\Span{\psi}\,.\label{radi}\ee  
\noindent Alternatively, a collection of ambient absolute rulers can be defined as a field $\gamma\in\bforma{\pl\Ker\psi\pr^*}$ on $\M$ of positive semi-definite covariant symmetric bilinear forms acting on tangent vectors annihilated by the absolute clock and such that the radical of $\gamma$ is spanned by the fundamental vector field \ie 
\be\Rad \gamma\cong\Span{\xi}\,.\label{radii}\ee  
}
\bprop{\label{aLstr} The above two definitions are equivalent. }
\proof{\noindent We only prove the implication $\gamma\Rightarrow h$, the converse statement can be obtained by similar means. Let $\gamma\in\bforma{\pl\Ker\psi\pr^*}$ be a field of positive semi-definite covariant symmetric bilinear forms on $\M$ satisfying $\Rad \gamma\cong\Span{\xi}$ and $\pset{\xi, e_1, \ldots, e_d}$ be a basis of $\Milneg$ satisfying
\bea
\gamma\pl e_i,e_j\pr=\delta_{ij},\ \forall i,j\in\pset{1,\ldots,d}. 
\eea
Such a basis will be referred to as an orthogonal basis of $\Milneg$. Note that the set of endomorphisms of $\Milneg$ mapping each orthogonal basis into another one forms a group isomorphic to the $d$-dimensional Euclidean group which acts regularly on the space of orthogonal bases via the group action:
\bea
\pset{\xi, e_i}\mapsto\pset{\xi, \bb_i\, \xi+\bR^{j}{}_{i}\, e_j}\label{eqchangeorthbases}
\eea
with $\bb\in\mR^d$ and ${\bR}\in O\pl d\pr$. 

\noindent Let us define $h\in\field{\vee^2\pl\Ann \xi\pr^*}$ by its action on a pair of 1-forms $\alpha, \beta\in\field{\Ann\xi}$ as:
\bea
h\pl\alpha,\beta\pr=\delta^{ij}\alpha\pl e_i\pr\beta\pl e_j\pr. 
\eea
From the group action \eqref{eqchangeorthbases}, one concludes that $h$ is invariant under a change of orthogonal basis. It can then be checked that $\Rad h\cong\Span\psi$. }

\noindent This notion of ambient absolute rulers allows to extend the \nr definition of a Leibnizian structure (\cf Definition 2.12 in \cite{Bekaert2014b}) to the following ambient analogue:
\bdefi{Ambient Leibnizian structure}{\label{defambientleib}An ambient Leibnizian structure $\mathscr L\pl\M,\xi,\psi,\gamma\pr$ is a quadruplet composed by the following elements:
\begin{itemize}
\renewcommand{\labelitemi}{$\bullet$} 
\item an ambient structure $(\M,\xi)$,
\item an absolute clock $\psi\in\field{\Ann \xi}$ on $\M$,
\item a collection of ambient absolute rulers  $\gamma\in\field{\vee^2\pl\Ker\psi\pr^*}$.
\end{itemize}
}
\noindent Note that the number of independent components in a Leibnizian structure on a $\pl d+2\pr$-dimensional manifold $\M$ is equal to $\frac{\pl d+2\pr\pl d+3\pr}{2}$ and thus equates the one of a Lorentzian structure on $\M$. 

\noindent Pursuing the analogy with the \nr case, we define the two following subclasses (\cf Table 1 in \cite{Bekaert2014b}):
\bi
\item An {\it ambient Aristotelian} structure will designate an ambient Leibnizian structure whose absolute clock satisfies the Frobenius integrability condition (\ie $\psi\w d\psi=0$).
\item An ambient Leibnizian structure with closed absolute clock (\ie $d\psi=0$) will be called an {\it ambient Augustinian} structure. 
\ei

\noindent The Frobenius integrability condition enjoyed by the absolute clock of any ambient Aristotelian structure $\mathscr L\pl\M,\xi,\psi,\gamma\pr$ ensures that the kernel of $\psi$ induces an \textit{involutive} distribution $\mathcal D=\pset{\mathcal D_x|\,\forall x\in\M}$, with $\mathcal{D}_x:= \Ker\psi_x$. 
The ambient spacetime $\M$ is thus foliated by codimension-one hypersurfaces. Each leaf $\tilde \M$ of the foliation is a maximal integral submanifold of $\M$ for the distribution $\mathcal D$ and is characterised by an immersion $i:\tilde\M\longhookrightarrow\M$ satisfying the following properties:
\begin{itemize}
\item $i_*\big(T\tilde\M\big)\cong\Ker\psi$
\item $\Ker i^*\cong\Span \psi$. 
\end{itemize}

\noindent The first isomorphism ensures that only vector fields $V\in\Milneg$ admit a well-defined projection $\tilde V\in \Gamma\big({T\tilde\M}\big)$ on the submanifold $\tilde\M$ as $i_*\tilde V:= V$. On the other hand, any $p$-form $\alpha\in\forma{p}{\M}$ projects on $\tilde\M$ as $\tilde\alpha:= i^*\alpha$, where $\tilde\alpha\in\forma{p}{\tilde\M}$. Note that the absolute clock is pulled back to zero, thanks to the second isomorphism. Projection of Koszul connections on $\tilde\M$ can also be defined in a way analogous to Diagram \eqref{diagdefibarnabla} by making the following diagram commute:
\bea
\xymatrix{
\big(\tilde X,\tilde Y\big) \ar@{^{(}->}[r]^{i_*}\ar[d]_{\tilde\nabla}&\ar[d]^{\nabla}\pl X,Y\pr \label{diagcarr}\\
\tilde\nabla_{\tilde X}\tilde Y \ar@{^{(}->}[r]^{i_*}&\nabla_XY
}
\eea
where $\tilde X,\tilde Y\in\Gamma\big({T\tilde\M}\big)$ are vector fields on $\tilde \M$ and $X:= i_*\tilde X$, $Y:= i_*\tilde Y$. A necessary and sufficient condition in order for $\nabla:\Gamma\big({T\M}\big)\to\End\pl{\Gamma\big({T\M}\big)}\pr$ to be projectable \ie to induce a well-defined Koszul connection $\tilde\nabla:\Gamma\big({T\tilde\M}\big)\to\End\pl{\Gamma\big({T\tilde\M}\big)}\pr$ on $\tilde \M$ is that $\psi\pl \nabla_VW\pr=0$ for all $V,W\in\Milneg$. A sufficient condition ensuring projectability is then $\nabla\psi=0$ or equivalently $X\crl\psi\pl Y\pr\crr=\psi\pl\nabla_XY\pr$ for all $X,Y\in\vf$. 
\vspace{2mm}

\noindent Any ambient Aristotelian structure $\mathscr L\pl\M,\xi,\psi,\gamma\pr$ then induces a Carrollian structure $\tilde{\mathscr C}\pl \tilde \M,\tilde\xi,\tilde \gamma\pr$ (\cf \Defi{defiCarrollstructure}) on each leaf of the foliation, where $i_*\tilde \xi:=\xi$ and $\tilde\gamma:= i^*\gamma$. 
Note that the class of Carrollian structures that can be induced in this setup is the most general one can define. In particular, the present construction allows to embed non-invariant Carrollian structures (\ie with $\Lag_{\tilde\xi}\tilde\gamma\neq0$) inside an ambient spacetime.

\noindent An ambient Leibnizian structure will be said projectable if both the absolute clock and rulers are projectable. Since both $\psi$ and $\gamma$ have been assumed to annihilate the fundamental vector field $\xi$, the only remaining condition consists in imposing their invariance, \ie $\Lag_\xi\psi=\Lag_\xi \gamma=0$. The following Proposition justifies further the terminology used: 
\bprop{\label{propbijaug}Let $\pl\M,\xi\pr$ be an ambient structure. Projectable ambient Leibnizian structures $\mathscr L\pl\M,\xi,\psi,\gamma\pr$ are in bijective correspondence with Leibnizian structures $\bar{\mathscr L}\pl\bar\M,\bar\psi,\bar\gamma\pr$ on the Platonic screen $\bar\M$. }
\noindent The proof is straightforward and follows from the fact that the pullback $\pi^*$ of the projection map $\pi:\M\to\bar\M$ defines the two isomorphisms $\field{T^*\bar\M} \cong\fieldinv{\Ann\xi}$ and $\field{\vee^2\pl\Ker\bar\psi\pr^*} \cong\fieldinv{\vee^2\pl\Ker\psi\pr^*}\cap\fieldinv{\vee^2\Ann\xi}$.

\bexa{Ambient Aristotle spacetime}{\label{exaambAristotle}The most simple example of an ambient Augustinian structure is given by an ambient spacetime $\M\cong{\mathbb R}^{d+2}$ with coordinates $(u,t,x^i)$
characterised by a fundamental vector field $\xi=\partial/\partial u$\,, a closed absolute clock $\psi=dt$ and flat ambient absolute rulers $\gamma=\delta_{ij}\,dx^i\vee dx^j$\,. On the one hand, this structure projects on the Platonic screen as the $\pl d+1\pr$-dimensional Aristotle spacetime $\bar\M\cong{\mathbb R}^{d+1}$ with coordinates $(t,x^i)$ (\cf Example 2.20 of \cite{Bekaert2014b}). On the other hand, the hyperplanes $t=$\,cst of the ambient Aristotle spacetime are Carroll spacetimes (\cf Example \ref{exaCarroll}) with coordinates $(u,x^i )$\,. 
}

\bpropp{Leibniz algebra}{The infinite-dimensional algebra $\leib_\infty\pl d+2\pr$ of infinitesimal isometries of the ambient Aristotle spacetime $\pl \M,\xi,\psi,\gamma\pr$, \ie the algebra of vector fields $X\in\vf$ satisfying
\bea
\Lag_X\xi=\Lag_X\psi=\Lag_X\gamma=0\label{eqleibalgebra}
\eea
is spanned by vector fields of the form
\bea
X=a\pl t,x^i\pr\frac{\partial}{\partial u}+\tau\, \frac{\partial}{\partial t}+\crl \lambda^i_{\ j}\pl t\pr x^j+b^i\pl t\pr\crr\frac{\partial}{\partial x^i}
\eea
where $a$ is an arbitrary function {of $t$ and $x^i$ while $\ensuremath{\boldsymbol \lambda}$ (resp. $\bb$) are arbitrary $\mathfrak{o}\pl d\pr$ (resp. $\mR^d$) valued functions of $t$ only, and $\tau\in\mathbb R$ is a constant.}

\noindent The finite-dimensional subalgebra {\rm\cite{footnote4}} $\leib\pl d+2\pr=\leib_\infty\pl d+2\pr\cap\mathfrak{igl}\pl d+2\pr$ of affine infinitesimal isometries of the ambient Aristotle spacetime $\pl \M,\xi,\psi,\gamma\pr$, \ie the algebra of vector fields $X\in\vf$ satisfying \eqref{eqleibalgebra} and whose second partial derivatives vanish, is generated by the following vector fields:
\begin{itemize}
\item Mass: $M:=\p_u$
\item Galilei Hamiltonian: $H:=\p_t$
\item Translations: $P_i:=\p_i$
\item Carroll Hamiltonian: $C:= t\, \p_u$
\item Carroll boosts: $D_i:= x_i\p_u$
\item {Galilei} boosts: $K_i:= x_i\p_u-t\p_i$
\item Rotations: $J_{ij}:= x_i\p_j-x_j\p_i$. 
\end{itemize}
The previous generators satisfy the {(schematic)} commutation relations:
\bea
\br{H}{C}\sim M,& \br{\textbf{P}}{\textbf{K}}\sim M,& \br{\textbf{P}}{\textbf{D}}\sim M\nn\\
\br{H}{\textbf{K}}\sim \textbf{P},& \br{\textbf{D}}{\textbf{K}}\sim C,& \br{\textbf{P}}{\textbf{J}}\sim \textbf{P}\label{Killingalgebra}\\
\br{\textbf{K}}{\textbf{J}}\sim \textbf{K},& \br{\textbf{D}}{\textbf{J}}\sim \textbf{D},& \br{\textbf{J}}{\textbf{J}}\sim \textbf{J}. \nn
\eea
}
\noindent It can be checked from the previous commutation relations that the finite-dimensional subalgebra $\leib\pl d+2\pr$ can be written as the semidirect sum \cite{footnote5}$$\leib\pl d+2\pr=\carr\pl d+1\pr\inplus \mR^{d+2}$$ with $\carr\pl d+1\pr\cong\Span\pset{C,D_i, K_i, J_{ij}}$ the (inhomogeneous) Carroll algebra and $\mR^{d+2}\cong\Span\pset{H, M,P_i}$ an Abelian ideal.

\noindent Any leaf $t=$cst of the ambient Aristotle spacetime is a Carroll spacetime (\cf Example \ref{exaambAristotle}). 
On any leaf of constant $t\neq0$, the Carroll Hamiltonian $C=t\partial_u$ and the difference between the Carroll and Galilei boosts $D_i-K_i=t\partial_i$ take respectively the interpretation of the Carrollian time and spatial translations generators }\cite{footnote6}.
~\\

\noindent The Bargmann algebra $\barg(d+2)$ is recovered as the subalgebra of $\leib\pl d+2\pr$ generated by vector fields \eqref{Killingalgebra} preserving the flat Minkowski metric (\ie satisfying the extra condition $\Lag_X\eta=0)$ and is given by  $\barg(d+2):=\pset{H,M,P_i,K_i,J_{ij}}$.

~\\Like its base counterpart, a collection of absolute rulers $\gamma$ (resp. $h$) is not defined on the whole (co)tangent space. One can circumvent this drawback and define, though in a non-canonical way, a covariant (resp. contravariant) metric by making use of additional structures, namely a field of observers (resp. an Ehresmann connection). We start by defining various projectors as follows:

\noindent Given a field of observers $N\in\FO$ (\ie $N\in\vf$ and $\psi\pl N\pr=1$), one can define a field of endomorphisms $\overset{N}{P}:\field{T\M}\to\field{\Ker \psi}$ as 
\bea 
\overset{N}{P}\pl X\pr=X-\psi\pl X\pr N ,\text{ with }X\in\vf\text{ a vector field on }\M\label{eqPN}
\eea 
together with its transpose denoted $\overset{N}{\bar P}:\ff\to \field{\Ann N}$
\bea
\overset{N}{\bar P}\pl \omega\pr=\omega-\omega\pl N\pr\psi,\text{ with }\omega\in\ff\text{ a 1-form on }\M\label{eqbarPN}. 
\eea
Their respective kernels are given by $\Ker \overset{N}{P}\cong\Spannn{N}$ and $\Ker \overset{N}{\bar P}\cong\Spannn{\psi}$. 

\noindent Finally, the gift of an Ehresmann connection $A\in\EC$ allows to define the field of endomorphisms $\ovA{P}{}:\field{T\M}\to\field{\Ker A}$ as 
\bea 
\ovA{P}{}\pl X\pr=X-A\pl X\pr \xi ,\text{ with }X\in\vf\text{ a vector field on }\M\label{eqPA}
\eea 
while its transpose reads $\ovbA{P}{}:\ff\to \field{\Ann \xi}$
\bea
\ovbA{P}{}\pl \omega\pr=\omega-\omega\pl \xi\pr A,\text{ with }\omega\in\ff\text{ a 1-form on }\M\label{eqbarPA}
\eea
whose respective kernels read $\Ker \ovA{P}{}\cong\Spannn{\xi}$ and $\Ker \ovbA{P}{}\cong\Spannn{A}$. 
\vspace{2mm}

\noindent We will now start discussing bases adapted to the ambient metric structure. While discussing bases at a given point, we will not write explicitly the point $x\in\M$. Via a slight abuse of notation, this allows to make the extension to field of bases. 
\bdefi{Leibnizian basis}{\label{defiLeibnizianbasis}Let $\mathscr L\pl\M,\xi,\psi,\gamma\pr$ be an ambient Leibnizian structure. A Leibnizian basis of the tangent space $T_x\M$ at a point $x\in\M$ is an ordered basis $L=\lbrace \xi, N, e_{1}, \ldots, e_{d}\rbrace$ with $\xi$ the fundamental vector, $N$ the tangent vector of an observer, $\lbrace \xi, e_{1}, \ldots, e_{d}\rbrace$ a basis of $\Ker \psi$, and $\lbrace e_{1}, \ldots, e_{d}\rbrace$ an orthonormal system with respect to $\gamma$. } 

\noindent Explicitly, the basis $L=\lbrace \xi, N, e_{1}, \ldots, e_{d}\rbrace$ must satisfy the conditions:
\vspace{2mm}
\begin{itemize}
\item[1.] $\psi\pl \xi\pr=0$, $\psi\pl N\pr=1$, $\psi\pl e_{i}\pr=0$
\item[2.] $\gamma\pl e_{i}, e_{j}\pr=\delta_{ij}$
\end{itemize}

\bprop{The set of automorphisms of $T_x\M$ preserving the collection of Leibnizian bases at a point forms a group isomorphic to the inhomogeneous Carroll group $\CARR\pl d+1\pr\subset GL\pl d+2,\mR\pr$ defined by the following set of matrices:
\bea
T=
\begin{pmatrix}
1&a&-\bbf^T\bR\\
0 & 1 &0 \\
0 & \bb&\bR\label{matrixLeibniz}
\end{pmatrix}
\eea
with $a\in\mR$, $\bb, \bbf\in\mR^d$ and ${\bR}\in O\pl d\pr$.}
\noindent The Carroll group \cite{Levy-Leblond1965,Duval2014e} admits a semidirect product structure decomposition as $\CARR\pl d+1\pr=O\pl d\pr \ltimes H\pl d\pr$ where $O\pl d\pr$ stands for the $d$-dimensional orthogonal group ($a=0$, $\bb=\mathbf{0}=\bbf$) and $H\pl d\pr$ for the $d$-dimensional Heisenberg group
($\bR=\mathds{1}$).
 The Carroll group acts regularly on the space of Leibnizian bases at $x$ via the group action:
\bea
\pset{\xi, N, e_{i}}\overset{T}{\mapsto}\pset{\xi, N+\bb^ie_{i}+a\, \xi, \bR^{j}{}_{i}\pl e_{j}-\bbf^T_j\xi\pr} \label{eqleibnizauto}
\eea
so that the space of Leibnizian bases at a point is a principal homogeneous space of the Carroll group. 
Consequently, the bundle of Leibnizian bases, denoted by $LB\pl \M,\xi,\psi,\gamma\pr$, is a $\CARR\pl d+1\pr$-structure on $\M$. A converse statement can be formulated by asserting that the base space of a $\CARR\pl d+1\pr$-structure carries an ambient Leibnizian structure. 

\noindent A basis of $T^*_x\M$ dual to $L=\lbrace \xi,N, e_{i}\rbrace$ can be defined as $L^*:=\pset{{A},\psi, \theta^i}$ where
\vspace{2mm}
\begin{enumerate}
\item[3.] $A\pl \xi\pr=1$, $A\pl N\pr=0$, $A\pl e_{i}\pr=0$
\item[4.] $\theta^i\pl \xi\pr=0$, $\theta^i\pl N\pr=0$, $\theta^i\pl e_{j}\pr=\delta^i_j$. 
\end{enumerate}
Remember that the condition $A\pl \xi\pr=1$ holds by definition for an Ehresmann connection. All together, the conditions 1-4 state that $L^*$ is indeed the dual basis to $L$.
Therefore it is clear that the Carroll group acts regularly on the space of dual Leibnizian bases at a point. More explicitly, the group action reads:
\bea
\pset{{A},\psi, \theta^i}\overset{T}{\mapsto}\pset{A+\bbf^T_i\theta^i-\pl a+\bbf^T_i\bb^i\pr\psi,\psi, (\bR^T){}^{i}_{\, \, j}\pl \theta^j-\bb^j\psi\pr}. \label{eqleibnizautodual}
\eea

\noindent As a last piece of terminology, we introduce the notion of Leibnizian pair:
\bdefi{Leibnizian pair}{\label{defiLeibnizianframe}Let $\mathscr L\pl\M,\xi,\psi,\gamma\pr$ be an ambient Leibnizian structure and ${LB\pl \M,\xi,\psi,\gamma\pr}/{O\pl d\pr}$ be the space of orbits for the action of $O\pl d\pr$ on the bundle of Leibnizian bases. A section of the former bundle will be called a Leibnizian pair. 
The space of Leibnizian pairs will be denoted $$\LF:=\field{\frac{LB\pl \M,\xi,\psi,\gamma\pr}{O\pl d\pr}}. $$
}
\noindent Explicitly, a Leibnizian pair is a couple $\pl N,A\pr$ with $N\in FO\pl\M,\psi\pr$ a field of observers and $A\in\EC$ an Ehresmann connection on $\M$ satisfying $A\pl N\pr=0$. The space of invariant Leibnizian pair (\ie satisfying the extra conditions $\Lag_\xi N=0=\Lag_\xi A$) will be denoted $\LFinv:=\fieldinv{\frac{LB\pl \M,\xi,\psi,\gamma\pr}{O\pl d\pr}}$. 

\noindent The typical fiber of the bundle ${LB\pl \M,\xi,\psi,\gamma\pr}/{O\pl d\pr}$ over $\M$ is the coset space $\CARR\pl d+1\pr/ O\pl d\pr\cong H\pl d\pr$ which is of dimension $2\, d+1$ and therefore matches the number of independent components of a Leibnizian pair: $2\pl d+1\pr-1$.

\noindent We will denote by $\mathscr H\pl d\pr$ the local Heisenberg group defined as the semi-direct product $\mathscr H\pl d\pr:=\big(\fV\oplus \field{\Ann \xi/\Span \psi}\big)\ltimes \fonc{\M}$
endowed with the composition law \bea\big( \crl V'\crr,\crl \alpha'\crr,a'\big)\cdot \big( \crl V\crr,\crl \alpha\crr,a\big)=\big( \crl V'+V\crr,\crl \alpha'+\alpha\crr,a'+a-\alpha\pl V'\pr\big) \nn\eea where $\crl  V\crr\in\fV$ denotes an equivalence class of vector fields $V\in\Milneg$ differing by a multiple of the fundamental vector field $\xi$ and $\crl \alpha\crr\in\field{\Ann \xi/\Span \psi}$ stands for an equivalence class of 1-forms $\alpha\in\field{\Ann \xi}$ differing by a multiple of the absolute clock $\psi$. 

\noindent The local Heisenberg group $\mathscr H\pl d\pr$ acts regularly on the space of Leibnizian pairs $\LF$ via the left action: 
\bea
&&\mathscr H\pl d\pr\times\LF \to \LF\label{eqGaction}\\
&&\big( \pl \crl V\crr,\crl \alpha\crr,a\pr,\pl N,A\pr\big)\mapsto\Big( N+\ovA{P}{}\pl V\pr+a\, \xi,A+\overset{N}{\bar P}\pl \alpha\pr-\pl a+\alpha\big( V\pr\big)\psi\Big)\nn
\eea
where $V\in\Milneg$ and $\alpha\in\field{\Ann \xi}$ are representatives of the equivalence classes $\crl  V\crr\in\fV$ and $\crl \alpha\crr\in\field{\Ann \xi/\Span \psi}$, respectively. Since the previous action of the local Heisenberg group $\mathscr H\pl d\pr$ is regular, the space of Leibnizian pairs $\LF$ is a principal homogeneous space. 
\vspace{2mm}
The notion of Leibnizian pair allows to define:
\bdefi{Spacelike projection of vector fields}{\label{defispacelikeprojection}Let $L:=\pl N,A\pr\in\LF$ be a Leibnizian pair. The field of endomorphisms $\overset{L}{P}:\field{T\M}\to\field{\Ker \psi\cap\Ker A}$ defined as \bea \overset{L}{P}\pl X\pr=X-\psi\pl X\pr N-A\pl X\pr\xi\label{eqPL}\eea with $X\in\vf$ a vector field on $\M$, is called a spacelike projector of vector fields. }
\noindent Note that $\Ker \overset{L}{P}\cong\Spann{\xi}{N}$. 
\noindent The transpose of $\overset{L}{P}$, denoted $\overset{L}{\bar P}:\ff\to \field{\Ann \xi\,\cap \Ann N}$
is 
given by
\bea
\overset{L}{\bar P}\pl \omega\pr=\omega-\omega\pl N\pr\psi-\omega\pl\xi\pr {A}\label{eqbarPL}
\eea
and $\Ker \overset{L}{\bar P}\cong\Spannn{\psi,{A}}$. 
\vspace{2mm}

\noindent Armed with these projectors, one can introduce a (non-canonical) generalisation of the metric $\gamma$ (respectively $h$) acting on the whole tangent space $T\M$ (respectively cotangent space $T^*\M$):
\bdefi{Ambient transverse metrics}{\label{defitransversemetricrel}Let $\mathscr L\pl\M,\xi,\psi,\gamma\pr$ be an ambient Leibnizian structure and $N\in\FO$ a field of observers. The covariant ambient transverse metric $\N{\gamma}\in\bforms$ is defined by its action on vector fields $X,Y\in\field{TM}$ as
\bea
\N{\gamma}\pl X,Y\pr=\gamma\pl \overset{N}{P}\pl X\pr, \overset{N}{P}\pl Y\pr\pr\label{eqtransversemetricrel}
\eea
where $\overset{N}{P}:\field{T\M}\to\field{\Ker \psi}$ stands for the projector on $\Ker\psi$ associated with the \fo $N$ (\cf eq.\eqref{eqPN}).

\noindent Conversely, the contravariant transverse metric $\ovA{h}\in\bform$ can be defined by its action on 1-forms $\alpha,\beta\in\ff$ as
\bea
\ovA{h}\pl \alpha,\beta\pr=h\pl \ovbA{P}{}\pl \alpha\pr, \ovbA{P}{}\pl \beta\pr\pr \label{eqtransversemetricrel2}
\eea
where $\ovbA{P}{}:\ff\to \field{\Ann \xi}$ is the projector on $\Ann \xi$ associated with the Ehresmann connection $A$ (\cf eq.\eqref{eqbarPA}).
}
\noindent The following relation follows from the previous definitions: $$\ovA{h}{}^{\mu\lambda}\N{\gamma}_{\lambda\nu}=\overset{L}{P}{}^\mu_\nu=\delta^\mu_\nu-\xi^\mu A_\nu-N^\mu\psi_\nu.$$ 

\noindent Under a change of Leibnizian pairs \eqref{eqGaction} parameterised by $\pl \crl V\crr,\crl \alpha\crr,a\pr\in\mathscr H\pl d\pr$, the previously introduced metrics transform as:
\bea
\N\gamma_{\mu\nu}&\overset{}{\mapsto}&\N\gamma_{\mu\nu}-2\, \psi_{(\mu}\N\gamma_{\nu)\lambda}V^\lambda+\gamma(V,V)\, \psi_\mu \psi_\nu\nn\\
\ovA h{}^{\mu\nu}&\overset{}{\mapsto}&\ovA h{}^{\mu\nu}-2\, \xi^{(\mu}\ovA h{}^{\nu)\lambda}\alpha_\lambda+h(\alpha,\alpha)\, \xi^\mu \xi^\nu.\nn
\eea

\pagebreak
\section{Ambient Galilean connections}\label{Ambient Galilean connections}
\noindent The previous Section introduced the notion of ambient Leibnizian structures. We discussed in particular the subclasses projecting as a Leibnizian structure on the Platonic screen (projectable ambient Leibnizian structures) and admitting a foliation by Carrollian structures (ambient Aristotelian structures). 
The next logical step consists in investigating how ambient Leibnizian structures can be endowed with a notion of parallelism, in the guise of a compatible connection and how these connections (called ambient Galilean) induce Galilean and Carrollian connections.

{More specifically, the present Section aims at providing solutions to the equivalence problem for connections compatible with ambient Leibnizian structures. We will do so by following throughout the systematic procedure introduced in \cite{Bekaert2014b} that we now briefly review:
\begin{itemize}
\item Let $\M$ be a manifold and denote $\mathscr D(\M)$ the space of (possibly torsional) Koszul connections on $\M$. As is well-known, $\mathscr D(\M)$ is an affine space modelled on the vector space $\mathscr V(\M):=\field{T^*\M\otimes T^*\M\otimes T\M}$. Explicitly, if $\Gamma\in\mathscr D(\M)$ is a connection on $\M$, then $\Gamma':=\Gamma+S$ with $S\in\mathscr V(\M)$ is also a connection on $\M$, \ie $\Gamma'\in\mathscr D(\M)$. 
\item Letting $\mathfrak m$ be a ``metric structure'' on $\M$ (defined loosely as a set of tensors on $\M$, possibly satisfying some compatibility relations), we denote $\mathscr D(\M,\mathfrak m)$ the subspace of Koszul connections compatible with $\mathfrak m$ (\ie satisfying $\nabla \mathfrak m=0$). The subspace $\mathscr D(\M,\mathfrak m)\subset\mathscr D(\M)$ possesses a structure of affine space modelled on a subspace of $\mathscr V(\M)$ denoted $\mathscr V(\M,\mathfrak m)\subset\mathscr V(\M)$. 

The princeps example is of course the case where $\mathfrak m:=g$ is a nondegenerate metric on $\M$. The space $\mathscr D(\M,g)$ of Koszul connections compatible with $g$ is then modelled on the vector space 
\bea
\GammaVr:=\pset{S\in\mathscr V(\M)\ /\ S^{(\lambda}_{\mu\nu}\, g^{\rho)\nu}_{}=0},\nn
\eea 
where we wrote the compatibility condition in terms of the inverse metric. 

The above condition on the tensor $S$ ensures that translating a compatible connection by $S$ preserves compatibility with $g$.  
\item The first non-trivial task to address in order to provide a classification of $\mathscr D(\M,\mathfrak m)$ consists in ``resolving'' the model space $\mathscr V(\M,\mathfrak m)$. Technically, this is done by introducing a vector space $\mathscr W(\M)$ whose definition does not rely \cite{footnote7}
 on the structure $\mathfrak m$ along with an explicit isomorphism \cite{footnote8} from $\mathscr V(\M,\mathfrak m)$ to $\mathscr W(\M)$. 

In the previous example, the vector space $\mathscr W(\M)$ is defined as $\mathscr W(\M):=\vectasymdu$ which can be shown to be isomorphic to $\GammaVr$ along the map:
\bea\varphi:\GammaVr\to\mathscr W(\M):S^\lambda_{\mu\nu}\mapsto T^\lambda_{\mu\nu}:=S^\lambda_{[\mu\nu]}\label{varphisiorel}\eea
whose inverse takes the form: 
\bea\varphi\un:\mathscr W(\M)\to\GammaVr:T^\lambda_{\mu\nu}\mapsto S^\lambda_{\mu\nu}:=g^{\lambda\rho}\crl T_{\mu|\rho\nu}+T_{\nu|\rho\mu}+T_{\rho|\mu\nu}\crr\label{varpphiisorelinv}\eea 
where $T^\lambda_{\mu\nu}:=g_{\lambda\rho}T^\rho_{\mu\nu}$. 

Note that the vector space $\mathscr W(\M)$ is defined intrinsically \ie without referring to the metric structure $g$ and hence qualifies as a resolution of $\mathscr V(\M,\mathfrak m)$. 
\item The second task consists in picking an origin for $\mathscr D(\M,\mathfrak m)$ in order to endow the latter with the richer structure of vector space. Note that, although any connection in $\mathscr D(\M,\mathfrak m)$ can play this role, depending on the type of structures $\mathfrak m$ preserved, some choices are more natural than others. In particular, we will rely on the two following criteria to motivate our choices:
\begin{enumerate}
\item The origin can be defined with the minimal amount of structure other than $\mathfrak m$. 
\item Whenever $\mathfrak m$ admits torsionfree compatible connections \cite{footnote9}, the origin is torsionfree. 
\end{enumerate}
Technically, choosing an origin involves the definition \cite{footnote10} of an affine map $\Theta:\mathscr D\pl\M,\mathfrak m\pr\to \mathscr W(\M)$ modelled on the isomorphism $\varphi:\GammaVr\to\mathscr W(\M)$, \ie the following condition is satisfied
\bea
\Theta\pl \Gamma'\pr-\Theta\pl \Gamma\pr=\varphi\pl \Gamma'-\Gamma\pr\nn
\eea
for all $\Gamma', \Gamma\in\mathscr D\pl\M,\mathfrak m\pr$. 

This condition together with the fact that $\varphi$ is an isomorphism of vector spaces ensures that there exists a (necessarily unique) element $\overset{0}{\Gamma}\in\Ker \Theta$ that then plays the role of origin of $\mathscr D(\M,\mathfrak m)$. 

In cases where $\mathfrak m$ admits torsionfree compatible connections, the second ``naturality'' condition for origins is satisfied if there exists a linear map 
$$j:\mathscr W(\M)_{T=0}\hookrightarrow\mathscr W(\M)$$ such that the following diagram commutes
\bea
 \xymatrix{
    \mathscr D(\M,\mathfrak m)_{T=0}\,  \ar@{^{(}->}[r]^{{i}} \ar[d]_{{\Theta}_{T=0}} & \mathscr D(\M,\mathfrak m) \ar[d]^{{\Theta}} \\
   \mathscr W(\M)_{T=0}\,   \ar@{^{(}->}[r]^{{j}}  & \mathscr W(\M) \label{diagclass}
  }
  \eea
  where $i:\mathscr D(\M,\mathfrak m)_{T=0}\hookrightarrow\mathscr D(\M,\mathfrak m)$ denotes the natural inclusion of torsionless compatible connections inside generic compatible connections.

Pursuing our on-going example of a nondegenerate metric structure $g$, the map $\Theta$ is defined as:
\bea
\Theta:\mathscr D\pl\M,g\pr\to\mathscr W(\M):\Gamma^\lambda_{\mu\nu}\mapsto  T^\lambda_{\mu\nu}:=\Gamma^\lambda_{[\mu\nu]}\label{mapThetarel}
\eea
which is obviously modelled on the isomorphism \eqref{varphisiorel}. The map \eqref{mapThetarel} thus associates with each compatible connection its torsion tensor. The origin $\overset{0}{\Gamma}$ singled out by this choice of map (\ie which spans the kernel of $\Theta$) is thus the unique compatible connection with zero torsion, namely the Levi-Civita connection whose components are the well-known Christoffel coefficients:
\bea
\overset{0}{\Gamma}{}^\lambda_{\mu\nu}=\half g^{\lambda\rho}\pl\p_\mu g_{\rho\nu}+\p_\nu g_{\rho\mu}-\p_\rho g_{\mu\nu}\pr. \label{Christoffel}
\eea
This choice of origin is natural in the sense previously defined \ie $\overset{0}{\Gamma}$ can be written solely in terms of $g$ (without the need of extra structure) and is torsionfree.  
\item Once an origin has been picked out, the latter can be used to ``represent'' any element of $\mathscr D\pl\M,\mathfrak m\pr$. Any compatible connection $\Gamma\in\mathscr D\pl\M,\mathfrak m\pr$ is indeed uniquely characterised by the element $\Theta(\Gamma)\in\mathscr W(\M)$ which characterises the difference between $\Gamma$ and the origin $\overset{0}{\Gamma}$. An explicit expression for $\Gamma$ is given by:
\bea
\Gamma=\overset{0}{\Gamma}+\varphi^{-1}\big(\Theta(\Gamma)\big). \label{eqKoszul}
\eea

Applying the above machinery to our example, we recover the well-known fact that any connection compatible with a nondegenerate metric $g$ is uniquely determined by its torsion tensor. Denoting the torsion tensor of an arbitrary compatible connection $\Gamma \in\mathscr D(\M,g)$ by $T\in \mathscr W(\M)$, \eqref{eqKoszul} together with \eqref{varpphiisorelinv} allows to recover the well-known Koszul formula:
\bea
\Gamma^\lambda_{\mu\nu}=\half g^{\lambda\rho}\crl\p_\mu g_{\rho\nu}+\p_\nu g_{\rho\mu}-\p_\rho g_{\mu\nu}+ T_{\mu|\rho\nu}+T_{\nu|\rho\mu}+T_{\rho|\mu\nu}\crr.\label{eqChristoffeltorsion}
\eea
We call $\mathscr W(\M)$ a {\it classification} \cite{footnote11} of the affine space $\mathscr D(\M,g)$ of connections compatible with $g$. 

\end{itemize}
We now aim at applying the previously sketched procedure to the study of ambient Galilean connections, defined as follows:
}

\bdefi{Ambient Galilean connections}{\label{defAmbient Galilean connections}An ambient Leibnizian structure supplemented with a compatible Koszul connection is called an ambient Galilean manifold. The compatible Koszul connection is then referred to as an ambient Galilean connection. }

\noindent Letting $\mathscr L\pl \M, \xi, \psi, \gamma\pr$ be an ambient Leibnizian structure, the compatibility conditions read 
{\it
\begin{enumerate}
\item $\nabla \xi=0$
\item $\nabla\psi=0$
\item $\nabla \gamma=0$. 
\end{enumerate}}
\noindent These three conditions can be more explicitly stated as:
{\it
\begin{enumerate}
\item $\nabla_X\xi=0$, for all $X\in\vf$
\item $X\crl\psi\pl Y\pr\crr=\psi\pl\nabla_XY\pr$, for all $X,Y\in\field{T\M}$ 
\item $X\crl\gamma\pl V,W\pr\crr=\gamma\pl \nabla_XV,W\pr+\gamma\pl V,\nabla_XW\pr$, for all $X\in\field{T\M}$ and $V,W\in\Milneg$. 
\end{enumerate}}
\noindent Note that the right-hand-side of equation {\it 3.} is well defined since $V\in\field{\Ker\psi}$ implies $\psi\pl\nabla_XV\pr=0$ (\cf Condition {\it 2.}) which in turn, ensures that $\nabla_XV\in\field{\Ker\psi}$, for all $X\in\field{T\M}$. 

\noindent When the absolute rulers are formulated in terms of a field $h\in\field{\vee^2(\Ann\,\xi)^*}$, Condition {\it 3.} can be restated as $\nabla h=0$ or equivalently: 
\bea
X\crl h\pl \alpha, \beta\pr\crr=h\pl\nabla_X\alpha, \beta\pr+h\pl \alpha,\nabla_X\beta\pr,\text{ for all }X\in\field{T\M} \text{ and }\alpha, \beta\in\field{\Ann \xi}. \nn
\eea
Again, we note that the right-hand side is well-defined since the expression $\pl\nabla_X\alpha\pr\pl\xi\pr=X\crl\,\alpha\pl\xi\pr\crr-\alpha\pl\nabla_X\xi\pr$ vanishes for all $\alpha\in\fAnnxi$ whenever Condition {\it 1.} holds, so that $\nabla_X\alpha\in\fAnnxi$. 

{
In Sections \ref{Torsionfree connections} and \ref{Torsional connections}, we give a solution to the equivalence problem for ambient Leibnizian structures by providing a classification of the space of ambient Galilean connections, first in the torsionfree case, then in full generality. In both cases, we make use of the obtained classifications to build a surjective (resp. injective) map from the space of ambient Galilean connections on $\M$ to the space of Galilean connections on the Platonic screen $\M/\mathbb R$ (resp. from the space of Carrollian connections on a wavefront worldvolume $\tilde \M\hookrightarrow\M$ to the space of ambient Galilean connections on $\M$).

}
\subsection{Torsionfree connections}\label{Torsionfree connections}

\bprop{\label{propconstrainttorsionless}Let $\mathscr L\pl\M,\xi,\psi,\gamma\pr$ be an ambient Leibnizian structure. Necessary and sufficient conditions for the existence of a torsionfree ambient Galilean connection compatible with $\mathscr L\pl\M,\xi,\psi,\gamma\pr$ are:
\bi
\item $d\psi=0$
\item $\Lag_\xi \gamma=0$.
\ei
In other words, $\mathscr L\pl\M,\xi,\psi,\gamma\pr$ must be a projectable ambient Augustinian structure in order to admit a torsionfree compatible connection. }
\noindent These two conditions are reminiscent of the nonrelativistic/ultrarelativistic requirements for the existence of a torsionfree connection compatible respectively with a given Leibnizian structure (\cf Proposition 3.2 in \cite{Bekaert2014b}) {or} Carrollian structure (\cf \Prop{propCarrolliancompatibility}). 
\noindent In the following, we will write $\mathscr S\pl\M,\xi,\psi,\gamma\pr$ for a projectable ambient Augustinian structure and denote $\mathscr D\pl\M,\xi,\psi,\gamma\pr$ the space of torsionfree ambient Galilean connections compatible with $\mathscr S\pl\M,\xi,\psi,\gamma\pr$. Since the ambient Augustinian structure $\mathscr S\pl\M,\xi,\psi,\gamma\pr$ is projectable, it induces a well-defined Augustinian structure $\bar{\mathscr S}\pl\bar\M,\bar\psi,\bar\gamma\pr$ on its Platonic screen (\cf Proposition \ref{propbijaug}). Furthermore, since $\psi$ is closed, the ambient spacetime $\M$ admits a foliation by codimension-one hypersurfaces endowed with Carrollian structures. We will denote $i:\tilde \M\longhookrightarrow \M$ one of the leaves and $\tilde{\mathscr C}\pl \tilde\M,\tilde\xi,\tilde\gamma\pr$ the induced Carrollian structure on $\tilde\M$. 

\bprop{\label{PropGalileanaffinespace}The space $\mathscr D\pl\M,\xi,\psi,\gamma\pr$ of torsionfree ambient Galilean connections possesses the structure of an affine space modelled on the vector space \bea\GammaV:=\Bigg\{S\in\field{\vee^2T^*\M\otimes T\M} \text{ satisfying conditions a)-c) }\Bigg\}\nn\eea
where:
\begin{enumerate}[a) ]
\item $S\pl X,\xi\pr=0$ for all $X\in\vf$
\item $\psi\big( S\pl X,Y\pr\big)=0$ for all $X,Y\in\vf$
\item $\gamma\pl S\pl X,V\pr,W\pr+\, \gamma\pl S\pl X,W\pr,V\pr=0\ \text{ for all } X\in\vf\text{ and }V,W\in\Milneg. $
\end{enumerate}
}
\blem{}{\label{Lemcanisot}The vector space $\GammaV$ is isomorphic to the vector space $\field{\otimes^2\Ann\xi}\cong\field{\w^2\Ann\xi}\oplus\field{\vee^2\Ann\xi}$. 
}
\noindent Explicitly, given an Ehresmann connection $A\in\EC$, one can construct the following non-canonical isomorphism:
\bea
\ovA{\varphi}&:&\mathscr V\pl\M,\xi,\psi,\gamma\pr\to \field{\w^2\Ann\xi}\oplus\field{\vee^2\Ann\xi}\label{eqisononcan}\\
&:&S\lmn\mapsto \pl {F}_{\mu\nu}=-2\N{\gamma}_{\lambda[\mu}S^\lambda_{\nu]\rho}N^\rho,{\Sigma}_{\mu\nu}=A_\lambda S\lmn\pr\nn
\eea
whose inverse takes the form
\bea
\ovA{\varphi}{}\un&:&\field{\w^2\Ann\xi}\oplus\field{\vee^2\Ann\xi}\to \mathscr V\pl\M,\xi,\psi,\gamma\pr\\
&:&\pl {F}_{\mu\nu}, {\Sigma}_{\mu\nu}\pr\mapsto S\lmn=\ovA{h}{}^{\lambda\rho}\psi_{(\mu}F_{\nu)\rho}+\xi^\lambda {\Sigma}_{\mu\nu}.\nn 
\eea
Note that the expression of the 2-form $F$ in eq.\eqref{eqisononcan} is independent of the choice of field of observers $N\in\FO$. In the previous expressions, $\N{\gamma}$ (respectively $\ovA{h}$) stands for the covariant (respectively contravariant) ambient transverse metric associated with the field of observers $N$ (respectively the Ehresmann connection $A$). 

\noindent Explicitly, under a Carroll boost $A\mapsto A+\alpha$, with $\alpha\in\fAnnxi$ (\cf Appendix \ref{sectionappendix}), the isomorphism $\ovA{\varphi}$ transforms as:
\bea
\ovAp{\varphi}\pl S\lmn\pr&=&\ovA{\varphi}\pl S\lmn\pr+\pl 0,\alpha_\lambda S\lmn\pr\\
\ovAp{\varphi}{}\un\pl {F}_{\mu\nu},\Sigma_{\mu\nu}\pr&=&\ovA{\varphi}{}\un\pl F_{\mu\nu},\Sigma_{\mu\nu}-\ovA{h}{}^{\lambda\rho}\alpha_\lambda\psi_{(\mu}{F}_{\nu)\rho}\pr. \label{eqtransvarphiNun}
\eea
\noindent We now construct the following map:
\bea
{\Theta}&:&\LFinv\times\mathscr D\pl\M,\xi,\psi,\gamma\pr\to\field{\w^2\Ann\xi}\oplus\field{\vee^2\Ann\xi}\nn\\
&:&\big( N,A,\Gamma\big)\mapsto\pl \N{F}_{\mu\nu}=-2\N{\gamma}_{\lambda[\mu}\nabla_{\nu]}N^\lambda,\ovA{\Sigma}_{\mu\nu}=-\nabla_{(\mu}A_{\nu)}\pr\,. 
\eea
For all invariant Leibnizian pairs $\pl N,A\pr\in\LFinv$, the map $\NA\Theta:= \Theta\big( \pl N,A\pr,\cdot\big):\mathscr D\pl\M,\xi,\psi,\gamma\pr\to\field{\w^2\Ann\xi}\oplus\field{\vee^2\Ann\xi}$ can be shown to be an affine map modelled on the linear map $\ovA{\varphi}$\,, \ie $\NA\Theta\pl\Gamma'\pr-\NA\Theta\pl\Gamma\pr=\ovA\varphi\pl \Gamma'-\Gamma\pr$ for all $\Gamma',\Gamma\in\mathscr D\pl\M,\xi,\psi,\gamma\pr$. The superscript acts here as a reminder of the fact that $\NA{\Theta}$ is not canonical. One can check that the fact that $\pl N,A\pr$ is an invariant Leibnizian pair ensures that $\N{F}\in \field{\w^2\Ann\xi}$ and $\ovA{\Sigma}\in\field{\vee^2\Ann\xi}$ \cite{footnote12}.

\noindent Furthermore, Proposition B.4 in \cite{Bekaert2014b} ensures that, since $\NA{\Theta}$ is an affine map modelled on the isomorphism of vector spaces $\ovA{\varphi}$, the kernel of $\NA{\Theta}$ is spanned by a unique $\NA{\Gamma}\in\mathscr D\pl\M,\xi,\psi,\gamma\pr$. 
We will refer to $\NA{\Gamma}\in\mathscr D\pl\M,\xi,\psi,\gamma\pr$ as the {\it torsionfree special ambient connection} associated with the invariant Leibnizian pair $\pl N,A\pr$. A Koszul formula for $\NA{\Gamma}$ reads as:
\bea
A\pl \NA\nabla_XY\pr&=&X\crl A\pl Y\pr\crr-\half dA\pl X,Y\pr\\
2\,\N\gamma\pl\NA\nabla_XY,V\pr&=&X\crl \N\gamma\pl Y,V\pr\crr+Y\crl \N\gamma\pl X,V\pr\crr-V\crl \N\gamma\pl X,Y\pr\crr\label{KformulaGalsp}\\
&&+\N\gamma\pl\br{X}{Y},V\pr-\N\gamma\pl\br{Y}{V},X\pr-\N\gamma\pl\br{X}{V},Y\pr\nn
\eea
for all $X,Y\in\vf$ and $V\in\Milneg$ while an explicit component expression for its coefficients is given by
\cite{footnote13}
:
\bea
\NA{\Gamma}\lmn=\xi^\lambda\p_{(\mu}A_{\nu)}+N^\lambda\p_{(\mu}\psi_{\nu)}+\half \ovA{h}{}^{\lambda\rho}\crl \p_\mu \N{\gamma}_{\rho\nu}+\p_\nu \N{\gamma}_{\rho\mu}-\p_\rho \N{\gamma}_{\mu\nu}\crr. \label{exptorsionlessambientspecial}
\eea

\noindent Given an invariant Leibnizian pair $\pl N,A\pr$, the associated torsionfree special ambient connection can thus be used in order to represent any torsionfree ambient Galilean connection ${\Gamma}\in\mathscr D\pl\M,\xi,\psi,\gamma\pr$ as $\Gamma=\NA{\Gamma}+\ovA{\varphi}{}\un\Big( \NA{\Theta}\pl \Gamma\pr\Big)$. The obtained Koszul formula reads as:
\bea
A\pl \nabla_XY\pr&=&X\crl A\pl Y\pr\crr-\half dA\pl X,Y\pr+\ovA\Sigma\pl X,Y\pr\label{KformulaGalA}\\
2\,\N\gamma\pl\nabla_XY,V\pr&=&X\crl \N\gamma\pl Y,V\pr\crr+Y\crl \N\gamma\pl X,V\pr\crr-V\crl \N\gamma\pl X,Y\pr\crr\nn\\
&&+\N\gamma\pl\br{X}{Y},V\pr-\N\gamma\pl\br{Y}{V},X\pr-\N\gamma\pl\br{X}{V},Y\pr\label{KformulaGal}\\
&&+\psi\pl X\pr\N F\pl Y,V\pr+\psi\pl Y\pr\N F\pl X,V\pr\nn
\eea
for all $X,Y\in\vf$ and $V\in\Milneg$. Note that the ``timelike'' part of $\Gamma$ is fully constrained by the compatibility condition of $\psi$ (\cf Condition \textit{2.} of \Defi{defAmbient Galilean connections}). 
Explicitly, the components of $\Gamma$ can be written as:
\bea
\Gamma\lmn&=&\xi^\lambda\p_{(\mu}A_{\nu)}+N^\lambda\p_{(\mu}\psi_{\nu)}+\half \ovA{h}{}^{\lambda\rho}\crl \p_\mu \N{\gamma}_{\rho\nu}+\p_\nu \N{\gamma}_{\rho\mu}-\p_\rho \N{\gamma}_{\mu\nu}\crr\label{TGconnect}\\
&&+\ovA{h}{}^{\lambda\rho}\psi_{(\mu}\N{F}_{\nu)\rho}+\xi^\lambda \ovA{\Sigma}_{\mu\nu}. \nn
\eea
This is the most general expression of a torsionfree ambient Galilean connection. The arbitrariness is encoded in both tensors $\N{F}\in \field{\w^2\Ann\xi}$ and $\ovA{\Sigma}\in\field{\vee^2\Ann\xi}$ corresponding respectively to the arbitrariness in a torsionfree Galilean connection (\cf Section 3 in \cite{Bekaert2014b}) and a torsionfree Carrollian connection (\cf Section \ref{Torsionfree Carrollian connections}). 
\vspace{2mm}

\noindent Before characterising further the space $\mathscr D\pl\M,\xi,\psi,\gamma\pr$, let us anticipate the forthcoming discussion regarding projection of connections on the Platonic screen by considering the subspace of \textit{invariant} torsionfree ambient Galilean connections, in the sense of \Prop{propinvKoszul}, denoted $\mathscr D_\inv\pl\M,\xi,\psi,\gamma\pr$. Note that in the torsionfree case, the invariance of such a Koszul connection $\nabla$ reduces to the condition that the parallel vector field $\xi$ is ``affine Killing'' for $\nabla$ \ie enjoys the property $\Lag_\xi \nabla=0$. While in the (pseudo)-Riemannian case, a Killing vector field is automatically affine Killing for the associated Levi-Civita connection, this property is lost in the degenerate case as embodied in the following Proposition: 
\bprop{\label{propprojtorfree}Let $\mathscr S\pl\M,\xi,\psi,\gamma\pr$ be a projectable ambient Augustinian structure and $\nabla\in\mathscr D\pl\M,\xi,\psi,\gamma\pr$ be a torsionfree ambient Galilean connection compatible with $\mathscr S\pl\M,\xi,\psi,\gamma\pr$. Let furthermore $\pl N,A\pr$ be an invariant Leibnizian pair and denote $\pl\N{F}, \ovA{\Sigma}\pr:=\NA{\Theta}\pl\nabla\pr$. Then $\nabla$ is invariant if and only if both $\N F$ and $\ovA \Sigma$ are invariant (\ie$\Lag_\xi\N F=0$ and $\Lag_\xi \ovA\Sigma=0$). 
}

\noindent Under a change of Leibnizian pair \eqref{eqGaction}, the map $\NA{\Theta}$ varies according to
\bea
\NAp{\Theta}\pl\Gamma\pr=\NA{\Theta}\pl\Gamma\pr+\pl2\p_{[\mu}\overset{N,V}{\Phi}_{\nu]}\,,\, -\half\Lag_{\alpha^\#{}}\N{\gamma}_{\mu\nu}+\ovA{h}{}^{\lambda\rho}\alpha_\lambda\psi_{(\mu}\N{F}_{\nu)\rho}+\p_{(\mu}\pl a+\alpha\pl V\pr\pr\psi_{\nu)}\pr\nn
\eea
where $\alpha^\#{}:= \ovA{h}\pl\alpha\pr\in\vf$ and $\overset{N,V}{\Phi}:= \N{\gamma}\pl V\pr-\half\gamma\pl V, V\pr\psi$. This expression can then be used to define an action of the group $\mathscr H_{\inv}\pl d\pr$ on the space $\mathscr R\pl\M,\xi,\psi,\gamma\pr:=\LFinv\times\big(\field{\w^2\Ann\xi}\oplus\field{\vee^2\Ann\xi}\big)$ as:
\bea
\begin{cases}
N'=N+\ovA{P}{}\pl V\pr+a\, \xi\\
A'=A+\overset{N}{\bar P}\pl\alpha\pr-\pl a+\alpha\pl V\pr\pr\psi\\
F'={F}+d\overset{N,V}{\Phi}\\
\Sigma'_{\mu\nu}={\Sigma}_{\mu\nu}-\half\Lag_{\alpha^\#{}}\N{\gamma}_{\mu\nu}+\ovA{h}{}^{\lambda\rho}\alpha_\lambda\psi_{(\mu}{F}_{\nu)\rho}+\p_{(\mu}\pl a+\alpha\pl V\pr\pr\psi_{\nu)} . 
\end{cases}\,
\eea

\noindent The previous group action can be seen to embody both nonrelativistic and ultrarelativistic group actions (denoted Milne and Carroll boosts) studied respectively in  \cite{Bekaert2014b} and Appendix \ref{sectionappendix}. Explicitly, in the invariant case, one can project the $\mathscr H_{\inv}\pl d\pr$-action on $\mathscr R_\inv\pl\M,\xi,\psi,\gamma\pr:=\LFinv\times\big(\fieldinv{\w^2\Ann\xi}\oplus\fieldinv{\vee^2\Ann\xi}\big)$ onto the Platonic screen $\bar\M$ so that to recover 
the action of the \nr Milne group $\field{\Ker\bar\psi}$ on $FO\pl \bar\M, \bar\psi\pr\times\Om^2\pl\bar\M\pr$ (\cf eq. (3.32) in \cite{Bekaert2014b}):
\bea
\pl \bar N,\bar F\pr\mapsto\pl \bar N',\bar F'\pr=\pl\bar N+\bar V,\bar F+d\overset{\bar N,\bar V}{\Phi}\pr\label{eqprojtrans1}
\eea
where $\bar N:=\pi_*N$, $\pi^*\bar F:= F$, $\bar V:= \pi_*V$ and $\overset{\bar N,\bar V}{\Phi}:= \bN{\gamma}\pl \bar V\pr-\half \bar\gamma\pl \bar V,\bar V\pr\bar\psi$. 
\vspace{2mm}

\noindent Similarly, one can recover the group action of $\field{\Ann \tilde \xi}$ on $EC\pl \tilde\M,\tilde\xi\pr\times\field{\vee^2\Ann\tilde\xi}$ $\Big($\cf eq.\eqref{groupactioncarr}$\Big)$ by pullback of the transformation relations of the couple $\pl A,\Sigma\pr$ on a leaf $i:\tilde\M\longhookrightarrow \M$:
\bea
\pl \tilde A,\tilde \Sigma\pr\mapsto \pl \tilde A+\tilde \alpha\,,\,\tilde \Sigma-\half \Lag_{\tilde \alpha^\#{}}\tilde \gamma\pr\label{eqprojtrans2}
\eea
where $\tilde A:= i^*A$, $\tilde \Sigma:= i^*\Sigma$, $\tilde \alpha:= i^*\alpha$ and ${\tilde \alpha^\#{}}:=\overset{\tilde A}{h}\pl\tilde\alpha\pr$. 
\bdefi{Ambient gravitational fieldstrength}{An $\mathscr H_{\inv}\pl d\pr$-orbit in $\mathscr R\pl\M,\xi,\psi,\gamma\pr$ is dubbed an ambient gravitational fieldstrength. The space of ambient gravitational fieldstrengths will be denoted $\mathscr F\pl\M,\xi,\psi,\gamma\pr:=\mathscr R\pl\M,\xi,\psi,\gamma\pr/\mathscr H_{\inv}\pl d\pr$. }
\noindent The subspace $\mathscr F_\inv\pl\M,\xi,\psi,\gamma\pr:=\mathscr R_\inv\pl\M,\xi,\psi,\gamma\pr/\mathscr H_{\inv}\pl d\pr$ of $\mathscr H_{\inv}\pl d\pr$-orbit in $\mathscr R_\inv\pl\M,\xi,\psi,\gamma\pr$ will be referred to as the space of invariant ambient gravitational fieldstrengths. 

\noindent Using this terminology, we can further characterise the affine space of torsionfree ambient Galilean connections. The following Proposition is a straightforward application of Proposition B.4 in \cite{Bekaert2014b}:

\bprop{\label{corGalilean}The space $\mathscr D\pl\M,\xi,\psi,\gamma\pr$ (resp. $\mathscr D_\inv\pl\M,\xi,\psi,\gamma\pr$) of (invariant) torsionfree ambient Galilean connections compatible with a given projectable ambient Augustinian structure $\mathscr S\pl\M,\xi,\psi,\gamma\pr$ possesses the structure of an affine space canonically isomorphic to the affine space $\mathscr F\pl\M,\xi,\psi,\gamma\pr$ (resp. $\mathscr F_\inv\pl\M,\xi,\psi,\gamma\pr$) of (invariant) ambient gravitational fieldstrengths. 
}
\noindent {According to the terminology introduced at the beginning of this Section, the affine space $\mathscr F\pl\M,\xi,\psi,\gamma\pr$ provides a classification of the affine space of torsionfree ambient Galilean connections. }This characterisation of torsionfree ambient Galilean connections in terms of ambient gravitational fieldstrengths mimics the construction of Proposition 3.14 in \cite{Bekaert2014b}. With the help of these two classifications, we establish the following fact:

\bccor{Let $\mathscr S\pl\M,\xi,\psi,\gamma\pr$ be a projectable ambient Augustinian structure and denote $\bar{\mathscr S}\pl\bar\M,\bar\psi,\bar\gamma\pr$ the induced Augustinian structure on the Platonic screen $\bar\M$. Let $\mathscr D_\inv\pl\M,\xi,\psi,\gamma\pr$ and $\bar{\mathscr D}\pl\bar\M,\bar\psi,\bar\gamma\pr$ denote the affine spaces of (invariant) torsionfree Galilean connections compatible with ${\mathscr S}$ and $\bar{\mathscr S}$, respectively. There is a surjective affine map $\pi:\mathscr D_\inv\pl\M,\xi,\psi,\gamma\pr\twoheadrightarrow\bar{\mathscr D}\pl\bar\M,\bar\psi,\bar\gamma\pr$. 
}
\noindent The proof takes advantage of the classification of (ambient) torsionfree Galilean connections in terms of (ambient) gravitational fieldstrengths, allowing to construct the map:
\bea
\pi:\mathscr F_\inv\pl\M,\xi,\psi,\gamma\pr\twoheadrightarrow\bar{\mathscr F}\pl\bar\M,\bar\psi,\bar\gamma\pr\,:\,
\crl N,A,\N F,\ovA\Sigma\crr\mapsto\crl \bar N,\bN{F}\crr\nn
\eea
where $\bar N:=\pi_*N$, $\pi^*\bN F:= \N F$. This map is obviously well-defined \big(\cf eq.\eqref{eqprojtrans1}\big) and surjective. The affine map $\pi:\mathscr D_\inv\pl\M,\xi,\psi,\gamma\pr\twoheadrightarrow\bar{\mathscr D}\pl\bar\M,\bar\psi,\bar\gamma\pr$ is then obtained by making the following diagram commute: 
\bea
\xymatrix{
\mathscr D_\inv\pl\M,\xi,\psi,\gamma\pr
\ar[r]^\sim
\ar@{->>}[d]_{\pi}&\mathscr F_\inv\pl\M,\xi,\psi,\gamma\pr\ar@{->>}[d]^{\pi}\\ \label{diagrammap}
\bar{\mathscr D}\pl\bar\M,\bar\psi,\bar\gamma\pr\ar[r]^\sim&\bar{\mathscr F}\pl\bar\M,\bar\psi,\bar\gamma\pr
}
\eea
\noindent In other words, any {torsionfree} Galilean manifold can be obtained as projection of a (class of) ambient {torsionfree} Galilean manifolds. Note that the presence of the left vertical arrow is justified since, given an invariant torsionfree ambient Galilean connection $\nabla\in\mathscr D_\inv\pl\M,\xi,\psi,\gamma\pr$, the torsionfree Galilean connection $\bar\nabla\in\bar{\mathscr D}\pl\bar\M,\bar\psi,\bar\gamma\pr$ defined on the Platonic screen via Diagram \eqref{diagrammap} coincides with the Koszul connection obtained by projection of $\nabla$ using Diagram \eqref{diagdefibarnabla} as can be checked by direct computation using Koszul formulas \eqref{KformulaGalA}-\eqref{KformulaGal} and comparing with eq.(4.51) in \cite{Bekaert2014b}. \noindent Componentwise, the projection of \eqref{TGconnect} takes the form:
\bea
&\bar\Gamma^{\bar\lambda}_{\bar\mu\bar\nu}=\bar N^{\bar\lambda}\p_{(\bar\mu}\bar\psi_{\bar\nu)}+\half\, \bar h{}^{\bar\lambda\bar\rho}\crl \p_{\bar\mu} \bN{\bar\gamma}_{\bar\rho\bar\nu}+\p_{\bar\nu} \bN{\bar\gamma}_{\bar\rho\bar\mu}-\p_{\bar\rho} \bN{\bar\gamma}_{\bar\mu\bar\nu}\crr+{\bar h}{}^{\bar\lambda\bar\rho}\bar\psi_{(\bar\mu}\bN{F}_{\bar\nu)\bar\rho}\,. &\nn
\eea 

\noindent The previous Corollary can be dualised to account for the embedding of Carrollian manifolds inside ambient Galilean manifolds:
\bcor{}{Let $\mathscr S\pl\M,\xi,\psi,\gamma\pr$ be a projectable ambient Augustinian structure and denote $i:\tilde \M\longhookrightarrow \M$ one of the leaves foliating $\M$ and $\tilde{\mathscr C}\pl \tilde\M,\tilde\xi,\tilde\gamma\pr$ the induced Carrollian structure. Let $\mathscr D\pl\M,\xi,\psi,\gamma\pr$ (resp. $\tilde{\mathscr D}\pl\tilde\M,\tilde\xi,\tilde\gamma\pr$) denote the affine spaces of torsionfree ambient Galilean connections (resp. torsionfree Carrollian connections) compatible with ${\mathscr S}$ (resp. $\tilde{\mathscr C}$). There is an injective affine map $i:\tilde{\mathscr D}\pl\tilde\M,\tilde\xi,\tilde\gamma\pr\hookrightarrow\mathscr D\pl\M,\xi,\psi,\gamma\pr$. 

}
\noindent The {injective map} is built in terms of the following injection of classifications:
\bea
i:\tilde{\mathscr F}\pl\tilde\M,\tilde\xi,\tilde\gamma\pr\hookrightarrow\mathscr F\pl\M,\xi,\psi,\gamma\pr\,:\,
\Big[ \tilde A,\overset{\tA}{\Sigma}\Big]\mapsto\crl N,A,\N F,\ovA\Sigma\crr\nn
\eea
where $A\in\EC$ is an Ehresmann connection on $\M$ satisfying $i^*A=\tilde A$ and $\ovA\Sigma\in\field{\vee^2\Ann\xi}$ satisfies $i^*\ovA\Sigma=\overset{\tA}{\Sigma}$. The couple $(N,\N{F})$ can be chosen arbitrarily. 

Again, this map can be checked to be well-defined using eq.\eqref{eqprojtrans2} so that it induces an injective map between the affine spaces of torsionfree ambient Galilean connections and torsionfree Carrollian connections as
\bea
\xymatrix{
\tilde{\mathscr D}\pl\tilde\M,\tilde \xi, \tilde\gamma\pr\ar@{^{(}->}[r]^{{i}}\ar[d]_\sim&\mathscr D\pl\M,\xi,\psi,\gamma\pr\ar[d]^\sim\\\label{diagrammap2}
\tilde{\mathscr F}\pl\tilde\M,\tilde\xi,\tilde\gamma\pr\ar@{^{(}->}[r]^{{i}}&\mathscr F\pl\M,\xi,\psi,\gamma\pr
}
\eea

\noindent Again, the present setup allows to embed the whole class of torsionfree Carrollian manifolds into torsionfree ambient Galilean manifolds. Similarly to the previous case, the embedding procedure described here matches the embedding procedure prescribed by Diagram \eqref{diagcarr}, hence the left vertical arrow. The component expression of the torsionfree Carrollian connection induced by \eqref{TGconnect} on the wavefront worldvolume $\tilde\M$ reads in components:
\bea
\tilde\Gamma^{\tilde\lambda}_{\tilde\mu\tilde\nu}=\tilde \xi^{\tilde\lambda}\p_{(\tilde\mu}\tilde A_{\tilde\nu)}+\half\, \overset{\tilde A}{h}{}^{\tilde\lambda\tilde\rho}\crl \p_{\tilde\mu} \tilde{\gamma}_{\tilde\rho\tilde\nu}+\p_{\tilde\nu} \tilde{\gamma}_{\tilde\rho\tilde\mu}-\p_{\tilde\rho} \tilde{\gamma}_{\tilde\mu\tilde\nu}\crr+\tilde\xi^{\tilde\lambda} \overset{\tilde A}{\Sigma}_{\tilde\mu\tilde\nu}\,.\nn 
\eea

\subsection{Torsional connections}
\label{Torsional connections}

\noindent We now address the issue of torsional ambient Galilean connections by mimicking the previous discussion. 
\bprop{\label{propconstrainttorsional}Let $\mathscr L\pl\M,\xi,\psi,\gamma\pr$ be an ambient Leibnizian structure. The torsion tensor $T\in\field{\w^2T^*\M\otimes T\M}$ of a torsional ambient Galilean connection compatible with $\mathscr L\pl\M,\xi,\psi,\gamma\pr$ satisfies the two following relations: 
\begin{enumerate}
\item $\psi\big( T\pl X,Y\pr\big)=d\psi\pl X,Y\pr$ for all $X,Y\in\vf$\,,
\item $\Lag_\xi \N\gamma\pl V,W\pr=\N\gamma\big( V,T\pl \xi,W\pr\big)+\N\gamma\big( W,T\pl \xi,V\pr\big)$ for all $N\in\FO$ and $V,W\in\Milneg$.
\end{enumerate}

\noindent Whenever the ambient absolute clock $\psi$ is projectable (\ie $\Lagxi\psi=0$), the second condition is equivalent to $\Lag_\xi\gamma\pl V,W\pr=\gamma\big( V,T\pl \xi,W\pr\big)+\gamma\big( W,T\pl \xi,V\pr\big)$ for all $V,W\in\Milneg$.

}

\noindent We emphasise that {\it any} ambient Leibnizian structure can be endowed with a compatible connection provided one allows the connection to have non-vanishing torsion. The previous Proposition displays the required conditions on the torsion tensor
which constrain \cite{footnote14} $\frac{\pl d+1\pr\pl d+2\pr}{2}+\frac{d\pl d+1\pr}{2}=\pl d+1\pr^2$ components.
This fact provides an {\it a posteriori} justification regarding the appearance of the tensors $\N F\in\field{\w^2\Ann\xi}$ and $\ovA\Sigma\in\field{\vee^2\Ann\xi}$ in Section \ref{Torsionfree connections} since the fibers of the vector bundle $\pl{\w^2\Ann\xi}\oplus{\vee^2\Ann\xi}\pr$ have dimension $\frac{d\pl d+1\pr}{2}+\frac{\pl d+1\pr\pl d+2\pr}{2}=\pl d+1\pr^2$. 

~\\
\noindent Let $\mathscr L\pl\M,\xi,\psi,\gamma\pr$ be an ambient Leibnizian structure and denote $\mathscr D\pl\M,\xi,\psi,\gamma\pr$ the set of compatible ambient Galilean connections \cite{footnote15}. 

\bprop{\label{PropGalileanaffinespacetor}The space $\mathscr D\pl\M,\xi,\psi,\gamma\pr$ possesses the structure of an affine space modelled on the vector space $\GammaV$ defined as:
\bea\GammaV:=\Bigg\{S\in\field{T^*\M\otimes T^*\M\otimes T\M} \text{ satisfying conditions a)-c) }\Bigg\}\nn\eea
where:
\begin{enumerate}[a) ]
\item $S\pl X,\xi\pr=0$ for all $X\in\vf$
\item $\psi\big( S\pl X,Y\pr\big)=0$ for all $X,Y\in\vf$
\item $\gamma\pl S\pl X,V\pr,W\pr+\, \gamma\pl S\pl X,W\pr,V\pr=0\ \text{ for all } X\in\vf\text{ and }V,W\in\Milneg. $
\end{enumerate}
}
\blem{}{\label{Lemcanisotor}The vector space $\GammaV$ is isomorphic to the vector space $\mathscr W\pl \M,\xi\pr:=\field{T^*\M\otimes\Ann\xi}\oplus\mathscr U\pl \M,\xi\pr$ where the vector space $\mathscr U\pl \M,\xi\pr$ is defined as
 \bea
\mathscr U\pl \M,\xi\pr:=\Bigg\{{\mathcal U}\in\field{\Ann\xi\otimes\w^2T^*\M}\text{ satisfying condition (*)}\Bigg\}\nn
\eea
where condition $\textit{(*)}$ reads:
\bea\textit{(*):  } {\mathcal U}\pl X|Y,\xi\pr+{\mathcal U}\pl Y|X,\xi\pr=0 \text{ for all }X,Y\in\vf \label{conditionast}.\eea
}
\noindent In components any element $\mathcal U_{\lambda|\mu\nu}\in\mathscr U\pl \M,\xi\pr$ thus satisfies the three following conditions: 
\bi
\item $\mathcal U_{\lambda|(\mu\nu)}=0$.
\item$\xi^\lambda\, \mathcal U_{\lambda|\mu\nu}=0$.
\item$\mathcal U_{(\mu|\nu)\lambda}\, \xi^\lambda=0$.
\ei

\noindent Explicitly, given an Ehresmann connection $A\in\EC$, one can construct the following non-canonical isomorphism \cite{footnote16}:
\bea
\ovA{\varphi}&:&\mathscr V\pl\M,\xi,\psi,\gamma\pr\to \mathscr W\pl \M,\xi\pr\label{varphiAleib}\\
&:&S\lmn\mapsto \Big(\,{\ovA\Sigma}_{\mu\nu}:=A_\lambda S\lmn\,,\, {\mathcal U}_{\lambda|\mu\nu}:=\N{\gamma}_{\lambda\rho}S^{\rho}_{[\mu\nu]}+\psi_\lambda\N{\gamma}_{\sigma[\mu}S^\sigma_{\nu]\rho}N^\rho\Big)\nn
\eea
where $N\in\FO$ is a field of observers \cite{footnote17}. The inverse isomorphism takes the form
\bea
\ovA{\varphi}{}\un&:&\mathscr W\pl \M,\xi\pr\to \mathscr V\pl\M,\xi,\psi,\gamma\pr\\
&:&\pl {\Sigma}_{\mu\nu}, {\mathcal U}_{\lambda|\mu\nu}\pr\mapsto \ovA S\lmn:=\xi^\lambda {\Sigma}_{\mu\nu}+\ovA{h}{}^{\lambda\rho}\crl {{\mathcal U}}_{\mu|\rho\nu}+{{\mathcal U}}_{\nu|\rho\mu}+{{\mathcal U}}_{\rho|\mu\nu}\crr\,\nn.
\eea

\noindent Applying the procedure followed in the torsionfree case, we now construct the following map:
\bea
{\Theta}&:&\LFinv\times\mathscr D\pl\M,\xi,\psi,\gamma\pr\to\mathscr W\pl \M,\xi\pr\label{mapthetaLeib}\\
&:&\big( N,A,\Gamma\big)\mapsto\Big(\ovA{\Sigma}_{\mu\nu}=-\nabla_{(\mu}A_{\nu)}+A_\lambda\, \Gamma\lmna,\nn\\ 
&& \hspace{2.6cm}\NA {\mathcal U}_{\lambda|\mu\nu}=\N{\gamma}_{\lambda\rho}\Gamma^{\rho}_{[\mu\nu]}-\half A_{[\mu}\Lag_\xi\N\gamma_{\nu]\lambda}+\psi_\lambda\N{\gamma}_{\sigma[\mu}\nabla_{\nu]}N^\sigma\Big)\nn. 
\eea
\noindent One can directly check, using the invariance of the Leibnizian pair $\pl N,A\pr$ as well as relation {\textit 2.} of \Prop{propconstrainttorsional}, that $\ovA\Sigma\in$ and $\NA {\mathcal U}$ are indeed elements of $\field{T^*\M\otimes\Ann\xi}$ and $\mathscr U\pl \M,\xi\pr$, respectively \cite{footnote18}. Furthermore, given any invariant Leibnizian pair $\pl N,A\pr\in\LFinv$, the map $\NA\Theta:= \Theta\big( \pl N,A\pr,\cdot\big):\mathscr D\pl\M,\xi,\psi,\gamma\pr\to\mathscr W\pl \M,\xi\pr$ can be shown to be an affine map modelled on the linear map $\ovA{\varphi}$ \ie 
\bea 
\NA\Theta\pl\Gamma'\pr-\NA\Theta\pl\Gamma\pr=\ovA{\varphi}\pl \Gamma'-\Gamma\pr\text{ for all }\Gamma',\Gamma\in\mathscr D\pl\M,\xi,\psi,\gamma\pr.
\eea 
\noindent 
We call the couple $\NA\Theta\pl \Gamma\pr:= \pl \ovA{\Sigma},\NA{{\mathcal U}}\pr$ the {\it ambient torsional gravitational fieldstrength associated with the ambient Galilean connection $\Gamma$ measured by the invariant Leibnizian pair $\pl N,A\pr$}. This piece of terminology as well as the fact that $\NA{\Theta}$ is an affine map modelled on the isomorphism of vector spaces $\ovA{\varphi}$ allows us to formulate the following Proposition:
\bpropp{Torsional special ambient connection}{\label{specialobstor}Given an invariant Leibnizian pair $\pl N,A\pr\in \LF$, there is a unique ambient Galilean connection $\NA{\Gamma}\in\mathscr D\pl\M,\xi,\psi,\gamma\pr$ compatible with the ambient Leibnizian structure $\mathscr L\pl\M,\xi,\psi,\gamma\pr$ such that the torsional ambient gravitational fieldstrength measured by $\pl N,A\pr$ with respect to $\NA{\Gamma}$ vanishes. The connection $\NA{\Gamma}$ will be called the torsional special ambient connection associated with $\pl N,A\pr$. }
\noindent In other words, $\Ker \NA{\Theta}\cong\Span\pset{\NA{\Gamma}}$. A Koszul formula for the torsional special ambient connection associated with a given invariant Leibnizian pair $\pl N,A\pr$ reads:
\bea
A\pl \NA\nabla_XY\pr&=&X\crl A\pl Y\pr\crr-\half dA\pl X,Y\pr\\
2\,\N\gamma\pl\NA\nabla_XY,V\pr&=&X\crl \N\gamma\pl Y,V\pr\crr+Y\crl \N\gamma\pl X,V\pr\crr-V\crl \N\gamma\pl X,Y\pr\crr\nn\\
&&+\N\gamma\pl\br{X}{Y},V\pr-\N\gamma\pl\br{Y}{V},X\pr-\N\gamma\pl\br{X}{V},Y\pr\label{KformulaGalsptor}\\
&&-A\pl Y\pr\Lag_\xi\N\gamma\pl X,V\pr\nn
\eea
for all $X,Y\in\vf$ and $V\in\Milneg$. An explicit expression for the components of $\NA{\Gamma}$ is given by \cite{footnote19}
\bea
\NA{\Gamma}\lmn&=&\xi^\lambda\p_{(\mu}A_{\nu)}+N^\lambda\p_{\mu}\psi_{\nu}+\half \ovA{h}{}^{\lambda\rho}\crl \p_\mu \N{\gamma}_{\rho\nu}+\p_\nu \N{\gamma}_{\rho\mu}-\p_\rho \N{\gamma}_{\mu\nu}\crr-\half \ovA{h}{}^{\lambda\rho}A_\nu\Lag_\xi\N\gamma_{\rho\mu}\nn. \\\label{torspecial}
\eea
\noindent Whenever the underlying ambient Leibnizian structure is a projectable Augustinian structure (and thus satisfies the two conditions of \Prop{propconstrainttorsionless}), then the previous expression reduces to expression \eqref{exptorsionlessambientspecial} of the torsionless special ambient connection associated with the invariant Leibnizian pair $\pl N,A\pr\in\LFinv$. 

\noindent Given an invariant Leibnizian pair $\pl N,A\pr$, the associated torsional special ambient connection can be used in order to represent any ambient Galilean connection ${\Gamma}\in\mathscr D\pl\M,\xi,\psi,\gamma\pr$ as $\Gamma=\NA{\Gamma}+\ovA{\varphi}{}\un\Big( \NA{\Theta}\pl \Gamma\pr\Big)$. The latter thus admits the following Koszul formula:
\bea
A\pl \nabla_XY\pr&=&X\crl A\pl Y\pr\crr-\half dA\pl X,Y\pr+\ovA\Sigma\pl X,Y\pr\label{KformulaGaltorA}\\
2\,\N\gamma\pl\nabla_XY,V\pr&=&X\crl \N\gamma\pl Y,V\pr\crr+Y\crl \N\gamma\pl X,V\pr\crr-V\crl \N\gamma\pl X,Y\pr\crr\nn\\
&&+\N\gamma\pl\br{X}{Y},V\pr-\N\gamma\pl\br{Y}{V},X\pr-\N\gamma\pl\br{X}{V},Y\pr\label{KformulaGaltor}\\
&&-A\pl Y\pr\Lag_\xi\N\gamma\pl X,V\pr\nn\\
&&+2\crl \NA {\mathcal U}\pl X|V,Y\pr+\NA {\mathcal U}\pl Y|V,X\pr+\NA {\mathcal U}\pl V|X,Y\pr\crr\nn
\eea
for all $X,Y\in\vf$ and $V\in\Milneg$. Explicitly, the components of $\Gamma$ can be written as:
\bea
{\Gamma}\lmn&=&\xi^\lambda\p_{(\mu}A_{\nu)}+N^\lambda\p_{\mu}\psi_{\nu}+\half\, \ovA{h}{}^{\lambda\rho}\crl \p_\mu \N{\gamma}_{\rho\nu}+\p_\nu \N{\gamma}_{\rho\mu}-\p_\rho \N{\gamma}_{\mu\nu}\crr-\half\, \ovA{h}{}^{\lambda\rho}A_\nu\, \Lag_\xi\N\gamma_{\mu\rho}\nn\\
&&+\, \xi^\lambda {\ovA\Sigma}_{\mu\nu}+\ovA h{}^{\lambda\rho}\Big[\NA{{\mathcal U}}_{\mu|\rho\nu}+\NA{{\mathcal U}}_{\nu|\rho\mu}+\NA{{\mathcal U}}_{\rho|\mu\nu}\Big]\,\label{eqtorsionalambientGalilean}. 
\eea
The Koszul formula \eqref{KformulaGaltorA}-\eqref{KformulaGaltor} or equivalently the component expression \eqref{eqtorsionalambientGalilean} constitute the most general expression for a generic torsional ambient Galilean connection compatible with the ambient Leibnizian structure $\mathscr L\pl\M,\xi,\psi,\gamma\pr$. The arbitrariness is encoded in the following couple:
\begin{itemize}
\item $\ovA\Sigma\in\field{T^*\M\otimes\Ann\xi}$
\item $\NA {\mathcal U}\in\mathscr U\pl \M,\xi\pr$. 
\end{itemize}
\noindent Note that the fibers of the vector bundle $\field{T^*\M\otimes\Ann\xi}\oplus\mathscr U\pl \M,\xi\pr$ have dimension: 
$\pl d+1\pr\pl d+2\pr+\frac{d\pl d+1\pr\pl d+2\pr}{2}=\frac{\pl d+1\pr\pl d+2\pr^2}{2}$. 
The amount of arbitrariness in the choice of a torsional compatible connection is thus the same for Lorentzian and ambient Leibnizian structures \cite{footnote20}. 

\noindent In order to make an explicit contact with the torsionfree case, we define the following vector space:
 \bea
\hspace{-1cm}\mathscr S\pl \M,\xi,N\pr:=\Bigg\{F\oplus U\in\Om^2(\M)\oplus\field{\Ann\xi\otimes\Ann N\otimes\w^2T^*\M}\text{ satisfying conditions 1.-3.}\Bigg\}\nn
\eea
where $N\in\FO$ is a fixed field of observers and conditions 1.-3. read:
\begin{enumerate}
\item $F(\xi,N)=0$
\item $U(V|\xi,N)=\half\, F(\xi,V)$ for all $V\in\Milneg$
\item $U(V|\xi,W)+U(W|\xi,V)=0$ for all $V,W\in\Milneg$.
\end{enumerate}
For all $N\in\FO$ the vector space $\mathscr S\pl \M,\xi,N\pr$ is isomorphic to $\mathscr U(\M,\xi)$ along the following canonical isomorphism:
\bea
\Psi&:&\mathscr U(\M,\xi)\to\mathscr S(\M,\xi,N)\nn\\
&:&\mathcal U_{\lambda|\mu\nu}\mapsto (F_{\mu\nu}:=-2\, N^\lambda\, \mathcal U_{\lambda|\mu\nu}\, ,\, U_{\lambda|\mu\nu}:=\N P{}^\rho_\lambda\, \mathcal U_{\rho|\mu\nu})\nn\\
\Psi\un&:&\mathscr S(\M,\xi,N)\to\mathscr U(\M,\xi)\nn\\
&:&(F_{\mu\nu}, U_{\lambda|\mu\nu})\mapsto \mathcal U_{\lambda|\mu\nu}:=U_{\lambda|\mu\nu}-\half\, \psi_\lambda\, F_{\mu\nu}\nn.
\eea
The isomorphism $\Psi$ can be trivially extended as an isomorphism:
\bea
\hat\Psi&:&\mathscr W\pl \M,\xi\pr\to \field{T^*\M\otimes\Ann\xi}\oplus\mathscr S(\M,\xi,N)\nn\\
&:&\big(\Sigma,\mathcal U)\mapsto\big(\Sigma,(F,U):=\Psi(\mathcal U)\big)\nn
\eea
together with its inverse
\bea
\hat\Psi\un&:& \field{T^*\M\otimes\Ann\xi}\oplus\mathscr S(\M,\xi,N)\to\mathscr W\pl \M,\xi\pr\nn\\
&:&(\Sigma,F,U)\mapsto\big(\Sigma,\mathcal U:=\Psi\un(F,U)\big)\nn.
\eea
Given a Leibnizian pair $\pl N,A\pr\in\LF$, any ambient Galilean connection $\Gamma\in\mathscr D\pl\M,\xi,\psi,\gamma\pr$ can be characterised uniquely by a triplet $(\Sigma,F,U)\in\field{T^*\M\otimes\Ann\xi}\oplus\mathscr S(\M,\xi,N)$ defined as:
\bea
(\ovA\Sigma,\N F,\NA U):=\big(\hat \Psi\circ \NA\Theta\big)(\Gamma).
\eea
The explicit expressions of these characteristic tensors read:
\bi
\item $\ovA\Sigma\pl X,Y\pr:=A\pl \nabla_XY\pr-X\crl A\pl Y\pr\crr+\half dA\pl X,Y\pr$ for all $X,Y\in\vf$
\item $\N F\in \Omega^{2}\pl{\M}\pr$ as $\N{F}\pl X,Y\pr:= \gamma\pl\nabla_XN, \overset{N}{P}\pl Y\pr\pr-\gamma\pl\nabla_YN, \overset{N}{P}\pl X\pr\pr$ for all $X,Y\in\vf$
\item $\NA U\in\field{\Ann\xi\otimes\w^2T^*\M}$ as 
\bea
\NA U\pl Z|X,Y\pr:=\half\Bigg[ \gamma\pl T\pl X,Y\pr,\overset{N}{P}\pl Z\pr\pr-\half\Big( A\pl X\pr\Lag_\xi\N\gamma\pl Y,Z\pr- A\pl Y\pr\Lag_\xi\N\gamma\pl X,Z\pr\Big)\Bigg]\nn
\eea
for all $X,Y,Z\in\vf$ 
\ei
or in components:
\bi
\item $\ovA{\Sigma}_{\mu\nu}=-\nabla_{(\mu}A_{\nu)}+A_\lambda\, \Gamma\lmna$
\item $\N F_{\mu\nu}=-2\N\gamma_{\lambda[\mu}\nabla_{\nu]}N^\lambda$
\item $\NA U_{\lambda|\mu\nu}=\N\gamma_{\lambda\rho}\Gamma^\rho_{[\mu\nu]}-\half A_{[\mu}\Lag_\xi\N\gamma_{\nu]\lambda}$.
\ei
\noindent The component expression of the ambient Galilean connection $\Gamma$ characterised by the triplet $(\ovA\Sigma,\N F,\NA U)$ is thus given by $\Gamma=\NA{\Gamma}+\big(\ovA{\varphi}{}\un\circ\hat \Psi\un\big)(\ovA\Sigma,\N F,\NA U)$ or explicitly:
\bea
{\Gamma}\lmn&=&\xi^\lambda\p_{(\mu}A_{\nu)}+N^\lambda\p_{\mu}\psi_{\nu}+\half \ovA{h}{}^{\lambda\rho}\crl \p_\mu \N{\gamma}_{\rho\nu}+\p_\nu \N{\gamma}_{\rho\mu}-\p_\rho \N{\gamma}_{\mu\nu}\crr-\half\, \ovA{h}{}^{\lambda\rho}A_\nu\Lag_\xi\N\gamma_{\mu\rho}\nn\\
&&+\, \ovA h{}^{\lambda\rho}\psi_{(\mu}\N F_{\nu)\rho}+\, \xi^\lambda {\ovA\Sigma}_{\mu\nu}+\ovA h{}^{\lambda\rho}\Big[\NA{{ U}}_{\mu|\rho\nu}+\NA{{ U}}_{\nu|\rho\mu}+\NA{{ U}}_{\rho|\mu\nu}\Big]\, . \label{eqambGalconnectioncomp}
\eea
Whenever the underlying ambient Leibnizian structure is a projectable Augustinian structure (so that $d\psi=0=\Lag_\xi\gamma$), contact with the torsionfree case can be made by setting $\ovA\Sigma_{[\mu\nu]}=0=\NA U_{\lambda|\mu\nu}$ so as to recover expression \eqref{TGconnect} for a generic torsionfree ambient Galilean connection \cite{footnote21}. 
The previous discussion provides an {\it a posteriori} justification for our choice of origin of the space of ambient Galilean connections \eqref{torspecial} (resp. \eqref{torspecialfull} whenever the invariance condition on the Leibnizian pair is relaxed), the latter reducing to the origin of the space of torsionless ambient Galilean connections \eqref{exptorsionlessambientspecial} \big(resp \eqref{exptorsionlessambientspecialfull}\big) whenever the underlying ambient Leibnizian structure is a projectable Augustinian structure. In other words, our choice satisfies the second naturality condition \big(\cf Diagram \eqref{diagclass}\big) \ie the following diagram commutes whenever $d\psi=0=\Lag_\xi\gamma$:
\bea
 \xymatrix{
    \mathscr D(\M,\xi,\psi,\gamma)_{T=0}  \ar@{^{(}->}[r]^{{i}} \ar[d]_{\NA{\Theta}_{T=0}} & \mathscr D(\M,\xi,\psi,\gamma) \ar[d]^{\NA{\Theta}} \\
    \field{\vee^2\Ann\xi}\oplus\field{\w^2\Ann\xi}  \ar@{^{(}->}[r]^{\hspace{1.6cm j}}  & \mathscr W(\M,\xi) \nn
  }
  \eea
  where $i:\mathscr D(\M,\xi,\psi,\gamma)_{T=0}\hookrightarrow\mathscr D(\M,\xi,\psi,\gamma):\nabla\mapsto\nabla$ is the natural injection and the map $j$ denotes the linear injection of the model space of $\mathscr D(\M,\xi,\psi,\gamma)_{T=0}$ into the one of $\mathscr D(\M,\xi,\psi,\gamma)$ and is defined as $j(F,\Sigma):=\hat\Psi\un(\Sigma,F,0)$ or explicitly
  \bea
  j&:&\field{\vee^2\Ann\xi}\oplus\field{\w^2\Ann\xi}\hookrightarrow\mathscr W(\M,\xi)\nn\\
  &:&(\Sigma_{\mu\nu},F_{\mu\nu})\mapsto(\Sigma_{\mu\nu}=\Sigma_{\mu\nu},\mathcal U_{\lambda|\mu\nu}=-\half\psi_\lambda F_{\mu\nu}).
  \eea

\vspace{2mm}

\noindent Proposition B.4 in \cite{Bekaert2014b} ensures that there is a canonical group action of the local Heisenberg group $\mathscr H_\inv\pl d\pr$ on the space $\mathscr R\pl\M,\xi,\psi,\gamma\pr:=\LFinv\times\mathscr W\pl \M,\xi\pr$ defined as
 \bea
\Bigg(
\Big(N,A\Big),\Big({\Sigma}_{\mu\nu}, {\mathcal U}_{\lambda|\mu\nu}\Big)\Bigg)
&\mapsto&\pl \Big(N',A'\Big),\overset{N',A'}{\Theta}\pl \NA\Theta{}\un\Big({\Sigma}_{\mu\nu}, {\mathcal U}_{\lambda|\mu\nu}\Big)\pr\pr\qquad\quad\nn
 \eea
for any element $\Big(\crl V\crr,\crl \alpha\crr,a\Big)\in\mathscr H_\inv\pl d\pr$ acting on $\LFinv$ as in \eqref{eqGaction}.
  
\bdefi{Torsional ambient gravitational fieldstrength}{An $\mathscr H_\inv\pl d\pr$-orbit in $\mathscr R\pl\M,\xi,\psi,\gamma\pr$ is dubbed a torsional ambient gravitational fieldstrength. The space of torsional ambient gravitational fieldstrengths will be denoted $\mathscr F\pl\M,\xi,\psi,\gamma\pr:=\mathscr R\pl\M,\xi,\psi,\gamma\pr/\mathscr H_\inv\pl d\pr$. }

\noindent Using this terminology, we can further characterise the affine space of ambient Galilean connections as follows:
\bprop{\label{corGalileantor}The space $\mathscr D\pl\M,\xi,\psi,\gamma\pr$ of ambient Galilean connections compatible with a given ambient Leibnizian structure possesses the structure of an affine space canonically isomorphic to the affine space $\mathscr F\pl\M,\xi,\psi,\gamma\pr$ of torsional ambient gravitational fieldstrengths. 
}
\noindent In other words, the affine space $\mathscr F\pl\M,\xi,\psi,\gamma\pr$ provides a classification of ambient Galilean connections. 
\noindent We now focus on the class of torsional Galilean manifolds admitting a well-defined projection of the Platonic screen. We let $\mathscr L\pl\M,\xi,\psi,\gamma\pr$ be a projectable ambient Leibnizian structure and denote $\bar{\mathscr L}\pl\bar\M,\bar\psi,\bar\gamma\pr$ the Leibnizian structure induced on $\bar\M$. We let $\nabla\in\mathscr D_\inv\pl\M,\xi,\psi,\gamma\pr$ be an invariant torsional ambient Galilean connection compatible with $\mathscr L\pl\M,\xi,\psi,\gamma\pr$ and $\pl N,A\pr\in\LFinv$ be an invariant Leibnizian pair.  We start the discussion by generalising \Prop{propprojtorfree} to the torsional case:
\bprop{\label{propinvariance}Let $\mathscr L\pl\M,\xi,\psi,\gamma\pr$ be a projectable ambient Leibnizian structure and $\nabla\in\mathscr D\pl\M,\xi,\psi,\gamma\pr$ be an ambient Galilean connection compatible with $\mathscr L\pl\M,\xi,\psi,\gamma\pr$. Let furthermore $\pl N,A\pr$ be an invariant Leibnizian pair and denote {\rm\cite{footnote22}} $\Big(\ovA{\Sigma},\N {\mathcal U}\Big):=\NA{\Theta}\pl\nabla\pr$. Then $\nabla$ is invariant if and only if 
\begin{itemize}
\item $\ovA\Sigma\pl \xi,X\pr=0$ for all $X\in\vf$
\item $\N {\mathcal U}\pl X|\xi,Y\pr=0$ for all $X,Y\in\vf$
\item $\Lag_\xi\, \ovA\Sigma=0$ and $\Lag_\xi\, \N {\mathcal U}=0$. 
\end{itemize}
}

\noindent In other words, the connection $\nabla$ is invariant if and only if $\ovA\Sigma\in\fieldinv{\otimes^2\Ann\xi}$ and $\N{\mathcal U}\in\fieldinv{\Ann\xi\otimes\w^2\Ann\xi}$. Note that the isomorphism $\fieldinv{\Ann\xi\otimes\w^2\Ann\xi}\cong\field{T^*\bar\M\otimes\w^2T^*\bar\M}$ ensures that $\N{\mathcal U}$ induces a well-defined $\bN{\mathcal U}$ on the Platonic screen. 
\noindent Accordingly, we will refer to an $\mathscr H_\inv\pl d\pr$-orbit in 
\bea
\mathscr R_\inv\pl\M,\xi,\psi,\gamma\pr:=\LFinv\times\fieldinv{\otimes^2\Ann\xi}\oplus\fieldinv{\Ann\xi\otimes\w^2\Ann\xi}\nn
\eea
as an invariant torsional ambient gravitational fieldstrength. The space of invariant torsional ambient gravitational fieldstrengths will be denoted \bea\mathscr F_\inv\pl\M,\xi,\psi,\gamma\pr:=\mathscr R_\inv\pl\M,\xi,\psi,\gamma\pr/\mathscr H_\inv\pl d\pr.\label{eqinvclassF}\eea

\noindent The next Corollary follows straightforwardly from Proposition \ref{propbijaug} using the classification of (ambient) Galilean connections in terms of (ambient) gravitational fieldstrengths (\cf \Prop{corGalileantor} above and Proposition 4.10 in \cite{Bekaert2014b}):
\bccor{Let $\mathscr L\pl\M,\xi,\psi,\gamma\pr$ be a projectable ambient Leibnizian structure and denote $\bar{\mathscr L}\pl\bar\M,\bar\psi,\bar\gamma\pr$ the induced Leibnizian structure on the Platonic screen $\bar\M$. Let $\mathscr D_\inv\pl\M,\xi,\psi,\gamma\pr$ and $\bar{\mathscr D}\pl\bar\M,\bar\psi,\bar\gamma\pr$ denote the affine spaces of (ambient invariant) Galilean connections compatible with ${\mathscr L}$ and $\bar{\mathscr L}$, respectively. There is a surjective affine map $\pi:\mathscr D_\inv\pl\M,\xi,\psi,\gamma\pr\twoheadrightarrow\bar{\mathscr D}\pl\bar\M,\bar\psi,\bar\gamma\pr$. 
}
\noindent Similarly to the torsionfree case, the {surjective affine map} is explicitly given by Diagram \eqref{diagrammap} using the surjectivity of the map:
\bea
\mathscr F_\inv\pl\M,\xi,\psi,\gamma\pr\twoheadrightarrow\bar{\mathscr F}\pl\bar\M,\bar\psi,\bar\gamma\pr\,:\,
\Big[ N,A,\ovA\Sigma,\N {\mathcal U}\Big]\mapsto\Big[ \bar N,\bN {\mathcal U}\Big]\label{eqinvclassF2}
\eea
where $\bar N:= \pi_*N$ and $\N {\mathcal U}=: \pi^*\, \bN {\mathcal U}$. Introducing the tensors $\bN {U}\in\field{\w^2 T^*\bar\M\otimes \Ker\bar\psi}$ and $\bN F\in\Om^2\pl\bar\M\pr$ as $\bN U{}^{\bar\lambda}_{\bar\mu\bar\nu}:= \bar h^{\bar\lambda\bar \rho}\, \bN {\mathcal U}_{\bar\lambda|\bar\mu\bar\nu}$ and $\bN F_{\bar\mu\bar\nu}:=-2\, \bar N^{\bar\lambda}\, \bN {\mathcal U}_{\bar\lambda|\bar\mu\bar\nu}$
allows to put the induced gravitational fieldstrength in the familiar form 
$\Big[ \bar N,\bN {\mathcal U}\Big]=\Big[ \bar N,\bN F,\bN U\Big]$. 

\noindent 
The previous map allows to define the affine surjective map $\pi:\mathscr D_\inv\pl\M,\xi,\psi,\gamma\pr\twoheadrightarrow\bar{\mathscr D}\pl\bar\M,\bar\psi,\bar\gamma\pr$ by making Diagram \eqref{diagrammap} commute. Again, given an invariant ambient Galilean connection $\nabla\in\mathscr D_\inv\pl\M,\xi,\psi,\gamma\pr$, the Galilean connection $\bar\nabla\in\bar{\mathscr D}\pl\bar\M,\bar\psi,\bar\gamma\pr$ defined on the Platonic screen via Diagram \eqref{diagrammap} can be checked to coincide with the Koszul connection obtained by projection of $\nabla$ using Diagram \eqref{diagdefibarnabla}. 
\noindent Componentwise, the projection of \eqref{eqambGalconnectioncomp} takes the form \cite{Bekaert2014b}:
\bea
\bar\Gamma^{\bar\lambda}_{\bar\mu\bar\nu}&=&\bar N^{\bar\lambda}\p_{(\bar\mu}\bar\psi_{\bar\nu)}+\half\, \bar h{}^{\bar\lambda\bar\rho}\Big[ \p_{\bar\mu} \bN{\gamma}_{\bar\rho\bar\nu}+\p_{\bar\nu} \bN{\gamma}_{\bar\rho\bar\mu}-\p_{\bar\rho} \bN{\gamma}_{\bar\mu\bar\nu}\Big]\nn\\&&+\, {\bar h}{}^{\bar\lambda\bar\rho}\bar\psi_{(\bar\mu}\bN{F}_{\bar\nu)\bar\rho}+\bar h{}^{\bar\lambda\bar\rho}\Big[\bN{{ U}}_{\bar\mu|\bar\rho\bar\nu}+\bN{{ U}}_{\bar\nu|\bar\rho\bar\mu}+\bN{{ U}}_{\bar\rho|\bar\mu\bar\nu}\Big]\nn
\eea 
where $\bN{{ U}}_{\bar\lambda|\bar\mu\bar\nu}:= \bN\gamma_{\bar\lambda\bar\rho}\, \bN U{}^{\bar\rho}_{\bar\mu\bar\nu}$.

\noindent A dual statement can be formulated in order to account for the embedding of the most general class of Carrollian manifolds inside ambient Galilean manifolds. The next Corollary provides such an embedding by making use of the classification of Carrollian manifolds introduced in Proposition \ref{corCarrolliantor}:
\bcor{}{Let $\mathscr L\pl\M,\xi,\psi,\gamma\pr$ be an ambient Aristotelian structure and denote $i:\tilde \M\longhookrightarrow \M$ one of the leaves foliating $\M$ and $\tilde{\mathscr C}\pl \tilde\M,\tilde\xi,\tilde\gamma\pr$ the induced Carrollian structure. Let $\mathscr D\pl\M,\xi,\psi,\gamma\pr$ (resp. $\tilde{\mathscr D}\pl\tilde\M,\tilde\xi,\tilde\gamma\pr$) denote the affine spaces of ambient Galilean connections (resp. Carrollian connections) compatible with ${\mathscr L}$ (resp. $\tilde{\mathscr C}$). There is an injective affine map $i:\tilde{\mathscr D}\pl\tilde\M,\tilde\xi,\tilde\gamma\pr\hookrightarrow\mathscr D\pl\M,\xi,\psi,\gamma\pr$. 
}
\noindent The map $i$
can be explicitly constructed by making Diagram \eqref{diagrammap2} commute using the following injective map:
\bea
\tilde{\mathscr F}\pl\tilde\M,\tilde\xi,\tilde\gamma\pr\hookrightarrow\mathscr F\pl\M,\xi,\psi,\gamma\pr\,:\,
\Big[ \tilde A,\overset{\tA}{\Sigma}, \overset{\tA}{U}\Big]\mapsto\crl N,A,\ovA\Sigma, \NA {\mathcal U}\crr\nn
\eea
where $A\in\EC$ is an Ehresmann connection on $\M$ satisfying $i^*A=\tilde A$, $\ovA\Sigma\in\field{\vee^2\Ann\xi}$ satisfies $i^*\ovA\Sigma=\overset{\tA}{\Sigma}$ and $\NA{\mathcal U}\in\mathscr U\pl \M,\xi\pr$ satisfies $i^*\NA{\mathcal U}=\overset{\tA}{U}$. 

\noindent This map in turn induces an injective map between the affine spaces of ambient Galilean connections and Carrollian connections by making Diagram \eqref{diagrammap2} commute. As in the Galilean case, the present setup allows to embed the whole class of Carrollian manifolds into ambient Galilean manifolds. Furthermore, the embedding procedure described here matches the one prescribed by Diagram \eqref{diagcarr}. The component expression of the Carrollian connection induced by \eqref{eqambGalconnectioncomp} on the wavefront worldvolume $\tilde\M$ reads:
\bea
\tilde\Gamma^{\tilde\lambda}_{\tilde\mu\tilde\nu}&=&\tilde \xi^{\tilde\lambda}\p_{(\tilde\mu}\tilde A_{\tilde\nu)}+\half\, \overset{\tilde A}{h}{}^{\tilde\lambda\tilde\rho}\crl \p_{\tilde\mu} \tilde{\gamma}_{\tilde\rho\tilde\nu}+\p_{\tilde\nu} \tilde{\gamma}_{\tilde\rho\tilde\mu}-\p_{\tilde\rho} \tilde{\gamma}_{\tilde\mu\tilde\nu}\crr-\half\, \overset{\tilde A}{h}{}^{\tilde\lambda\tilde\rho}\tilde A_{\tilde\nu}\Lag_{\tilde\xi}\tilde\gamma_{\tilde\mu\tilde\rho}\nn\\
&&+\tilde\xi^{\tilde\lambda} \overset{\tilde A}{\Sigma}_{\tilde\mu\tilde\nu}+\overset{\tilde A}{h}{}^{\tilde\lambda\tilde\rho}\Big( \overset{\tilde A}{{ U}}_{\tilde\mu|\tilde\rho\tilde\nu}+ \overset{\tilde A}{{ U}}_{\tilde\nu|\tilde\rho\tilde\mu}+\overset{\tilde A}{{ U}}_{\tilde\rho|\tilde\mu\tilde\nu}\Big)\nn\,\nn 
\eea 
thus reproducing the most general torsionful Carrollian connection \eqref{Carrconntor2}.
\section{Conclusion}

\noindent We argued that the notion of ambient Leibnizian structures provides a non-Lorentzian ambient framework unifying both Galilean and Carrollian structures. This unification has been shown to work at different levels:
\bi
\item \textbf{Algebraic}: The Leibniz algebra $\leib(d+2)$ defined in \eqref{Killingalgebra} contains both the Bargmann and Carroll algebra as subalgebras. 
\item \textbf{Metric}: Ambient Leibnizian structures (\cf \Defi{defambientleib}) unify both nonrelativistic Leibnizian and ultrarelativistic Carrollian metric structures. 
\item  \textbf{Parallelism}: Ambient Galilean connections (\cf \Defi{defAmbient Galilean connections}) unify both Galilean and Carrollian connections, both in the torsionfree and torsional case \big(\cf explicit component formulas \eqref{TGconnect} and \eqref{eqambGalconnectioncomp}\big). 
\ei
In other words, the most general Galilean (resp. Carrollian) manifold can be obtained as projection (resp. embedded into)  an ambient Galilean manifold. 
\noindent In a forthcoming work \cite{Morand2017}, we will show that similar results hold for the subclass of ambient Galilean manifolds admitting a compatible Lorentzian metric (\ie Bargmann structures). 
\noindent This class of Lorentzian manifolds will be shown to possess the sufficient arbitrariness to embed the whole class of Galilean and Carrollian manifolds. In particular, 
{\it torsionfree} Galilean connections (with generic force field $\bar F$) will be shown to arise as projection of {\it torsional} (Lorentzian metric compatible) connections where the force field is inherited from the ``light-like part''  of the ambient torsion. 

\vspace{8mm}

\noindent \underline{Note added:} A first version of the present paper appeared on arXiv as the preprint \barxiv{1505.03739}. Shortly after, the paper \cite{Hartong:2015xda} appeared on arXiv as the preprint \barxiv{1505.05011}. While these works have been developed independently, there is a strong overlap between sections 2 and 3 of \cite{Hartong:2015xda} and our appendix A. 

\noindent Since the present paper is a substantially reorganized version of the first version of the preprint \barxiv{1505.03739}, let us mention that the present subsection A.3 was added in a later version, in order to mirror the presentation given in \cite{Bekaert2014b} of Newtonian connections, but some considerations (such as an analogue of the Carrollian potential and the invariants of Table 2) were already present in \cite{Hartong:2015xda}.

\section*{Acknowledgements}

\noindent We are grateful to Claude Barrab\`es for useful exchanges about null hypersurfaces. K.M. thanks the {\it Institut des Hautes \'Etudes Scientifiques} (IH\'ES, Bures-sur-Yvette) for hospitality where part of this work was completed. The work of K.M. is supported by the Chilean Fondecyt Postdoc Project $\text{N}^\circ$3160325.
\pagebreak
\appendix

\section{Through the Looking Glass: a compendium on Carrollian manifolds}
\label{sectionappendix}
\noindent The present Appendix can be seen as a mirror image of the work \cite{Bekaert2014b} where the principal definitions and results regarding intrinsic \nr Galilean geometry are dualised to the Carrollian case. 

\subsection{Carrollian structures}\label{Carrollian_structures}

\bdefi{Carrollian metric}{Let $\pl\M,\xi\pr$ be an ambient structure. A Carrollian metric on $\pl\M,\xi\pr$ is a positive semi-definite covariant metric $\gamma\in\bforms$ whose radical is spanned by the fundamental vector field $\xi$. Alternatively, a Carrollian metric can be defined as a field $h\in\bforma{\pl\Ann\xi\pr^*}$ on $\pl\M,\xi\pr$ of positive-definite contravariant symmetric bilinear forms acting on 1-forms annihilating the fundamental vector field $\xi$. 
}

\bdefi{Carrollian structure \cite{Duval2014e}}{\label{defiCarrollstructure}
A Carrollian structure consists of a triplet $\mathscr C\pl \M, \xi, \gamma\pr$ composed by the following elements: 
\begin{itemize}
\renewcommand{\labelitemi}{$\bullet$} 
\item an ambient structure $\pl\M,\xi\pr$
\item a Carrollian metric $\gamma$. 
\end{itemize}
}
\noindent A Carrollian structure such that $\Lag_\xi \gamma=0$ will be said {\it invariant}. 
\vspace{2mm}

\noindent Let $\mathscr C\pl \M, \xi, \gamma\pr$ be a Carrollian structure. The space of Ehresmann connections $\EC$ (\cf \Defi{defiEhresmannconnection}) possesses the structure of an affine space modelled on $\fAnnxi$. The action of $\fAnnxi$ on $\EC$ as $A\mapsto A+\alpha$ with $A\in\EC$ will be referred to as a {\it Carroll boost} parameterised by the 1-form $\alpha\in\fAnnxi$. 

\bexa{Carroll spacetime}{\label{exaCarroll}The most simple example of a Carrollian structure is given by the Carroll spacetime {$\M\cong{\mathbb R}^{d+1}$ with coordinates $\pl u,x^i\pr$ characterised by the following fundamental vector field and (flat)} Carrollian metric:
\bcase{
\xi=\frac{\partial}{\partial u}\\
\gamma=\delta_{ij}\,dx^i\vee dx^j\nn
}
where $i,j\in\pset{1,\dots,d}$ and $\delta_{ij}$ is the Kronecker delta. 
{Equivalently, one may consider the following contravariant metric: $h=\delta^{ij}\,\frac{\partial}{\partial x^i}\vee\frac{\partial}{\partial x^j}$.} 
}

\bdefi{Transverse cometric}{\label{defitransversemetric}Let $\mathscr C\pl \M, \xi, \gamma\pr$ be a Carrollian structure and  $A\in\EC$ an Ehresmann connection on $\M$. The transverse cometric $\ovA{h}\in\bform$ is defined by its action on 1-forms $\alpha,\beta\in\ff$ as
\bea
\ovA{h}\pl \alpha,\beta\pr=h\pl \ovbA{P}{}\pl \alpha\pr, \ovbA{P}{}\pl \beta\pr\pr. \label{eqtransversemetric}
\eea
}
\noindent The right-hand side of eq.\eqref{eqtransversemetric} is well-defined since the image of the projector $\ovbA{P}{}$ lies in $\fAnnxi$. The transverse cometric $\ovA{h}$ can be easily shown to satisfy the two relations:
\bea
\begin{cases}
\overset{A}{h}{}^{\mu\nu}\, A_\nu=0\\ 
\overset{A}{h}{}^{\mu\lambda}\, \gamma_{\lambda\nu}=\delta^\mu_\nu-\xi^\mu A_\nu\label{eqstransversemetriccond}. 
\end{cases}
\eea
In fact, given $A\in\EC$, there is a unique contravariant metric $\ovA{h}\in\bform$ satisfying the conditions \eqref{eqstransversemetriccond}. 
\bprop{\label{propCarrollvsLeibniz}Let ${\mathscr C}\pl \M, \xi, \gamma\pr$ be a Carrollian structure and  $A\in\EC$ an Ehresmann connection on $\M$. These data define a Leibnizian structure ${\mathscr L}\big(\M,A,\overset{A}{h}\big)$ whose absolute clock is the Ehresmann connection $A$ and whose collection of absolute rulers is the transverse cometric $\ovA{h}$.}
\noindent Dually:
\bprop{\label{propLeibnizvsCarroll}Let ${\mathscr L}\pl \M, \psi, h\pr$ be a Leibnizian structure and  $N\in\FO$ a field of observers on $\M$. These data define a Carrollian structure ${\mathscr C}\big(\M,N,\overset{N}{\gamma}\big)$ whose fundamental vector field is the field of observers $N$ and whose Carrollian metric is the transverse metric $\overset{N}{\gamma}$.}

\bdefi{Carrollian basis}{\label{defiCarrollianbasis}Let $\mathscr C\pl \M, \xi, \gamma\pr$ be a Carrollian structure. A Carrollian basis of the dual tangent space $T^*_x\M$ at a point $x\in\M$ is an ordered basis $C^*_x=\lbrace A_x, \theta^{1}|_{x}, \ldots, \theta^{d}|_{x}\rbrace$ with $A_x$ satisfying $A_x\pl \xi_x\pr=1$ and $\lbrace \theta^{1}|_{x}, \ldots, \theta^{d}|_{x}\rbrace$ a basis of $\Ann\xi_x$ which is orthonormal with respect to $h_x$. }
\noindent Explicitly, the basis $C^*_x=\lbrace A_x, \theta^{1}|_{x}, \ldots, \theta^{d}|_{x}\rbrace$ must satisfy the conditions:
\vspace{2mm}
\begin{enumerate}
\item $A_x\pl \xi_x\pr=1$
\item$\theta^{i}|_{x}\pl\xi_x\pr=0\,  ,\forall\, i\in \lbrace1, \ldots, d\rbrace$
\item$h_x\pl \theta^{i}|_{x}, \theta^{j}|_{x}\pr=\delta_{ij}\, ,\forall\, i,j\in\pset{ 1, \ldots, d}$. 
\end{enumerate}
\vspace{2mm}
\noindent A basis of $T_x\M$ dual to $C^*_x=\lbrace A_x, \theta^{i}|_{x}\rbrace$ is given by $C_x=\lbrace \xi_x, e_{i}|_{x}\rbrace$, where the $d$ vectors $e_i|_x$ satisfy the requirement: $\theta^i|_x\pl e_{j}|_{x}\pr=\delta^i_j$. {For instance, in the above case of the Carroll spacetime (Example \ref{exaCarroll}): 
$A=du$ and $\theta^i=dx^i$ define a Carrollian basis at any point.}
\bprop{\label{propCarrollianbasis}At each point $x\in\M$, the set of automorphisms of $T^*_x\M$ mapping each Carrollian basis into another one forms a group isomorphic to the homogeneous Carroll group.}
\noindent Explicitly, the automorphisms preserving the collection of Carrollian bases can be represented as the following matrices
\bea
T=
\begin{pmatrix}
1 & \bb^T \\
0& \bR^T\label{matrixCarroll}
\end{pmatrix}
\eea
with $\bb\in\mR^d$ and $\bR\in O\pl d\pr$. This set of matrices forms a subgroup of $GL\pl d+1,\mR\pr$ isomorphic to the homogeneous Carroll group $\CARRZ\pl d+1\pr$. The homogeneous Carroll group therefore acts regularly on the space of Carrollian bases via the group action:
\bea
\pset{A, \theta^i}\mapsto\pset{A+\bb^T_i\theta^i,\bR^T{}^i_j\theta^j}. \label{eqgalileiauto}
\eea
\noindent Given $C^*=\pset{A,\theta^i}$, one can define a dual basis $C=\pset{\xi,e_i}$, with $\theta^i\pl e_j\pr=\delta^i_j$ on which the homogeneous Carroll group acts transitively as:
\bea
\pset{\xi,e_i}\mapsto\pset{\xi, \bR^j_i\pl e_j-\bb^T_j\xi\pr}. 
\eea

\bdefi{Carrollian manifold}{A Carrollian structure supplemented with a Koszul connection compatible with the fundamental vector field and Carrollian metric is called a Carrollian manifold. The Koszul connection is then referred to as a Carrollian connection. 
}

\noindent If we let $\mathscr C\pl \M, \xi, \gamma\pr$ be a Carrollian structure, the compatibility conditions explicitly read: 
\begin{enumerate}
\item $\nabla\xi=0$
\item $\nabla \gamma=0$. 
\end{enumerate}
\bprop{\label{propCarrolliancompatibility}Let $\mathscr C\pl \M, \xi, \gamma, \nabla\pr$ be a Carrollian manifold and denote $T\in\field{\w^2T^*\M\otimes T\M}$ the torsion tensor associated with the Carrollian connection $\nabla$. The following relation holds:
\bea
\Lag_\xi\gamma\pl X,Y\pr=\gamma\pl X,T\pl \xi,Y\pr\pr+\gamma\pl Y,T\pl \xi,X\pr\pr\text{ for all }X,Y\in\vf. 
\label{eqtorsionconstraint}
\eea
}
In particular, one can conclude from the previous relation that only invariant Carrollian structures admit torsionfree compatible connections. This is just a particular instance of the standard requirement that the radical of a degenerate metric preserved by a torsionfree connection must be a Killing distribution (\cf \eg Theorem 5.1 in \cite{Duggal1996}). Table \ref{TableCarroll} sums up the previously discussed Carrollian structures by mirroring them with their Galilei duals.

\begin{table}
\begin{center}
\begin{tabular}{|c|c|c|c|}
\hline
\multicolumn{2}{|c|}{\textbf{Galilei}}&\multicolumn{2}{c|}{\textbf{Carroll}}\\
\hline
\multicolumn{2}{|c|}{Absolute clock $\psi\in\ff$}&\multicolumn{2}{c|}{Fundamental vector field $\xi\in\vf$}\\
\hline
\begin{tabular}{c}
Collection of\\
absolute rulers
 \end{tabular}
&\begin{tabular}{c}
$h\in\field{\vee^2T\M}$\\
$\Rad h\cong\Span \psi$\\
\hline
$\gamma\in\bforma{\pl\Ker\psi\pr^*}$
 \end{tabular}&Carrollian metric&
 \begin{tabular}{c}
$\gamma\in\field{\vee^2T^*\M}$\\
$\Rad \gamma\cong\Span \xi$\\
\hline
$h\in\bforma{\pl\Ann\xi\pr^*}$
 \end{tabular}
\\
\hline
\begin{tabular}{c}
Field of\\
observers
 \end{tabular}
&$N\in\FO$&\begin{tabular}{c}
Ehresmann\\
connection
 \end{tabular}&$A\in\EC$
\\
\hline
\begin{tabular}{c}
Milne boosts
 \end{tabular}
&
\begin{tabular}{c}
$N\mapsto N+V$\\
with $V\in\Milneg$
 \end{tabular}
&\begin{tabular}{c}
Carroll boosts
 \end{tabular}&
 \begin{tabular}{c}
$A\mapsto A+\alpha$\\
with $\alpha\in\fAnnxi$
 \end{tabular}
\\
\hline
\multicolumn{2}{|c|}{\begin{tabular}{c}
Transverse metric\\
\hline
$\N{\gamma}\in\field{\vee^2T^*\M}$\\
$\N{\gamma}\pl X,Y\pr:=\gamma\pl \overset{N}{P}\pl X\pr,\overset{N}{P}\pl Y\pr\pr$\\
for all $X,Y\in\vf$
 \end{tabular}}
&\multicolumn{2}{c|}{
\begin{tabular}{c}
Transverse cometric\\
\hline
$\ovA{h}\in\field{\vee^2T\M}$\\
$\ovA{h}\pl \alpha,\beta\pr:= h\pl \ovbA{P}{}\pl \alpha\pr, \ovbA{P}{}\pl \beta\pr\pr$\\
for all $\alpha,\beta\in\ff$
 \end{tabular}}\\
 \hline
\multicolumn{4}{|c|}{Torsion constraint}\\
\multicolumn{2}{|c}{
\begin{tabular}{c}
$\psi\pl T\pl X,Y\pr\pr=d\psi\pl X,Y\pr$\\
 for all $X,Y\in \field{T\M}$
 \end{tabular}
}
&
\multicolumn{2}{c|}{
\begin{tabular}{c}
$\Lag_\xi\gamma\pl X,Y\pr=\gamma\pl X,T\pl \xi,Y\pr\pr+\gamma\pl Y,T\pl \xi,X\pr\pr$\\
  for all $X,Y\in\vf$
 \end{tabular}
}\\
\hline
\end{tabular}
\caption{Duality between Galilei and Carroll structures\label{TableCarroll}}
\end{center}
\end{table}

\pagebreak
\subsection{Torsionfree Carrollian connections}
\label{Torsionfree Carrollian connections}

\noindent Let $\mathscr C\pl \M, \xi, \gamma\pr$ be an invariant Carrollian structure. The space of torsionfree connections compatible with $\mathscr C\pl \M, \xi, \gamma\pr$ will be denoted $\mathscr D\pl \M, \xi, \gamma\pr$. 

\bprop{\label{PropCarrollianaffinespace}The space $\mathscr D\pl\M,\xi,\gamma\pr$ of torsionfree Carrollian connections possesses the structure of an affine space modelled on the vector space
\bea\mathscr V\pl\M,\xi,\gamma\pr:=\Bigg\{S\in\field{\vee^2T^*\M\otimes T\M} \text{ satisfying conditions a)-b) }\Bigg\}\nn\eea
where:
\begin{enumerate}[a) ]
\item $S\pl X,\xi\pr=0$ for all $X\in\vf$
 \item $\gamma\pl S\pl X,Y\pr,Z\pr+\gamma\pl S\pl X,Z\pr,Y\pr=0$ for all $X,Y,Z\in\vf$\,. 
\end{enumerate}
}
\noindent Note that condition {\it b)} as well as the symmetry of $S$ ensures that $\gamma\pl S\pl X,Y\pr,Z\pr=0$ for all $X,Y,Z\in\vf$. In other words, $S\pl X,Y\pr\in\Span\xi$ for all $X,Y\in\vf$. 
\blem{}{\label{Lemcanisocarr}The vector space $\mathscr V\pl\M,\xi,\gamma\pr$ is isomorphic to the vector space $\field{\vee^2\Ann\xi}$. 
}
\noindent Explicitly, given an Ehresmann connection $A\in\EC$, one can construct the following canonical isomorphism:
\bea
{\varphi}\,:\,\mathscr V\pl\M,\xi,\gamma\pr\to\field{\vee^2\Ann\xi}\,:\,S\lmn\mapsto {\Sigma}_{\mu\nu}=A_\lambda S\lmn
\eea
whose inverse takes the form
\bea
{\varphi}{}\un\,:\,\field{\vee^2\Ann\xi}\to \mathscr V\pl\M,\xi,\gamma\pr\,:\,\Sigma_{\mu\nu}\mapsto S\lmn=\xi^\lambda {\Sigma}_{\mu\nu}\,.
\eea

\noindent We now construct the following map:
\bea
{\Theta}\,:\,\PC\times\mathscr D\pl\M,\xi,\gamma\pr\to\field{\vee^2\Ann\xi}\,:\,\pl A,\Gamma\pr\mapsto\ovA{\Sigma}_{\mu\nu}=-\nabla_{(\mu}A_{\nu)}\nn. 
\eea
For all principal connection $A\in\PC$, the map $\ovA\Theta:= \Theta\pl A,\cdot\pr:\mathscr D\pl\M,\xi,\gamma\pr\to\field{\vee^2\Ann\xi}$ can be shown to be an affine map modelled on the linear map ${\varphi}$ \ie $\ovA\Theta\pl\Gamma'\pr-\ovA\Theta\pl\Gamma\pr=\varphi\pl \Gamma'-\Gamma\pr$. The superscript acts again as a reminder of the fact that $\ovA{\Theta}$ is not canonical. Note that the fact that $A$ is a principal connection (so that $\Lag_\xi A=0$) ensures that $\ovA{\Sigma}\in\field{\vee^2\Ann\xi}$. 

\noindent The fact that $\ovA{\Theta}$ is an affine map modelled on the isomorphism of vector spaces $\varphi$ ensures that the kernel of $\ovA{\Theta}$ is spanned by a unique $\ovA{\Gamma}\in\mathscr D\pl\M,\xi,\gamma\pr$. 
We call $\ovA{\Gamma}$ the {\it torsionfree special Carrollian connection} associated with the principal connection $A\in\PC$ whose Koszul formula is given by:
\bea
A\pl \ovA\nabla_XY\pr&=&X\crl A\pl Y\pr\crr-\half dA\pl X,Y\pr\\
2\,\gamma\pl\ovA\nabla_XY,Z\pr&=&X\crl \gamma\pl Y,Z\pr\crr+Y\crl \gamma\pl X,Z\pr\crr-Z\crl \gamma\pl X,Y\pr\crr\label{KformulaCarrollsp}\\
&&+\gamma\pl\br{X}{Y},Z\pr-\gamma\pl\br{Y}{Z},X\pr-\gamma\pl\br{X}{Z},Y\pr\nn
\eea
for all $X,Y,Z\in\vf$. In components, the coefficients of the torsionfree special Carrollian connection associated with $A$ read explicitly:
\bea
\ovA{\Gamma}\lmn=\xi^\lambda\p_{(\mu}A_{\nu)}+\half \ovA{h}{}^{\lambda\rho}\crl \p_\mu \gamma_{\rho\nu}+\p_\nu \gamma_{\rho\mu}-\p_\rho \gamma_{\mu\nu}\crr. \label{torsionlessspecialcarroll}
\eea

\noindent Given a principal connection $A$, the associated torsionfree special Carrollian connection can thus be used in order to represent any torsionfree Carrollian connection ${\Gamma}\in\mathscr D\pl\M,\xi,\gamma\pr$ as $\Gamma=\ovA{\Gamma}+\varphi{}\un\Big( \ovA{\Theta}\pl \Gamma\pr\Big)$. The Koszul formula for a generic torsionfree Carrollian connection then reads as:
\bea
A\pl \nabla_XY\pr&=&X\crl A\pl Y\pr\crr-\half dA\pl X,Y\pr+\ovA{\Sigma}\pl X,Y\pr\\
2\,\gamma\pl\nabla_XY,Z\pr&=&X\crl \gamma\pl Y,Z\pr\crr+Y\crl \gamma\pl X,Z\pr\crr-Z\crl \gamma\pl X,Y\pr\crr\label{KformulaCarroll}\\
&&+\gamma\pl\br{X}{Y},Z\pr-\gamma\pl\br{Y}{Z},X\pr-\gamma\pl\br{X}{Z},Y\pr\nn
\eea
for all $X,Y,Z\in\vf$. The components of $\Gamma$ take the form \cite{footnote23}:
\bea
\Gamma\lmn=\xi^\lambda\p_{(\mu}A_{\nu)}+\half \ovA{h}{}^{\lambda\rho}\crl \p_\mu \gamma_{\rho\nu}+\p_\nu \gamma_{\rho\mu}-\p_\rho \gamma_{\mu\nu}\crr+\xi^\lambda \ovA{\Sigma}_{\mu\nu}. \nn
\eea
\noindent This is the most general expression of a torsionfree Carrollian connection compatible with the Carrollian structure $\mathscr C\pl \M, \xi, \gamma\pr$, the arbitrariness being encoded in the twice-covariant symmetric tensor $\ovA{\Sigma}\in\field{\vee^2\Ann\xi}$ (\cf \cite{Bergshoeff2014b} for a conjecture of this fact in the context of gauging procedures). 

\noindent Under a Carroll boost parameterised by the invariant 1-form $\alpha\in\fieldinv{\Ann\xi}$, the map $\ovA{\Theta}:\mathscr D\pl\M,\xi,\gamma\pr\to\field{\vee^2\Ann\xi}$ varies according to
\bea
\ovAp{\Theta}\pl\Gamma\pr=\ovA{\Theta}\pl\Gamma\pr-\half\Lag_{\alpha^\#{}}\gamma\nn
\eea
for all $\Gamma\in\mathscr D\pl\M,\xi,\gamma\pr$, where $\alpha^\#{}:= \ovA{h}\pl\alpha\pr\in\fieldinv{T\M}$. This expression can then be used to define a $\fieldinv{\Ann\xi}$-action on the space $\PC\times\field{\vee^2\Ann\xi}$ as:
\bea
&&\fieldinv{\Ann\xi}\times\Big(\PC\times\field{\vee^2\Ann\xi}\Big)\to\PC\times\field{\vee^2\Ann\xi}\nn\\
&&\qquad \Big(\alpha,\pl A,\Sigma\pr\Big)\mapsto\pl A'=A+\alpha,\Sigma'=\Sigma-\half\Lag_{\alpha^\#{}}\gamma\pr\,.\label{groupactioncarr} 
\eea
\noindent We will denote $\tilde {\mathscr F}(\tilde\M,\tilde\xi,\tilde\gamma)$ the affine space of orbits $\tilde {\mathscr F}(\tilde\M,\tilde\xi,\tilde\gamma):=\frac{\PC\times\field{\vee^2\Ann\xi}}{\fieldinv{\Ann\xi}}$.
\bprop{\label{corCarrollian}The space of torsionfree Carrollian connections compatible with a given invariant Carrollian structure possesses the structure of an affine space canonically isomorphic to the affine space of orbits: 
$$\mathscr D\pl \M, \xi, \gamma\pr\cong\tilde {\mathscr F}(\tilde\M,\tilde\xi,\tilde\gamma)\,.$$ 
}
\subsection{Newton-Carroll connection}
\noindent We conclude the investigation of torsionfree Carrollian connections by discussing a subset of the latter class, dubbed {\it Newton-Carroll connections}, in analogy with the subclass of torsionfree Newtonian connections in the Galilean case. The present discussion will thus mirror the one of Sections 3.3-3.4 in \cite{Bekaert2014b}. 

\noindent We start by defining the notion of Carrollian potential, as follows:
\bdefi{Carrollian potential}{Let $\mathscr C\pl \M, \xi, \gamma\pr$ be a Carrollian structure. A Carrollian potential is a map
\bea
N:\EC&\to&\vf\nn\\
A\mapsto \ovA N
\eea
such that for all $A',A\in\EC$, $A'=A+\alpha$ with $\alpha\in\field{\Ann\xi}$, the following relation holds:
\bea 
\overset{A'}{N}=\ovA{N}+\ovA h(\alpha)-\half\, \ovA h(\alpha,\alpha)\, \xi. 
\eea
}
\noindent The notion of Carrollian potential allows to articulate the definition of a Newton-Carroll connection as follows:
\bdefi{Newton-Carroll connection}{Let $\mathscr C\pl \M, \xi, \gamma\pr$ be an invariant Carrollian structure. A Newton-Carroll connection compatible with $\mathscr C\pl \M, \xi, \gamma\pr$ is a torsionfree compatible connection such that there exists a Carrollian potential $N$ such that $\ovA\Sigma=-\half \Lag_{\ovA N}\gamma$. }
\noindent Explicitly, the coefficients associated with a Newton-Carroll connection takes the form:
\bea
\Gamma\lmn=\xi^\lambda\p_{(\mu}A_{\nu)}+\half \ovA{h}{}^{\lambda\rho}\crl \p_\mu \gamma_{\rho\nu}+\p_\nu \gamma_{\rho\mu}-\p_\rho \gamma_{\mu\nu}\crr-\half\xi^\lambda \Lag_{\ovA N}\gamma_{\mu\nu}\label{eqcoeffcarrnewt}
\eea
where $A\in\PC$. 

\noindent Note that the transformation property of the vector field $\ovA N$ under a Carroll boost ensures that the above expression is Carroll boost-invariant. Furthermore, the following transformation: 
\bea
\ovA N\overset{f}{\mapsto} \ovA N+ f\, \xi, \text{ with }f\in\fonc{\tilde\M}
\eea
referred to as {\it Maxwell-gauge transformation}, also preserves expression \eqref{eqcoeffcarrnewt}. 

\noindent The Carroll-bost invariance of the Newton-Carroll connection \eqref{eqcoeffcarrnewt} can be made more explicit by 
defining the following Carroll boost-invariant quantities: 
\vspace{2mm}
\begin{center}
\begin{tabular}{|M{2.9cm}
|M{4.2cm}
|M{4.2cm}|}
   \hline
\rule[-0.5cm]{0cm}{1cm}Type&
\rule[-0.5cm]{0cm}{1cm}Definition&
\rule[-0.5cm]{0cm}{1cm}Gauge transformation law
\tabularnewline\hline
\rule[-0.5cm]{0cm}{1cm}$E\in \EC$&
\rule[-0.5cm]{0cm}{1cm}$E:=A-\gamma( \overset{A}{N})$&
\rule[-0.5cm]{0cm}{1cm}$E\mapsto E$
\tabularnewline
   \hline 
\rule[-0.5cm]{0cm}{1cm}$\lambda\in \fonc{\M}$&
\rule[-0.5cm]{0cm}{1cm}$\lambda:= 2\, A(\overset{A}{N})-\gamma( \overset{A}{N},\overset{A}{N})$&
\rule[-0.5cm]{0cm}{1cm}$\lambda\mapsto \lambda +2\, f$
\tabularnewline\hline
\rule[-0.5cm]{0cm}{1cm}$g\in\field{\vee^2T\M}$&
\rule[-0.5cm]{0cm}{1cm}$g:= \overset{A}{h}+2\, \xi\, \vee\, \overset{A}{N}$&
\rule[-0.5cm]{0cm}{1cm}$g\mapsto g +2\, f\, \xi\otimes \xi$
\tabularnewline\hline
\end{tabular}
\end{center}
\captionof{table}{Carroll boost-invariant objects\label{tableCarrollinvariant}}
\vspace{4mm}
Note that the objects of Table \ref{tableCarrollinvariant} are related by the following (Maxwell gauge invariant) relation:
\bea
g^{\mu\nu}E_\nu=\lambda\,  \xi^\mu. 
\eea

These Carroll-boost invariant objects can thus be used in order to reexpress the coefficients \eqref{eqcoeffcarrnewt} in a manifestly Carroll-boost invariant way as follows:
\bea
\Gamma^\lambda_{\mu\nu}=\xi^\lambda\p_{(\mu}E_{\nu)}+\half g^{\lambda\rho}\crl\p_\mu \gamma_{\rho\nu}+\p_\nu \gamma_{\rho\mu}-\p_\rho \gamma_{\mu\nu}\crr. \label{eqcarrnewt2}
\eea
\bprop{The following affine spaces are isomorphic \textnormal{\cite{footnote24prime}}:	\label{proposition:Newton-Carroll}
\begin{enumerate}
\item The space of Newton-Carroll connections. 
\item The space of torsionfree special connections. 
\item The space of principal connections.
\end{enumerate}
}
Before addressing the issue of torsional Carrollian connections, 
we display two dual Propositions regarding the intersection between Newton-Carroll and Newton-Cartan manifolds which are the connection analogue of the Propositions \ref{propCarrollvsLeibniz} and \ref{propLeibnizvsCarroll} on metric structures (\cf \cite{Dereli:2004je} for similar statements): 
\bprop{Let $\C(\M,\xi,\gamma)$ be an invariant Carrollian structure endowed with the principal connection $A\in \PC$. We denote $\ovA\Gamma\in\mathscr D(\M,\xi,\gamma)$ the associated torsionfree special connection and $\ovA h\in\field{\vee^2\Ker A}$ the transverse cometric associated with $A$. 
\bi
\item The quadruplet $(\M,A,\ovA h, \ovA\Gamma)$ is a Newtonian manifold if and only if the principal connection is flat ($dA=0$). 
\item If the principal connection $A$ is flat, the torsionfree special connection $\ovA\Gamma$ is the Levi-Civita connection associated with the (non-canonical) nondegenerate metric $\ovA g_c:=\gamma+c\, A\otimes A$ with inverse $\ovA g{}\un_c:=\ovA h+c\un\, \xi\otimes\xi$ where $c\in\mathbb R\setminus\pset{0}$ is a non-vanishing constant. 
\ei
}
Dually:
\bprop{Let $\mathscr L(\M,\psi,h)$ be an Augustinian structure endowed with the field of observers $N\in \FO$. We denote $\N\Gamma\in\mathscr D(\M,\psi,h)$ the associated torsionfree special connection and $\N\gamma\in\field{\vee^2\Ann N}$ the transverse metric associated with $N$. 
\bi
\item The quadruplet $(\M,N,\N\gamma,\N\Gamma)$ is a Newton-Carroll manifold if and only if the field of observers $N$ is Killing for $\N\gamma$ (\ie $\Lag_N\N\gamma=0$). 
\item If $\Lag_N\N\gamma=0$, the torsionfree special connection $\N\Gamma$ is the Levi-Civita connection associated with the (non-canonical) nondegenerate metric $\N g_c:=\N\gamma+c\, \psi\otimes \psi$ with inverse $\N g{}\un_c:= h+c\un\, N\otimes N$ where $c\in\mathbb R\setminus\pset{0}$ is a non-vanishing constant. 
\ei
}

 \subsection{Torsional Carrollian connections}
 \label{Torsional Carrollian connections}
 \noindent We let $\mathscr C\pl \M, \xi, \gamma\pr$ be a Carrollian structure and denote $\mathscr D\pl\M,\xi,\gamma\pr$ the space of compatible Carrollian connections. 
  
 \bprop{\label{PropCarrollianaffinespacetor}The space $\mathscr D\pl\M,\xi,\gamma\pr$ of compatible Carrollian connections possesses the structure of an affine space modelled on the vector space $\mathscr V\pl\M,\xi,\gamma\pr$ defined as:
 \bea\mathscr V\pl\M,\xi,\gamma\pr:=\Bigg\{S\in\field{T^*\M\otimes T^*\M\otimes T\M} \text{ satisfying conditions a)-b) }\Bigg\}\nn\eea
 where
 \begin{enumerate}[a) ]
 \item $S\pl X,\xi\pr=0$ for all $X\in\vf$
 \item $\gamma\pl S\pl X,Y\pr,Z\pr+\gamma\pl S\pl X,Z\pr,Y\pr=0$ for all $X,Y,Z\in\vf$. 
 \end{enumerate}
 }
 \blem{}{\label{Lemcanisotorcar}The vector space $\mathscr V\pl\M,\xi,\gamma\pr$ is isomorphic to the vector space $\mathscr W\pl \M,\xi\pr:=\field{T^*\M\otimes\Ann\xi}\oplus\mathscr U\pl \M,\xi\pr$ where the vector space $\mathscr U\pl \M,\xi\pr$ is defined as
 \bea
\mathscr U\pl \M,\xi\pr:=\Bigg\{{ U}\in\field{\Ann\xi\otimes\w^2T^*\M}\text{ satisfying condition (*)}\Bigg\}\nn
\eea
where condition $\textit{(*)}$ reads:
\bea\textit{(*):  } { U}\pl X|\xi,Y\pr+{ U}\pl Y|\xi,X\pr=0 \text{ for all }X,Y\in\vf. \label{conditionast2}\eea
} 

 \noindent Explicitly, given an Ehresmann connection $A\in\EC$, one can construct the following non-canonical isomorphism \cite{footnote24}:
 \bea
 \ovA{\varphi}&:&\mathscr V\pl\M,\xi,\gamma\pr\to  \mathscr W\pl \M,\xi\pr\nn\\
 &:&S\lmn\mapsto \pl {\ovA\Sigma}_{\mu\nu}=A_\lambda\, S\lmn, U_{\lambda|\mu\nu}=\gamma_{\lambda\rho}\, S^\rho_{[\mu\nu]}\pr\nn
 \eea
 whose inverse takes the form
 \bea
 \ovA{\varphi}{}\un&:& \mathscr W\pl \M,\xi\pr\to \mathscr V\pl\M,\xi,\gamma\pr\nn\\
 &:&\pl {\Sigma}_{\mu\nu}, U_{\lambda|\mu\nu}\pr\mapsto \ovA S\lmn=\xi^\lambda {\Sigma}_{\mu\nu}+\ovA{h}{}^{\lambda \rho}\pl U_{\mu|\rho\nu}+U_{\nu|\rho\mu}+U_{\rho|\mu\nu}\pr\nn. 
  \eea

 \noindent We now construct the following map:
 \bea
 {\Theta}&:&\PC\times\mathscr D\pl\M,\xi,\gamma\pr\to \mathscr W\pl \M,\xi\pr\nn\\
 &:&\big( A,\Gamma\big)\mapsto\pl {\ovA{\Sigma}}_{\mu\nu}=-\nabla_{(\mu}A_{\nu)}+A_\lambda\, \Gamma\lmna, \ovA{U}_{\lambda|\mu\nu}=\gamma_{\lambda\rho}\, \Gamma^\rho_{[\mu\nu]}-\half A_{[\mu}\Lag_\xi\gamma_{\nu]\lambda}\pr\nn. 
 \eea
Note that the invariance of $A$ (\ie $\Lag_\xi A=0$) guarantees \cite{footnote25} that $\ovA{\Sigma}\in\field{T^*\M\otimes\Ann\xi}$ while the fact that $\ovA U\in\mathscr U\pl \M,\xi\pr$ is ensured by relation \eqref{eqtorsionconstraint}.  
Given a principal connection $A\in\PC$, the map $\ovA\Theta:= \Theta\big( A,\cdot\big):\mathscr D\pl\M,\xi,\gamma\pr\to \mathscr W\pl \M,\xi\pr$ can be shown to be an affine map modelled on the linear map $\ovA{\varphi}$. This fact ensures that $\Ker \ovA\Theta$ is spanned by a unique $\ovA{\Gamma}\in\mathscr D\pl\M,\xi,\gamma\pr$, called the {\it torsional special Carrollian connection} associated with $A$, whose Koszul formula reads:
 \bea
A\pl \ovA\nabla_XY\pr&=&X\crl A\pl Y\pr\crr-\half\, dA\pl X,Y\pr\\
2\,\gamma\pl\ovA\nabla_XY,Z\pr&=&X\crl \gamma\pl Y,Z\pr\crr+Y\crl \gamma\pl X,Z\pr\crr-Z\crl \gamma\pl X,Y\pr\crr\label{KformulaCarrollsptor}\\
&&+\gamma\pl\br{X}{Y},Z\pr-\gamma\pl\br{Y}{Z},X\pr-\gamma\pl\br{X}{Z},Y\pr\nn\\
&&-A\pl Y\pr\Lag_\xi\gamma\pl X,Z\pr\nn
\eea
for all $X,Y,Z\in\vf$ while an explicit expression of the connection coefficients in components is given by: 
 \bea
\ovA{\Gamma}{}\lmn=\xi^\lambda\p_{(\mu}A_{\nu)}+\half \ovA{h}{}^{\lambda\rho}\crl \p_\mu \gamma_{\rho\nu}+\p_\nu \gamma_{\rho\mu}-\p_\rho \gamma_{\mu\nu}\crr-\half \ovA{h}{}^{\lambda\rho}A_{\nu}\Lag_\xi \gamma_{\mu\rho}. \nn
\eea
\noindent Whenever the underlying Carrollian structure is invariant (\ie $\Lag_\xi\gamma=0$), the last term vanishes and one recovers the expression \eqref{torsionlessspecialcarroll} of the torsionless special Carrollian connection associated with the principal connection $A\in\PC$. 

\noindent Given a principal connection $A\in\PC$, the associated torsional special Carrollian connection can be chosen as origin of $\mathscr D\pl\M,\xi,\gamma\pr$ and can thus be used in order to represent any torsional Carrollian {connection: 
 \bea
A\pl \nabla_XY\pr&=&X\crl A\pl Y\pr\crr-\half\, dA\pl X,Y\pr+\ovA\Sigma\pl X,Y\pr\\
2\,\gamma\pl\nabla_XY,Z\pr&=&X\crl \gamma\pl Y,Z\pr\crr+Y\crl \gamma\pl X,Z\pr\crr-Z\crl \gamma\pl X,Y\pr\crr\label{KformulaCarrolltor}\\
&&+\gamma\pl\br{X}{Y},Z\pr-\gamma\pl\br{Y}{Z},X\pr-\gamma\pl\br{X}{Z},Y\pr\nn\\
&&-A\pl Y\pr\Lag_\xi\gamma\pl X,Z\pr\nn\\
&&+2\crl \ovA U\pl X|Z,Y\pr+\ovA U\pl Y|Z,X\pr+\ovA U\pl Z|X,Y\pr\crr\nn
\eea
for all $X,Y,Z\in\vf$. The arbitrariness is encoded in both tensors $\ovA\Sigma\in\field{T^*\M\otimes\Ann\xi}$ and $\ovA U\in\mathscr U\pl \M,\xi\pr$. The tensor $\ovA\Sigma$ characterises the obstruction to the preservation of the principal connection $A$ by the Carrollian connection $\nabla$ while the tensor $\ovA U$ can be thought of as the part of the torsion tensor that is not constrained by relation \eqref{eqtorsionconstraint}. 

\noindent In components, the most general torsional Carrollian connection compatible with $\mathscr C\pl \M, \xi, \gamma\pr$ thus {reads \cite{footnote26}:
 \bea
{\Gamma}{}\lmn&=&\xi^\lambda\p_{(\mu}A_{\nu)}+\half \ovA{h}{}^{\lambda\rho}\crl \p_\mu \gamma_{\rho\nu}+\p_\nu \gamma_{\rho\mu}-\p_\rho \gamma_{\mu\nu}\crr-\half \ovA{h}{}^{\lambda\rho}A_{\nu}\Lag_\xi \gamma_{\mu\rho}\nn\\
&&+\, \xi^\lambda {\ovA\Sigma}_{\mu\nu}+\ovA{h}{}^{\lambda \rho}\pl \ovA{U}_{\mu|\rho\nu}+\ovA{U}_{\nu|\rho\mu}+\ovA{U}_{\rho|\mu\nu}\pr\label{Carrconntor2}. 
\eea
The} transformation law of the map $\ovA{\Theta}$ under a Carroll boost induces an action of $\fieldinv{\Ann \xi}$ on $\PC\times\Big( \mathscr W\pl \M,\xi\pr\Big)$ as:
\bea
\begin{cases}
A'=A+\alpha\nn\\
\ovA\Sigma{}'_{\mu\nu}=\ovA\Sigma_{\mu\nu}-\half \Lag_{\alpha^\#{}}\gamma_{\mu\nu}-\half\alpha^\#{}^\rho A_\nu\Lag_\xi\gamma_{\mu\rho}+\alpha^\#{}^\rho\pl \ovA{U}_{\mu|\rho\nu}+\ovA{U}_{\nu|\rho\mu}+\ovA{U}_{\rho|\mu\nu}\pr\\
\ovA{U}{}'_{\lambda|\mu\nu}=\ovA{U}_{\lambda|\mu\nu}-\half \alpha_{[\mu}\Lag_\xi\gamma_{\nu]\lambda}
\end{cases}\,
\eea 
where $\alpha^\#{}:= \ovA{h}\pl\alpha\pr\in\field{T\M}$.

\bprop{\label{corCarrolliantor}The space of torsional Carrollian connections compatible with a given invariant Carrollian structure possesses the structure of an affine space canonically isomorphic to the affine space of orbits 
$$\frac{\PC\times\Big( \mathscr W\pl \M,\xi\pr\Big)}{\fieldinv{\Ann\xi}}\cong \mathscr D\pl \M, \xi, \gamma\pr\,.$$ 
}

\pagebreak

\bibliographystyle{unsrt.bst}

\begin{thebibliography}{100}
\bibitem{footnote3} {The term (Galilean, Carrollian, \etcp) ``manifold'' will stand for a smooth manifold endowed with a metric structure and a compatible connection. }
\bibitem{footnote4} {We denote $\mathfrak{igl}\pl d+2\pr=\mathfrak{gl}\pl d+2\pr\inplus\mathbb R^{d+2}$ the affine Lie algebra in $d+2$ dimensions. }
\bibitem{footnote6} {The leaf $t=0$ is special in that only the homogeneous Carroll algebra acts faithfully.}
\bibitem{footnote7} {Actually, when dealing with non-Riemannian geometries, the definition of the vector space $\mathscr W(\M)$ will sometimes necessitate a piece of the metric structure (\eg $\mathscr W(\M,\psi)$ for Galilean connections and $\mathscr W(\M,\xi)$ for Carrollian and ambient Galilean connections). }
\bibitem{footnote9} {Indeed, an important discrepancy between Riemannian and non-Riemannian geometries lies in the fact that, in the latter case, only a subset of ``metric'' structures admits torsionfree compatible connections (\eg Leibnizian structures with closed absolute clock in the Galilean case or invariant Carroll structures in the Carrollian case). }
\bibitem{footnote16} {One can go one step further and package tensors $\Sigma$ and $\mathcal U$ into a single object $T$ possessing the symmetries of a torsion tensor, just like in the relativistic case. Explicitly, one constructs an isomorphism between $\mathscr W\pl \M,\xi\pr$ and $\field{T^*\M\otimes \w^2T^*\M}$ as:
\bea
\ovA{\phi}&:&\mathscr W\pl \M,\xi\pr\to \field{T^*\M\otimes \w^2T^*\M}\label{varphiAleib32}\\
&:&\pl {\Sigma}_{\mu\nu}, {\mathcal U}_{\lambda|\mu\nu}\pr\mapsto \ovA T_{\lambda|\mu\nu}:=\mathcal U_{\lambda|\mu\nu}+A_\lambda \Sigma_{[\mu\nu]}+A_{[\mu}\Sigma_{\nu]\lambda}\nn
\eea
whose inverse isomorphism reads
\bea
\ovA{\phi}{}\un&:&\field{T^*\M\otimes \w^2T^*\M}\to \mathscr W\pl \M,\xi\pr\\
&:&T_{\lambda|\mu\nu}\mapsto  \Big(\,{\ovA\Sigma}_{\mu\nu}:=\xi^\lambda\crl {T}_{\mu|\lambda\nu}+{T}_{\nu|\lambda\mu}+{T}_{\lambda|\mu\nu}\crr, \nn\\
&& \hspace{2cm}\ovA{\mathcal U}_{\lambda|\mu\nu}:=T_{\lambda|\mu\nu}-\xi^\rho\crl A_\lambda T_{\rho|\mu\nu}+A_{[\mu}T_{\nu]|\rho\lambda}+A_{[\mu}T_{\rho|\nu]\lambda}-A_{[\mu}T_{\lambda|\nu]\rho}\crr\Big)\nn. 
\eea

\noindent Composing $\ovA\varphi$ and $\ovA\phi$ leads to a resolution of $\mathscr V\pl\M,\xi,\psi,\gamma\pr$. Explicitly, we define the isomorphism $\ovA{\Phi}:=\ovA\phi\circ \ovA\varphi$  reading as
\bea
\ovA{\Phi}&:&\mathscr V\pl\M,\xi,\psi,\gamma\pr\to \field{T^*\M\otimes \w^2T^*\M}\label{varphiAleib2}\\
&:&S\lmn\mapsto \ovA T_{\lambda|\mu\nu}:=(\N\gamma_{\lambda \rho}+A_\lambda A_\rho)S^\rho_{[\mu\nu]}+A_\rho A_{[\mu}S^\rho_{\nu]\lambda}+\psi_\lambda \N\gamma_{\sigma[\mu}S^\sigma_{\nu]\rho}N^\rho\nn
\eea
where $N\in\FO$ is an arbitrary field of observers. The inverse isomorphism takes the form
\bea
\ovA{\Phi}{}\un&:&\field{T^*\M\otimes \w^2T^*\M}\to \mathscr V\pl\M,\xi,\psi,\gamma\pr\\
&:&T_{\lambda|\mu\nu}\mapsto \ovA S\lmn=\ovA P{}^\sigma_\nu\, (\ovA h{}^{\lambda\rho}+\xi^\lambda\xi^\rho)\crl {T}_{\mu|\rho\sigma}+{T}_{\sigma|\rho\mu}+{T}_{\rho|\mu\sigma}\crr\,\nn.
\eea
}
\bibitem{footnote17} {Note that the tensor ${\mathcal U}_{\lambda|\mu\nu}$ in \eqref{varphiAleib} is independent of the choice of field of observers $N$, hence the absence of superscript on $\mathcal U$. }
\bibitem{footnote18} {Similarly to the torsionfree case, one can relax the invariance condition on the Leibnizian pair and define the following affine map: 
\bea
\hspace{-2cm}
{\Theta}&:&\LF\times\mathscr D\pl\M,\xi,\psi,\gamma\pr\to\mathscr W\pl \M,\xi\pr\nn\\
&:&\big( N,A,\Gamma\big)\mapsto\Big(\ovA{\Sigma}_{\mu\nu}=-\nabla_{(\mu}A_{\nu)}+A_\lambda\, \Gamma\lmna+A_{(\mu}\Lag_\xi A_{\nu)}, \label{Thetanoninv}\\
&&\hspace{2.25cm}\NA {\mathcal U}_{\lambda|\mu\nu}=\N{\gamma}_{\lambda\rho}\Gamma^{\rho}_{[\mu\nu]}-\half A_{[\mu}\overset{L}{P}{}^{\rho}_{\nu]} \overset{L}{P}{}^{\sigma}_\lambda\Lag_\xi\N\gamma_{\rho\sigma}+\psi_\lambda\crl\N{\gamma}_{\sigma[\mu}\nabla_{\nu]}N^\sigma-\N{\gamma}_{\rho[\mu}A_{\nu]}\Lag_\xi N^\rho\crr \Big)\nn. 
\eea

}
\bibitem{footnote20} {Recall that in the Lorentzian case, the arbitrariness is encoded in the unconstrained torsion tensor \ie in a section of the vector bundle $\pl{\w^2T^*\M\otimes T\M}\pr$ whose fibers have dimension $\frac{\pl d+1\pr\pl d+2\pr^2}{2}$. }
\bibitem{footnote22} {Note that the invariance of $\gamma$ and $N$ ensure $\Lag_\xi\N\gamma=0$, so that the second component of $\NA\Theta$ does not depend on $A$ and will then be denoted $\N {\mathcal U}$. }
\bibitem{footnote23} {If one considers a generic Ehresmann connection $A\in\EC$, the map $\ovA\Theta$ becomes $\ovA\Theta:\Gamma\mapsto\ovA\Sigma_{\mu\nu}=-\nabla_{(\mu}A_{\nu)}+A_{(\mu}\Lagxi A_{\nu)}$ while the generic torsionfree Carrollian connection reads:
\bea
\Gamma\lmn=\xi^\lambda\p_{(\mu}A_{\nu)}+\half \ovA{h}{}^{\lambda\rho}\crl \p_\mu \gamma_{\rho\nu}+\p_\nu \gamma_{\rho\mu}-\p_\rho \gamma_{\mu\nu}\crr-\xi^\lambda A_{(\mu}\Lagxi A_{\nu)}+\xi^\lambda \ovA{\Sigma}_{\mu\nu}. \nn
\eea
}
\bibitem{footnote24} {\noindent One can go one step further and package tensors $\Sigma$ and $U$ in a single object $T$ by constructing an isomorphism between $\mathscr W\pl \M,\xi\pr$ and $\field{T^*\M\otimes \w^2T^*\M}$ as:
\bea
\ovA{\phi}&:&\mathscr W\pl \M,\xi\pr\to \field{T^*\M\otimes \w^2T^*\M}\label{varphiAleib3}\\
&:&\pl {\Sigma}_{\mu\nu}, {U}_{\lambda|\mu\nu}\pr\mapsto \ovA T_{\lambda|\mu\nu}:=U_{\lambda|\mu\nu}+A_\lambda \Sigma_{[\mu\nu]}+A_{[\mu}\Sigma_{\nu]\lambda}\nn
\eea
whose inverse isomorphism reads
\bea
\ovA{\phi}{}\un&:&\field{T^*\M\otimes \w^2T^*\M}\to \mathscr W\pl \M,\xi\pr\\
&:&T_{\lambda|\mu\nu}\mapsto  \Big(\,{\ovA\Sigma}_{\mu\nu}:=\xi^\lambda\crl {T}_{\mu|\lambda\nu}+{T}_{\nu|\lambda\mu}+{T}_{\lambda|\mu\nu}\crr\nn\\
&& \hspace{2cm}\ovA{U}_{\lambda|\mu\nu}:=T_{\lambda|\mu\nu}-\xi^\rho\crl A_\lambda T_{\rho|\mu\nu}+A_{[\mu}T_{\nu]|\rho\lambda}+A_{[\mu}T_{\rho|\nu]\lambda}-A_{[\mu}T_{\lambda|\nu]\rho}\crr\Big)\nn. 
\eea

Composing $\ovA\varphi$ and $\ovA\phi$ leads to a resolution of $\mathscr V\pl\M,\xi,\gamma\pr$. Explicitly, we define the isomorphism $\ovA{\Phi}:=\ovA\phi\circ \ovA\varphi$  reading as:
\bea
\ovA{\Phi}&:&\mathscr V\pl\M,\xi,\gamma\pr\to \field{T^*\M\otimes \w^2T^*\M}\label{varphiAleib22}\\
&:&S\lmn\mapsto \ovA T_{\lambda|\mu\nu}:=(\N\gamma_{\lambda \rho}+A_\lambda A_\rho)S^\rho_{[\mu\nu]}+A_\rho A_{[\mu}S^\rho_{\nu]\lambda}\nn
\eea
while the inverse isomorphism takes the form:
\bea
\ovA{\Phi}{}\un&:&\field{T^*\M\otimes \w^2T^*\M}\to \mathscr V\pl\M,\xi,\gamma\pr\\
&:&T_{\lambda|\mu\nu}\mapsto \ovA S\lmn=\ovA P{}^\sigma_\nu\, (\ovA h{}^{\lambda\rho}+\xi^\lambda\xi^\rho)\crl {T}_{\mu|\rho\sigma}+{T}_{\sigma|\rho\mu}+{T}_{\rho|\mu\sigma}\crr\,\nn.
\eea
 
 }
\bibitem{footnote25} {If one insists in dealing with a generic (\ie not necessarily invariant) Ehresmann connection, the previous map can be modified as:
 \bea
 {\Theta}&:&\EC\times\mathscr D\pl\M,\xi,\gamma\pr\to \mathscr W\pl \M,\xi\pr\nn\\
 &:&\big( A,\Gamma\big)\mapsto\pl {\ovA{\Sigma}}_{\mu\nu}=-\nabla_{(\mu}A_{\nu)}+A_\lambda\, \Gamma\lmna+A_{(\mu}\Lag_\xi A_{\nu)}, \ovA{U}_{\lambda|\mu\nu}=\gamma_{\lambda\rho}\, \Gamma^\rho_{[\mu\nu]}-\half A_{[\mu}\Lag_\xi\gamma_{\nu]\lambda}\pr\nn. 
 \eea
}
\bibitem{footnote26} {If one relaxes the invariance condition on the Ehresmann connection $A$, the most general torsional Carrollian connection compatible with $\mathscr C\pl \M, \xi, \gamma\pr$ can be expressed as
 \bea
{\Gamma}{}\lmn&=&\xi^\lambda\p_{(\mu}A_{\nu)}+\half \ovA{h}{}^{\lambda\rho}\crl \p_\mu \gamma_{\rho\nu}+\p_\nu \gamma_{\rho\mu}-\p_\rho \gamma_{\mu\nu}\crr-\xi^\lambda A_{(\mu}\Lag_\xi A_{\nu)}-\half \ovA{h}{}^{\lambda\rho}A_{\nu}\Lag_\xi \gamma_{\mu\rho}\nn\\
&&+\, \xi^\lambda {\ovA\Sigma}_{\mu\nu}+\ovA{h}{}^{\lambda \rho}\pl \ovA{U}_{\mu|\rho\nu}+\ovA{U}_{\nu|\rho\mu}+\ovA{U}_{\rho|\mu\nu}\pr\nn. 
\eea
 }
\end{thebibliography}

\providecommand{\href}[2]{#2}\begingroup\raggedright

\newpage{\pagestyle{empty}\cleardoublepage}

\end{document}